\documentclass[aip,jmp,12pt,floatfix]{revtex4-1}

\usepackage{graphicx}
\usepackage{xcolor}
\usepackage{amsmath,amssymb}
\usepackage{ulem}

\newcommand{\Ham}{H}
\newcommand{\Heff}{\Ham_\mathrm{eff}}
\newcommand{\Geff}{G_\mathrm{eff}}
\newcommand{\define}{:=}
\newcommand{\bra}{\langle}
\newcommand{\ket}{\rangle}

\begin{document}

\title{Time-reversal symmetric resolution of unity  without background integrals in open quantum systems}
\author{Naomichi Hatano}
\email{hatano@iis.u-tokyo.ac.jp}
\affiliation{Institute of Industrial Science, University of Tokyo, 4-6-1 Komaba, Meguro, Tokyo 153-8505, Japan}
\affiliation{Department of Physics, Washington University, One Brookings Drive, Saint Louis, Missouri 63130}
\author{Gonzalo Ordonez}
\email{gordonez@butler.edu}
\affiliation{Department of Physics and Astronomy, Butler University, Gallahue Hall, 4600 Sunset Ave, Indianapolis, Indiana 46208}
\date{May 23rd, 2014}

\begin{abstract}
We present a new  complete set of states for a class of open quantum systems, to be used in expansion of the Green's function and the time-evolution operator.
A remarkable feature of the  complete set is that it observes  time-reversal symmetry in the sense that it contains decaying states (resonant states) and growing states (anti-resonant states) parallelly.
We can thereby pinpoint the occurrence  of the breaking of time-reversal symmetry at the choice of 
whether we solve  Schr\"{o}dinger equation as an initial-condition problem or a  terminal-condition problem.
Another feature of the complete set is that in the subspace of the central scattering area of the system,  it consists of contributions of all states with point spectra but does not contain any background integrals.
In computing the time evolution, we can clearly see contribution of which point spectrum produces which time dependence.
In the whole  infinite state space, the  complete set does contain an integral but it is over \textit{un}perturbed eigenstates of the environmental area of the system and hence can be calculated analytically.
We demonstrate the usefulness of the  complete set by computing explicitly the survival probability and the escaping probability   as well as the dynamics of wave packets.
The origin of each term of  matrix elements is clear in our formulation, particularly the exponential decays due to the resonance poles.
\end{abstract}

\keywords{Open quantum system; Resonant state; Complete set; Green's function expansion; Time-reversal symmetry breaking; Arrow of time; Tight-binding model; Quadratic eigenvalue problem}

\maketitle

\section{Introduction}
\label{sec1}

Over the past few decades there has been significant progress in the understanding of  dissipative phenomena, both in classical and quantum mechanics. 
This has been  particularly achieved through the realization that dissipative parameters such as the lifetime of an unstable particle or the diffusion coefficient of Brownian motion are ultimately connected, at the most basic level, to complex eigenvalues of the Hamiltonian or the Liouville operators \cite{Nakanishi58,Prigogine73,Sudarshan78,Bohm89,Tasaki91,Petrosky96,Petrosky97,Hatano10,Klaiman11}. 
These eigenvalues correspond to generalized eigenfunctions. Alternatively, the complex eigenvalues can be associated with eigenfunctions of an effective Hamiltonian or effective Liouvillian that are non-Hermitian operators \cite{Livshits56,Feshbach58,Feshbach62,Rotter91,Petrosky96,Petrosky97,Rotter09,Nakano11,Hatano13}.
(In some references\cite{Petrosky96,Petrosky97}, these non-Hermitian operators were called collision operators.)
Either way, the conclusion is that dissipative phenomena need \textit{not} be formulated as  approximations or coarse graining of basic dynamics;
 note that perturbation approximation often breaks the unitarity of the time-evolution operator and thereby breaks time-reversal symmetry.
Instead, they can be formulated in terms of complex eigenvalues of the basic dynamical operators, without resorting to approximations.

A deeper question is the origin of time-reversal symmetry breaking or irreversibility and how it is connected to time-reversible dynamics. 
Previous work has addressed this problem by introducing time-reversal symmetry breaking in the complete set of eigenfunctions of the dynamical operator (for example the Hamiltonian in quantum mechanics). 
This was achieved by starting with a set of eigenfunctions with continuous real eigenvalues, and then deforming the integration contour of the complete set to include complex eigenvalues on either the lower half plane  (resonant states  corresponding to future-oriented evolution) or on the upper half plane (anti-resonant states, corresponding to past-oriented evolution) \cite{Nakanishi58,Sudarshan78,Tasaki91,Ordonez01,Petrosky01}. 
The contour deformation, however, is not unique.
One can select certain complex eigenvalues by encircling them, while ignoring other complex eigenvalues. 
The remainder contour becomes   a ``background integral''.  
The resulting  spectrum of eigenvalues  thus includes both discrete  complex eigenvalues  and a continuum coming from the background integral. Since this separation into discrete eigenvalues and background integral is not unique, the physical interpretation of this construction is not very clear.

Another common view on irreversibility is that it is connected to the initial conditions of the system in question, or even the whole universe. 
It is assumed that the initial state is a state of very low entropy; 
the second law of thermodynamics then explains irreversibility. 
This view, however, does not make a precise connection with dynamics and again it relies on approximations or coarse graining in order to derive the second law.

In this paper we synthesize the views described above by formulating a complete set of eigenfunctions of the Hamiltonian for a class of quantum mechanical systems that include both resonant and anti-resonant states. 
Our complete set is explicitly time-reversal symmetric.  
We avoid the introduction of any arbitrary background integral, as we obtain a complete set of eigenfunctions corresponding to {\it all} the real and complex discrete eigenvalues (\textit{i.e.}\ all point spectra) of the Hamiltonian. 
We find that for a time-reversal symmetric condition at time $t=0$, time-reversal symmetry is broken for $t\ne 0$ depending on  the sign of $t$. 
Mathematically, this corresponds to choosing whether we solve an initial-condition problem or a terminal-condition problem, which may favor resonant or anti-resonant states, respectively. 

Moreover, we find that for certain specifically prepared conditions which are \textit{not} time-reversal symmetric, anti-resonant states dominate during a part of  the time evolution and resonant states dominate during another part. 
For example, by performing a momentum inversion on a wave function emitted from a quantum dot, the wave function will collect itself back into the quantum dot (like a movie played backwards, showing water waves collecting themselves toward a point where a rock was dropped). 
During this period anti-resonant states dominate. 
Subsequently the wave function  is re-emitted, a process during which resonant states dominate.   
By maintaining time-reversal symmetry in the set of eigenfunctions, the selection of future-oriented resonant states or past-oriented anti-resonant states is uniquely determined by the overlap between either the initial or terminal conditions and the discrete eigenfunctions of the Hamiltonian.  

Our set of time-reversal symmetric eigenstates is limited to a class of quantum systems with a tight-binding Hamiltonian. 
In different contexts, sets of time-reversal symmetric eigenstates have been presented  for scattering problems\cite{GarciaCalderon10}  and for the Friedrichs-Lee model\cite{Kim14}. 
However, our formulation, based on the solution of a quadratic eigenvalue problem, is general enough that it can be extended to other systems, including systems considered in non-equilibrium statistical mechanics. 
Such extensions are left for future work. 
In the present paper we will focus on the tight-binding systems because we can then present the main ideas in enough detail that they can be subsequently generalized.

The outline of the present paper is as follows.
Sections~\ref{sec2} and~\ref{sec3} introduce basic concepts and models which we use throughout the paper.
Section~\ref{sec4} presents an overview of the results of the paper.
Sections~\ref{sec5}--\ref{sec10} give the derivation of the results step by step in details.
Using the results, we calculate the survival probability in Sec.~\ref{sec11} and the escaping probability in Sec.~\ref{sec12}.
Finally in Secs.~\ref{sec13} and~\ref{sec14}, we analyze dynamics of wave packets, breaking them down into resonant and anti-resonant states.
Section~\ref{sec15} summarizes the paper.

\section{Resonant state as an eigenstate that breaks the time reversal symmetry: a short review}
\label{sec2}

In this section, we present a concise review of the resonant state as an eigenstate of the Schr\"{o}dinger equation.
Let us consider for the moment the standard Schr\"{o}dinger equation in a one-dimensional space with a real potential on a compact support around the origin:
\begin{align}\label{eq2-10}
\left(-\frac{d^2}{dx^2}+V(x)\right)\psi(x)=E\psi(x),
\end{align}
where $V(x)\in\mathbb{R}$ and $V(x)=0$ for $|x|>L$.
(We will switch to the tight-binding model on a discretized space in the next section.)
Note that Eq.~\eqref{eq2-10} observes the time-reversal symmetry;
external magnetic fields are absent.
The wave function $\psi(x)$ for a real eigenvalue therefore can be made a real function.

\begin{figure}
\centering
\includegraphics[width=0.3\textwidth]{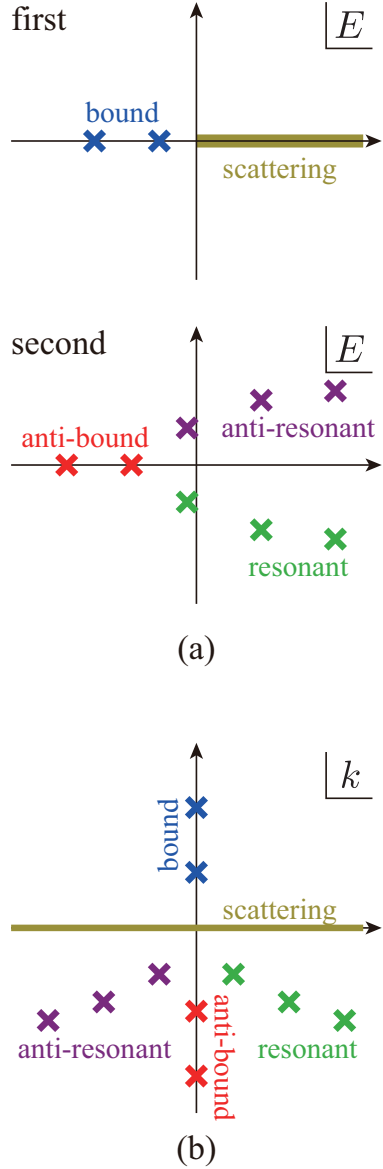}
\caption{The positions of the discrete and continuous eigenvalues in (a) the first and second Riemann sheets of the complex $E$ plane and (b) the complex $k$ plane.}
\label{fig1}
\end{figure}

The Schr\"{o}dinger equation~\eqref{eq2-10} has eigenvalues with point spectra and those with a continuous spectrum; see Fig.~\ref{fig1}.
Let us refer to the former as the discrete eigenstates and the latter as the continuous eigenstates.
The discrete eigenstates have four types, namely the bound states, the anti-bound states, the resonant states and the anti-resonant states. 
All of them are given by the wave function under the Siegert boundary condition\cite{Gamow28,Siegert39,Peierls59,leCouteur60,Zeldovich60,Hokkyo65,Romo68,Berggren70,Gyarmati71,Landau77,Romo80,Berggren82,Berggren96,Madrid05,Hatano08,GarciaCalderon10}
\begin{align}\label{eq2-30}
\psi(x)\propto e^{ik|x|}
\quad\mbox{for $|x|>L$}
\end{align}
with the eigenvalue $E=k^2$.
This is indeed equivalent to seeking the poles of the $S$-matrix.\cite{Landau77,Hatano08}

The Schr\"{o}dinger equation~\eqref{eq2-10}, although its Hamiltonian appears to be Hermitian, nonetheless can harbor complex eigenenergies because the system is open.
The Hamiltonian is Hermitian inside the Hilbert space but not outside it.\cite{Hatano08,Hatano13}
The eigenfunctions for the complex energy eigenvalues indeed diverge spatially and hence reside outside the Hilbert space.

Out of the four types of the discrete states, the bound states are located on the positive imaginary axis of the complex $k$ plane, \textit{i.e.}\ $k=i\kappa$ with $\kappa>0$, and hence are on the negative real axis of the first Riemann sheet of the complex $E$ plane.
Their eigenfunction decay exponentially in the form $\exp(-\kappa|x|)$, which can be confirmed by inserting $k=i\kappa$ into Eq.~\eqref{eq2-30}.
The bound states are the only discrete eigenstates that are inside the Hilbert space.

\begin{figure}
\includegraphics[width=0.4\textwidth]{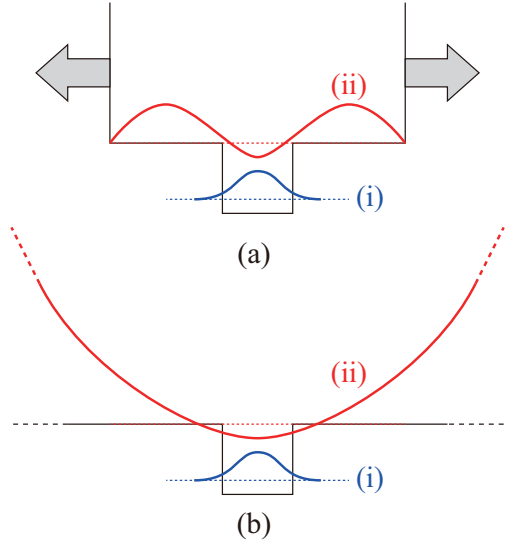}
\caption{(a) A closed system with a bound state~(i)  inside the potential and a bound state~(ii) outside it.
(b) When the boundaries are taken away so that the system may be open, the bound state~(i) remains a bound state, but the bound state~(ii) diverges spatially, turning into an anti-bound state.}
\label{fig2}
\end{figure}

The anti-bound states are on the negative imaginary axis of the complex $k$ plane, \textit{i.e.}\ $k=i\kappa$ with $\kappa<0$, and hence on the negative real axis of the second Riemann sheet of the complex $E$ plane.
Their eigenfunctions diverge spatially in the form $\exp(|\kappa||x|)$.
We may understand the origin of these states in the following way.
Consider a closed system where we have an attractive potential around the origin with infinitely high walls on the right and left boundaries (Fig.~\ref{fig2}).
We would have only bound states, some of which would be inside the range of the attractive potential but some outside, analogously to the bonding and anti-bonding orbitals of a chemical bonding.
If we move away the boundary walls to make the system open, the states outside the potential range would turn into the anti-bound states.
This is why the wave functions of the anti-bound states spatially diverge.

The resonant states are located in the fourth quadrant of the complex $k$ plane and hence in the lower half of the second Riemann sheet of the complex $E$ plane.
Note that the term ``resonant states" here refers to eigenstates of the time-independent Schr\"{o}dinger equation~\eqref{eq2-10};
it does not refer to resonant phenomena found in the time evolution of an incoming wave packet scattered by a trapping potential as a solution of the time-dependent Schr\"{o}dinger equation.
(We will analyze the time evolution of wave packets in Secs.~\ref{sec13} and~\ref{sec14}.)

\begin{figure}
\includegraphics[width=0.35\textwidth]{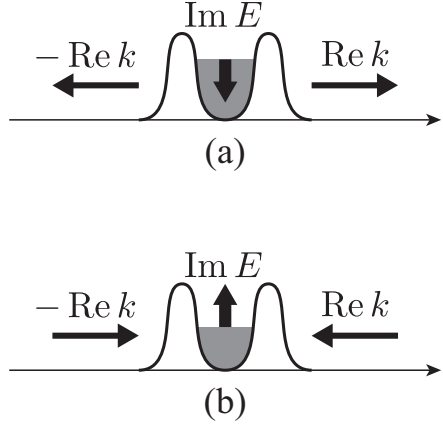}
\caption{(a) Visualization of a resonant state as an eigenstate of the static Schr\"{o}dinger equation.
The number of particles decreases exponentially in the trapping potential according to the negative imaginary part of the eigenenergy.
The corresponding amount of the particles leak from the potential according to the positive real part of the eigen-wave number.
(b) Visualization of an anti-resonant state, which is the time reversal of a resonant state.
Particles gather into the trapping potential according to the negative real part of the eigen-wave number.
The number of the particles thereby increases exponentially according to the positive imaginary part of the eigenenergy.}
\label{fig3}
\end{figure}

We can visualize the resonant states as eigenstates of the static Schr\"{o}dinger equation as shown in Fig.~\ref{fig3}(a).
On one hand, the state decays exponentially in time because the imaginary part of its eigenenergy $E$ is negative in the exponent of the time-evolution factor $e^{-iEt}$.
On the other hand, a resonant state has only out-going waves according to Eq.~\eqref{eq2-30} because the real part of its eigen-wave number $k$ is positive.
These two facts combined yield the view that the particles escape away from the trapping potential.
We can indeed prove the particle-number conservation on the basis of this view by noting the proportionality between the imaginary part of the eigenenergy and the real part of the eigen-wave number.\cite{Hatano08,Hatano09,Goldzak10}
Because the imaginary part of the eigen-wave number is negative, the eigenfunction spatially diverges.
This makes possible for the seemingly Hermitian Hamiltonian to have complex eigenvalues, as we explained above.
From a physical point of view, the divergence in space means that the particles eventually escape away from the potential.

The anti-resonant states are the time reversal of the resonant states as visualized in Fig.~\ref{fig3}(b).
An anti-resonant state has only in-coming waves because the real part of its eigen-wave number $k$ is negative.
The state grows exponentially in time because the imaginary part of its eigenenergy $E$ is positive in the exponent of $e^{-iEt}$.
We thereby have the view that the particles are injected into the trapping potential.

We stress here that each of resonant and anti-resonant states has an arrow of time, breaking the time-reversal symmetry.
A resonant state and an anti-resonant state always appear as a complex conjugate pair, together recovering the time-reversal symmetry that the original Schr\"{o}dinger equation observes.
We can therefore regard the appearance of the pair of time-reversal asymmetric states out of a time-reversal symmetric equation as the seed of spontaneous breaking of time reversal symmetry.

A key feature of our expansion of the Green's function and the time-evolution operator, which we will present below, is that it contains the resonant and anti-resonant states parallelly, thereby retaining the time-reversal symmetry until the last moment.
Mathematically, the time-reversal symmetry is finally broken when we choose to observe the time evolution from an initial condition or the one towards a terminal condition.

\section{System in question}
\label{sec3}

\begin{figure}
\centering
\includegraphics[width=0.45\textwidth]{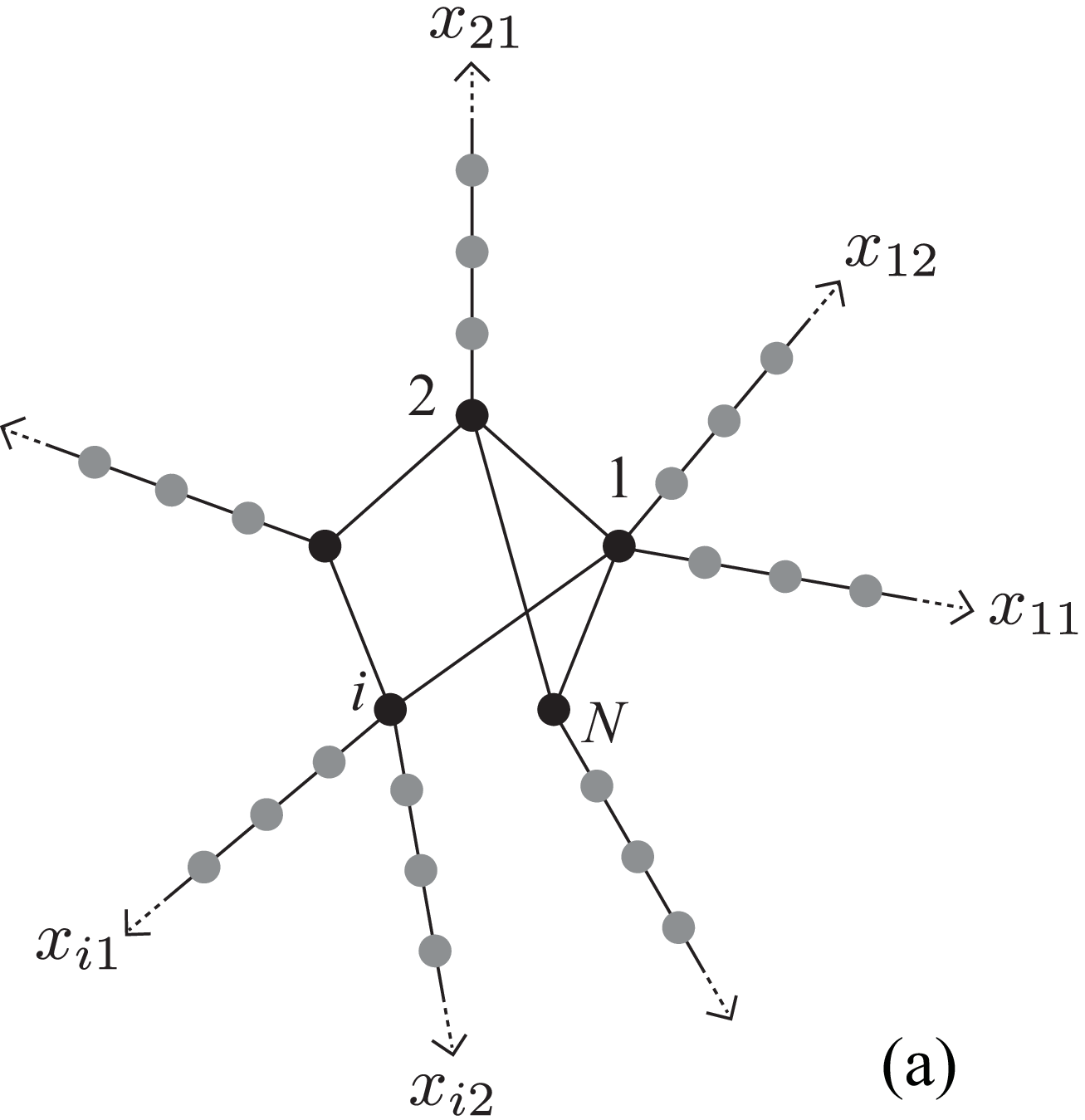}
\\
\vspace{\baselineskip}
\includegraphics[width=0.6\textwidth]{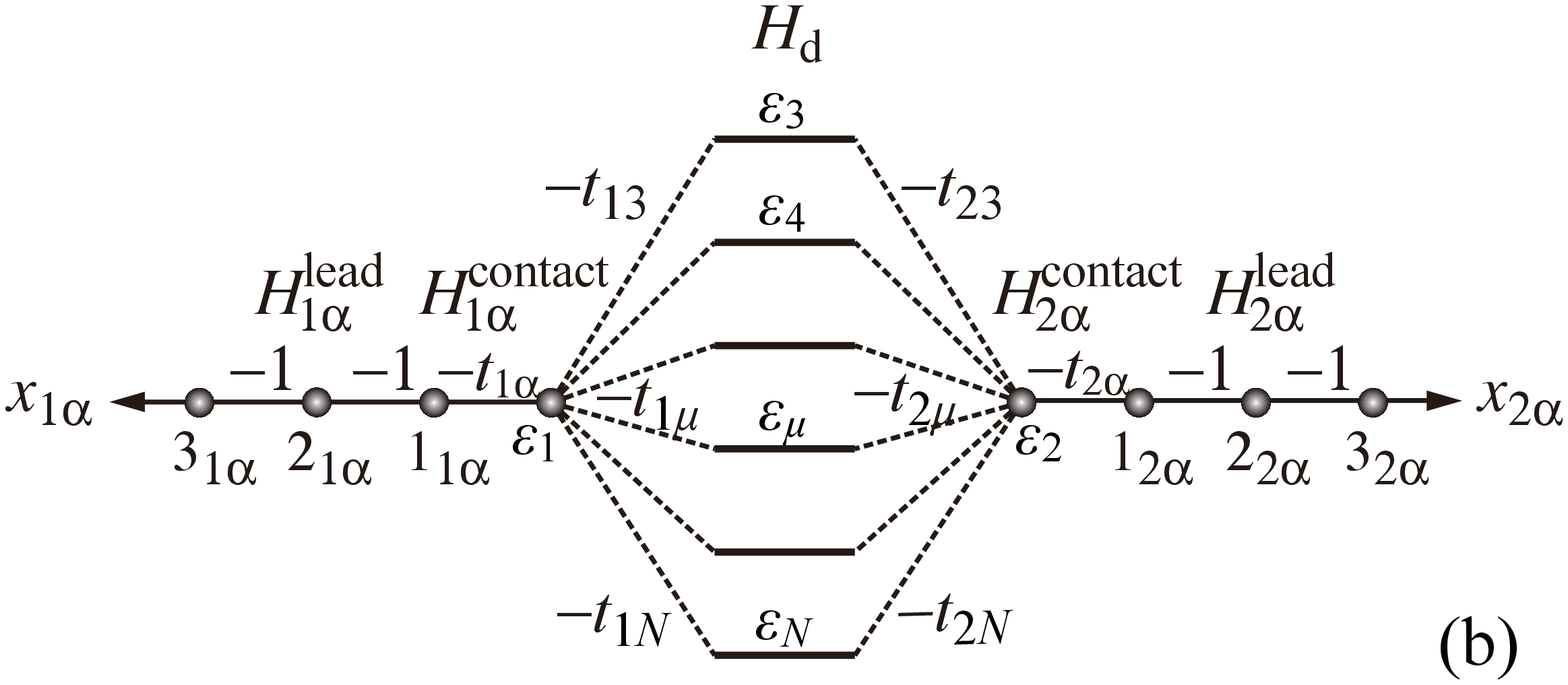}
\caption{(a) A schematic view of the general system which we consider in the present  paper.
(b) An $(N-2)$-level quantum dot with two quantum wires. The model (a) can be a prototype of the system (b).}
\label{fig4}
\end{figure}

In this section, we introduce the system in question.
We consider a class of open quantum systems consisting of a dot of $N$ sites with  semi-infinite leads attached to some of the dot sites (Fig~\ref{fig4}(a)). 
Note that the leads must be semi-infinite in order for us to have resonant states;
otherwise the leak shown in Fig.~\ref{fig3}(a) would come back to the trapping potential and destroy the resonant states.
Any number of leads can be attached to any sites of the dot.
Let the label $i\alpha$ denote the $\alpha$th lead attached to the $i$th site of the dot, $d_i$.
The Hamiltonian which we consider here consists of three parts; namely, the dot Hamiltonian, the lead Hamiltonian and the contact Hamiltonian:
\begin{align}\label{eq10}
\Ham\define\Ham_\mathrm{d}+\Ham_\mathrm{leads}+\Ham_\mathrm{contacts}.
\end{align}

The first term is the tight-binding Hamiltonian inside the dot:
\begin{align}\label{eq40}
\Ham_\mathrm{d}\define
-\sum_{\substack{i,j=1 \\ i\neq j}}^Nt_{ij}|d_i\ket\bra d_j| + \sum_{i=1}^N \varepsilon_i |d_i\ket\bra d_i|,
\end{align}
where $|d_i\ket$ denotes the site basis of the $i$th site of the dot and $t_{ij}=t_{ji}\in\mathbb{R}$, so that we have no magnetic fields and the time reversal symmetry is not explicitly broken.
The chemical potentials $\varepsilon_i$ at the dot sites are all real, $\varepsilon_i\in\mathbb{R}$, and hence the dot Hamiltonian $\Ham_\mathrm{d}$ is real symmetric under the basis set of $\{|d_i\ket|i=1,2,\ldots,N\}$.

The second term of~\eqref{eq10} is the tight-binding Hamiltonian for the semi-infinite leads:
\begin{align}
\Ham_\mathrm{leads}&\define\sum_{i=1}^N\sum_{{\alpha}=1}^{n_i}\Ham_{i\alpha}^{\mathrm{lead}},
\\
\Ham_{i\alpha}^{\mathrm{lead}}&\define
-t_\mathrm{lead}\sum_{x_{i\alpha}=1}^\infty \left(
|x_{i\alpha}+1\ket\bra x_{i\alpha}|+|x_{i\alpha}\ket\bra x_{i\alpha}+1|\right)
\end{align}
where $\Ham_{i\alpha}^\mathrm{lead}$ is the tight-binding Hamiltonian of the $\alpha$th of all $n_i$ leads that are attached to the $i$th site of the dot, $|x_{i\alpha}\ket$  with $1\leq x_{i\alpha}<\infty$ denotes the basis of a site on the lead and the hopping amplitude  $t_\mathrm{lead}\in\mathbb{R}$ is common to all leads.
We hereafter put 
\begin{align}
t_\mathrm{lead}=1,
\end{align}
making it the unit of the energy.
We have also put the lattice constant to unity.
The dispersion relation on a lead is therefore given by
\begin{align}\label{eq60}
E=-(e^{ik}+e^{-ik})=-2\cos k,
\end{align}
where $k$ is the wave number limited to the first Brillouin zone
\begin{align}
-\pi\leq k\leq \pi
\end{align}
because the leads consist of regular lattices.

The last term of~\eqref{eq10} is the tight-binding coupling Hamiltonian between the dot and the leads:
\begin{align}
\Ham_\mathrm{contacts}&\define\sum_{i=1}^N\sum_{\alpha=1}^{n_i}\Ham_{i\alpha}^{\mathrm{contact}},
\\
\Ham_{i\alpha}^{\mathrm{contact}}&\define
-t_{i\alpha}\left(|1_{i\alpha}\ket\bra d_i|+|d_i\ket\bra 1_{i\alpha}|\right),
\end{align}
where $1_{i\alpha}$ denotes the end site of the lead which directly couples to the dot site $d_i$ and $t_{i\alpha}\in\mathbb{R}$.

This Hamiltonian can be a prototype model of various open quantum systems.
For example, the model in Fig.~\ref{fig4}(a) can describe an $(N-2)$-level quantum dot which is connected to two quantum wires;\cite{Sasada11} see Fig.~\ref{fig4}(b).

We next define projection operators which separate the dot space and the lead space:
\begin{align}\label{eq80}
P&\define \sum_{i=1}^N|d_i\ket\bra d_i|,
\\
Q&\define I_\infty-P=\sum_{i=1}^N\sum_{\alpha=1}^{n_i}\sum_{x_{i\alpha}=1}^\infty|x_{i\alpha}\ket\bra x_{i\alpha}|,
\end{align}
where $I_\infty$ is the identity operator in the whole space spanned by all site bases.
Note the relations
\begin{align}
PHP&=H_\mathrm{d},
\\
PHQ+QHP&=H_\mathrm{contacts},
\\
QHQ&=H_\mathrm{leads}.
\end{align}
As long as we restrict ourselves to the $P$ subspace, which is spanned by the dot sites $\{|d_i\ket\}$, the $P$ operator~\eqref{eq80} is the $N\times N$ identity matrix, and therefore we will refer to it as $I_N$ where appropriate.

As we will show below in Secs.~\ref{sec6} and~\ref{sec7}, this open system generally has $2N$ pieces of discrete states with point spectra, which contain bound, anti-bound, resonant and anti-resonant states, as was reviewed in Sec.~\ref{sec2}.
The locations of the discrete states are modified from the ones indicated in Fig.~\ref{fig1} for the spatially continuous models to the ones indicated in Table~\ref{tab1} for the  tight-binding models.
\begin{table}
\caption{The positions of the discrete states (the states with point spectra) in the complex $E$ plane, the complex $k$ plane and the complex $\lambda$ plane for  tight-binding systems.
See Fig.~\ref{fig5} for the symbols.}
\label{tab1}
\vspace{\baselineskip}
\begin{tabular}{lc}
\hline
state & $E_n$, $k_n$ and $\lambda_n$ \\
\hline\hline
bound & on the lines AC or BD \\
\hline
anti-bound & on the lines AE or BF \\
\hline
resonant & in the plane $\delta$ \\
\hline
anti-resonant & in the plane $\gamma$ \\
\hline
\end{tabular}
\end{table}
\begin{figure}
\centering
\includegraphics[width=0.3\textwidth]{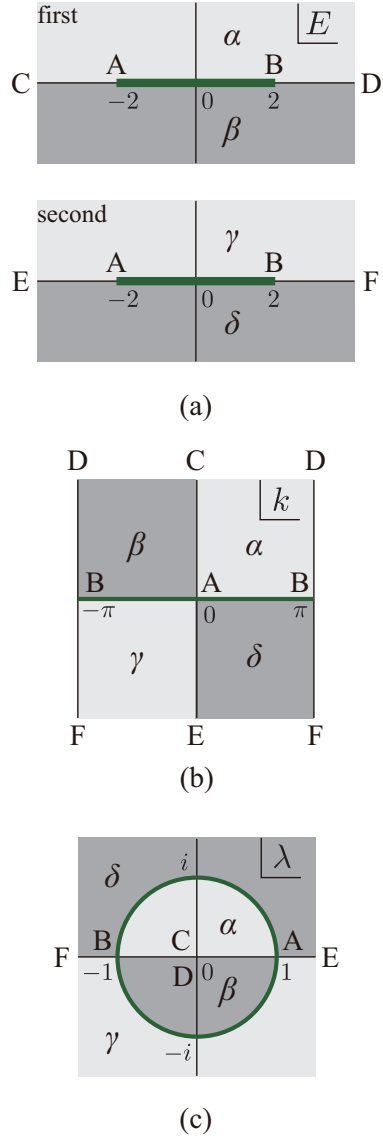}
\caption{The correspondence among the three complex planes: (a) the complex $E$ plane with the first and second Riemann sheets; (b) the complex $k$ plane; (c) the complex $\lambda$ plane.
Parts of the planes with the same index $\alpha$, $\beta$, $\gamma$, or $\delta$ correspond to each other.
The points with the same index A, B, C, D, E, or F correspond to each other.
The segment AB is the cosine energy band, which connects the two Riemann sheets as a branch cut.
The band edges A and B are the branch points.}
\label{fig5}
\end{figure}

Because the present system has the time-reversal symmetry, every term in the Hamiltonian~\eqref{eq10} can be expressed as a real symmetric matrix.
We can thereby obtain the following relations for the discrete eigenstates.
We can express the states with real eigenvalues, namely the bound and anti-bound states, as real vectors.
In other words, these states do not break the time-reversal symmetry, which the Schr\"{o}dinger equation observes.
We can therefore transpose the Schr\"{o}dinger equation $H|\psi_n\ket=E_n|\psi_n\ket$ to have
\begin{align}\label{eq171}
|\psi_n\ket^T H=E_n|\psi_n\ket^T.
\end{align}
We thereby find that the left-eigenvector is not only the Hermitian conjugate but the real transpose of the right-eigenvector:
\begin{align}
\bra\tilde{\psi}_n|\define|\psi_n\ket^T=|\psi_n\ket^\dag=:\bra\psi_n|
\quad\mbox{for $n\in$ bound or anti-bound states.}
\end{align}
We here unnecessarily added the tilde symbol to the left-eigenvector on the left in order to unify the symbol for the resonant and anti-resonant states, for which the left-eigenvectors are not the Hermitian conjugate of the right-eigenvector, as we will show now.
Wherever appropriate, we also use the symbol without the tilde for the left-eigenvectors, which we let denote the Hermitian conjugate of the right-eigenvectors.

Indeed, the right-eigenvectors for the resonant and anti-resonant states are not generally real vectors because their eigenvalues are complex.
In other words, these states break the time-reversal symmetry.
Since the anti-resonant state is the time reversal of the corresponding resonant state, we have 
\begin{align}\label{eq:n-nbar}
E_{\bar{n}}={E_n}^\ast
\quad\mbox{and}\quad
|\psi_{\bar{n}}\ket=|\psi_n\ket^\ast
\quad\mbox{for $n\in$ resonant states and $\bar{n}\in$ anti-resonant states,}
\end{align}
where we let the subscript $\bar{n}$ denote the anti-resonant state that corresponds to the resonant state labeled by $n$.
The pair of a resonant state and its complex conjugate anti-resonant state $(n,\bar{n})$ recovers the time-reversal symmetry of the whole system of solutions.
Because of Eq.~\eqref{eq171}, the left-eigenvector of each resonant or anti-resonant state is real transpose but \textit{not} the Hermitian conjugate of the right-eigenvector:\cite{Zeldovich60,Hokkyo65,Romo68,Moiseyev11}
\begin{align}\label{eq201}
\bra\tilde{\psi}_n|=|\psi_n\ket^T\neq|\psi_n\ket^\dag
\quad\mbox{for $n\in$ resonant or anti-resonant.}
\end{align}
We thereby arrive at the relations
\begin{align}\label{eq:n-nbar1}
\bra\tilde{\psi}_n|&=|\psi_{\bar{n}}\ket^\dag,
\\\label{eq:n-nbar2}
\bra\tilde{\psi}_{\bar{n}}|&=|\psi_n\ket^\dag
\quad\mbox{for $n\in$ resonant states and $\bar{n}\in$ anti-resonant states;}
\end{align}
see Appendix~A of Ref.~\cite{Sasada11} for the relations with and without magnetic fields.

\section{Overview of the results}
\label{sec4}

Let us here present the results that we will prove below  for the class of open quantum systems defined in Sec.~\ref{sec3}.
The main result in the present paper is  the new resolution of unity in the form
\begin{align}\label{eq179}
I_N&=\sum_{n=1}^{2N}P|\psi_n\ket\bra\tilde{\psi}_n|P.
\end{align}
The subscript $n$ in Eq.~\eqref{eq179} denotes each of all $2N$ discrete eigenstates including resonant, anti-resonant, bound and anti-bound states.
The bra and ket vectors $\bra\tilde{\psi}_n|$ and $|\psi_n\ket$ are their left- and right-eigenvectors, respectively.

We can use this new complete set of $2N$ states to expand the Green's function $G(E)$ and the time-evolution operator $e^{-iHt}$ in the forms
\begin{align}\label{eq140}
PG(E)P&=P\frac{1}{EI_\infty-H}P=\sum_{n=1}^{2N}
P|\psi_n\ket\frac{\lambda\lambda_n}{\lambda-\lambda_n}\bra\tilde{\psi}_n|P,
\\\label{eq141}
Pe^{-iHt}P&=
\frac{1}{2\pi i}\sum_{n=1}^{2N}\int_{C_2}
\exp\left[i\left(\lambda+\frac{1}{\lambda}\right)t\right]
P|\psi_n\ket\frac{\lambda_n}{\lambda-\lambda_n}\bra\tilde{\psi}_n|P
\left(-\lambda+\frac{1}{\lambda}\right)d\lambda,
\end{align}
where
\begin{align}\label{eq145}
\lambda=e^{ik}
\end{align}
converts the dispersion relation~\eqref{eq60} into the form
\begin{align}\label{eq146}
E(\lambda)=-\lambda-\frac{1}{\lambda}.
\end{align}
The $2N$ pieces of discrete eigenvalues $E_n$ are here represented by $\lambda_n$ defined in

\begin{align}\label{eq170}
E_n=-2\cos k_n=E(\lambda_n)=-\lambda_n-\frac{1}{\lambda_n};
\end{align}
we choose an appropriate one of $k_n$ and $\lambda_n$ from the two solutions of Eq.~\eqref{eq170} on the basis of Table~\ref{tab1}.
The integration contour $C_2$ in Eq.~\eqref{eq141} is to be specified below in Fig.~\ref{fig7}(c).


The expansion of the Green's function in the $P$ subspace, Eq.~\eqref{eq140}, as well as the basic idea of the following proofs first appeared in Ref.~\cite{Klaiman11} for a one-dimensional open quantum system.
The Green's function in the $P$ subspace is particularly important because it gives the transmission coefficient from a lead $i\alpha$ to another lead $j\beta$ in the form\cite{Fisher81,Datta95}
\begin{align}
T_{i\alpha,j\beta}(E)=(t_{i\alpha}t_{j\beta})^2\sin^2k\left|\bra d_i|G(E)|d_j\ket\right|^2,
\end{align}
and hence the Landauer formula~\cite{Landauer57,Datta95} can convert it to the electric conductance as
\begin{align}
\mathcal{G}_{i\alpha,j\beta}=\frac{2e^2}{h}T_{i\alpha,j\beta},
\end{align}
where $e$ is the charge of an electron and $h$ is the Planck constant.
The time evolution in the $P$ subspace, Eq.~\eqref{eq141}, is useful in computing the survival probability of a particle in an excited state, which we will do in Sec.~\ref{sec11}.

A remarkable point of the expansions~\eqref{eq179}--\eqref{eq141} is the absence of the background integral.
In the conventional analysis, the resonant states were taken into account by modifying the contour of the background integral as follows.
R.~Newton~\cite{Newton60,Newton82} proved that the bound states and the continuum scattering eigenstates form a complete set of the open quantum system.
We can straightforwardly convert the proof to the  tight-binding system to have
\begin{align}\label{eq220}
I_\infty=&\sum_{n\in\mbox{\scriptsize bound}}|\phi_n\ket\bra\phi_n|
+\int_{-\pi}^\pi \frac{dk}{2\pi}|\phi_k\ket\bra\phi_k|,
\\\label{eq230}
\frac{1}{EI_\infty-H}=&\sum_{n\in\mbox{\scriptsize bound}}|\phi_n\ket\frac{1}{E-E_n}\bra\phi_n|
+\int_{-\pi}^\pi \frac{dk}{2\pi}|\phi_k\ket\frac{1}{E-E_k}\bra\phi_k|,
\end{align}
where the first summation runs over all bound states and $|\phi_k\ket$ denotes the continuum scattering eigenstate with the dispersion relation~\eqref{eq60}.
We here denoted the eigenstates by $|\phi_n\ket$ intentionally because its normalization is different from the one of $|\psi_n\ket$ in Eqs.~ \eqref{eq179}--\eqref{eq141} as will be given below in Eqs.~\eqref{eq841} and~\eqref{eq842}.
Note that for the bound and scattering eigenstates, the left-eigenvectors are the Hermitian conjugate of the right-eigenvectors and hence we omitted the tilde symbols from the left-eigenvectors.

\begin{figure}
\centering
\includegraphics[width=0.4\textwidth]{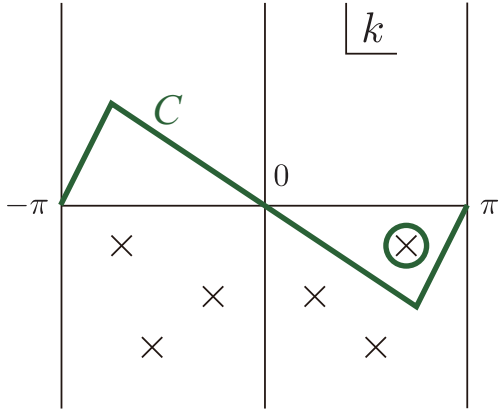}
\caption{Modification of the integration contour in Eqs.~\eqref{eq220} and~\eqref{eq230} over the range $-\pi\leq k \leq \pi$ on the real axis.
The portion $0\leq k\leq \pi$ is lowered to include some of the resonant states in the fourth quadrant whereas the other resonant states are excluded.
The portion $-\pi \leq k\leq 0$ is raised to keep the symmetry $-k\leftrightarrow k$.
All anti-resonant states in the third quadrant are therefore excluded.}
\label{fig6}
\end{figure}

We could extract the contributions of \textit{some} of the resonant states by modifying the contour on the real axis into the fourth quadrant of the complex $k$ plane:~\cite{Berggren68,Berggren70,Berggren96}
\begin{align}\label{eq240}
\int_{-\pi}^\pi \frac{dk}{2\pi}|\phi_k\ket\bra\phi_k|
=\sum_{n\in\mbox{\scriptsize some res.}}|\phi_n\ket\bra\tilde{\phi}_n|
+\int_C \frac{dk}{2\pi}|\phi_k\ket\bra\phi_k|;
\end{align}
see Fig.~\ref{fig6}.
We refer to the second term on the right-hand side as the background integral.

The approach has two drawbacks.
First, it is arbitrary to choose which resonant states we include and which we exclude.
In other words, we are  splitting the left-hand side of Eq.~\eqref{eq240} arbitrarily into the resonant contributions and the background integral.
Therefore, in general, the background integral has no clear physical meaning.
Second, the formulation explicitly breaks the time-reversal symmetry.
It extracts the contributions of resonant states but not those of the time-reversal anti-resonant states.

In contrast, our approach produces expansions which perfectly maintain the time-reversal symmetry.
Furthermore, the expansions do not have any background integrals as long as they are considered in the $P$ subspace.
An integral indeed appears when we include the $Q$ subspace but with a critical difference.
For example, the resolution of unity takes the form
\begin{align}\label{eq245}
I_\infty
= \sum_{n=1}^{2N}P|\psi_n\ket\bra\tilde{\psi}_n|P
 + \sum_{i\alpha}\int \frac{dk_{i\alpha}}{2\pi} \, Q|k_{i\alpha}\ket\bra k_{i\alpha} |Q
\end{align}
as we will show below in Eq.~\eqref{eqIinfty}, where $|k_{i\alpha}\ket$ denotes the scattering eigenstate~\eqref{plainwave} of the \textit{un}perturbed lead Hamiltonian $H_\mathrm{leads}=QHQ$.
Since the integral in the second term on the right-hand side is with respect to the unperturbed states, we can generally carry it out rigorously for specific matrix elements and thereby eliminate the background integral.
It is remarkable that the perturbations $H_\mathrm{contacts}=PHQ+QHP$ affect only the first term in the $P$ subspace in our formulation.
This is because we factor out the $Q$ subspace as we will show in the next section.

\section{Effective Hamiltonian}
\label{sec5}

Throughout the three sections~\ref{sec5}--\ref{sec7}, we will map the eigenvalue problem of the open quantum systems in  infinite dimensions, first into a nonlinear eigenvalue problem in  $N$ dimensions, and then into a generalized linear eigenvalue problem in $2N$ dimensions.
When we compute physical quantities in Secs.~\ref{sec8}--\ref{sec12}, we start from the $2N$-dimensional space, trace back first into the $N$-dimensional space and then into the infinite-dimensional space.

In the present section, we show how we factor out the $Q$ subspace and focus on the $N$-dimensional $P$ subspace.
We utilize the effective Hamiltonian for an open quantum system defined in the $P$ subspace:\cite{Livshits56,Feshbach58,Feshbach62,Rotter91,Albeverio96,Petrosky96,Petrosky97,Fyodorov97,Dittes00,Pichugin01,Sadreev03,Okolowicz03,Kunz06,Kunz08,Rotter09,Sasada11}
\begin{align}\label{eq250}
\Heff(E)\define PHP+PHQ\frac{1}{E-QHQ}QHP.
\end{align}
It has the same discrete eigenvalues as those of the full Hamiltonian $H$; see Appendix~\ref{appA} for derivation.
Equivalently, the Green's function of the full Hamiltonian
\begin{align}
G(E)\define\frac{1}{EI_\infty-H}
\end{align}
is equal to that of the effective Hamiltonian in the $P$ subspace:\cite{Sasada11}
\begin{align}\label{eq260}
PG(E)P=
P\frac{1}{EI_\infty-H}P=\frac{1}{EI_N-\Heff(E)}=:G_\mathrm{eff}(E);
\end{align}
see Appendix~\ref{appB} for a proof.

For the specific open quantum system defined in Fig.~\ref{fig4}(a), we can easily write down the effective Hamiltonian as follows:
\begin{align}\label{eq70}
\Ham_\mathrm{eff}(E)&=\Ham_\mathrm{d}+\Sigma(E),
\\\label{eq71}
\Sigma(E)&\define -e^{ik}PHQHP=\sum_{i=1}^N\sum_{\alpha=1}^{n_i}\Sigma_{i\alpha}(E),
\end{align}
where Eq.~\eqref{eqb280} gives the self-energy for the ($i\alpha$)th lead in the form~\cite{Sasada08,Sasada11}
\begin{align}\label{eq72}
\Sigma_{i\alpha}(E)\define
-\left(t_{i\alpha}\right)^2e^{ik}|d_i\ket\bra d_i|
=-\left(t_{i\alpha}\right)^2\lambda|d_i\ket\bra d_i|
\end{align}
with the wave number $k$ being related to the particle energy $E$ as in the dispersion relation~\eqref{eq60} and $\lambda$ being defined in Eq.~\eqref{eq145}.
Note that the effective Hamiltonian~\eqref{eq70} is an $N\times N$ matrix in the basis set $\{|d_i\ket | i=1,2,\ldots,N\}$ of the $P$ subspace.

The eigenvalue problem therefore reads as follows:
\begin{align}\label{eq100}
\left(E_nI_N-\Ham_\mathrm{eff}(E)\right)P|\psi_n\ket=0,
\end{align}
where $P|\psi_n\ket$ is an $N$-dimensional column vector.
We can always reproduce the vector in the whole space from the solution $P|\psi_n\ket$ of Eq.~\eqref{eq100} as follows:\cite{Livshits56,Feshbach58,Feshbach62,Rotter91,Albeverio96,Petrosky96,Petrosky97,Fyodorov97,Dittes00,Pichugin01,Sadreev03,Okolowicz03,Kunz06,Kunz08,Rotter09,Sasada11}
\begin{align}\label{eq300}
|\psi_n\ket&=P|\psi_n\ket+Q|\psi_n\ket
\nonumber\\
&=P|\psi_n\ket+\frac{1}{E_n-QHQ}QHP|\psi_n\ket
\nonumber\\
&=\left(P+\frac{1}{E_n-QHQ}QHP\right)(P|\psi_n\ket),
\end{align}
where we used Eq.~\eqref{eqa40}.
Using Eq.~\eqref{eqb240}, we can compute it explicitly for the specific open quantum system in Fig.~\ref{fig4}(a) in the form~\cite{Sasada08,Sasada11}
\begin{align}\label{eq320}
\bra x_{i\alpha} | \psi_n \ket &= \bra x_{i\alpha} | \frac{1}{E_n-QHQ}QHP | d_i \ket \bra d_i | \psi_n \ket
\nonumber\\
&=t_{i\alpha}e^{ik_nx_{i\alpha}}\bra d_i|\psi_n\ket
=t_{i\alpha}{\lambda_n}^{x_{i\alpha}}\bra d_i|\psi_n\ket,
\end{align}
where $k_n$ and $\lambda_n$ are given in Eq.~\eqref{eq170}.
This is consistent with the fact that all eigenstates with point spectra are given under the Siegert boundary condition~\eqref{eq2-30}.\cite{Gamow28,Siegert39,Peierls59,leCouteur60,Zeldovich60,Hokkyo65,Romo68,Berggren70,Gyarmati71,Landau77,Romo80,Berggren82,Berggren96,Madrid05,Hatano08,GarciaCalderon10}

To summarize this section, we have mapped the infinite-dimensional eigenvalue problem $H|\psi\ket=E|\psi\ket$ to the $N$-dimensional eigenvalue problem~\eqref{eq100}.
We, however, cannot use Eq.~\eqref{eq100} for the expansion of the Green's function as it is;
it is not a standard eigenvalue problem because the effective Hamiltonian $\Ham_\mathrm{eff}(E)$ itself depends on the energy.
Indeed, we will see below that the eigenvalue problem has $2N$ pieces of eigenvalues, not $N$ pieces.

\section{Quadratic eigenvalue problem for the effective Hamiltonian}
\label{sec6}

In the present section, we will formulate the eigenvalue problem of the effective Hamiltonian, Eq.~\eqref{eq100}, as a quadratic eigenvalue problem in the $N$-dimensional space.
We then map the problem into a generalized (linear) eigenvalue problem in an expanded $2N$-dimensional space.

Let us rewrite the eigenvalue equation~\eqref{eq100} using $\lambda$ defined in Eq.~\eqref{eq145}.
Since the particle energy now is given by Eq.~\eqref{eq146},
we have\cite{Klaiman11}
\begin{align}\label{eq120}
\left(-\lambda I_N-\frac{1}{\lambda}I_N-\Ham_\mathrm{d}+\lambda\Theta\right)P|\psi\ket=0,
\end{align}
where
\begin{align}
\Theta\define -\frac{1}{\lambda}\Sigma(E)=PHQHP
\end{align}
is an $N\times N$ diagonal matrix with the constant diagonal elements
\begin{align}\label{eq350}
\Theta_{ii}\define\sum_{\alpha=1}^{n_i}\left(t_{i\alpha}\right)^2
\end{align}
for $1\leq i\leq N$.
It is indeed related to the matrix $\Gamma$ in Ref.~\cite{Sasada11} as
\begin{align}\label{eq135}
\Gamma=\Theta\sin k.
\end{align}
We can further rewrite Eq.~\eqref{eq120} as
\begin{align}\label{eq150}
Z(\lambda) P|\psi\ket=0
\end{align}
with
\begin{align}\label{eq160}
Z(\lambda)\define \lambda^2(I_N-\Theta)+\lambda\Ham_\mathrm{d}+I_N,
\end{align}
which is a quadratic eigenvalue problem~\cite{Tisseur01} in the sense that $Z(\lambda)$ is quadratic in $\lambda$ instead of a linear function in the standard eigenvalue problem.
Note that the term in the original eigenvalue function~\eqref{eq100} is recovered by
\begin{align}\label{eq180}
E(\lambda)I_N-\Ham_\mathrm{eff}(E(\lambda))=-\frac{1}{\lambda} Z(\lambda).
\end{align}
Writing down the eigenvalue equation~\eqref{eq100} in the form of the quadratic eigenvalue equation~\eqref{eq150} as well as further algebra leading to the expansion of the Green's function, Eq.~\eqref{eq800} below, was first done in Ref.~\cite{Klaiman11} for a one-dimensional open quantum system.

Following the standard treatment of the quadratic eigenvalue problem~\cite{Tisseur01}, we linearize Eq.~\eqref{eq150} as follows:
\begin{align}\label{eq190}
\left(A-\lambda B\right)|\Psi\ket=0,
\end{align}
where $A$ and $B$ are $\lambda$-independent $2N\times 2N$ real symmetric matrices given by
\begin{align}\label{eq410}
A&\define
\begin{pmatrix}
0 & I_N \\
I_N & \Ham_\mathrm{d}
\end{pmatrix},
\\\label{eq420}
B&\define
\begin{pmatrix}
I_N & 0 \\
0 & -I_N+\Theta
\end{pmatrix},
\end{align}
and $0$ here means the $N\times N$ zero matrix, yielding
\begin{align}
A-\lambda B&=\begin{pmatrix}
-\lambda I_N & I_N \\
I_N & \Ham_\mathrm{d}+\lambda(I_N-\Theta)
\end{pmatrix},
\end{align}
while $|\Psi\ket$ is a $\lambda$-dependent $2N$-dimensional column vector given by
\begin{align}
|\Psi\ket\define
\begin{pmatrix}
P|\psi\ket \\
\lambda P|\psi\ket
\end{pmatrix}.
\end{align}
Note that the matrix $B$ is a diagonal matrix.

Equation~\eqref{eq190} is a generalized linear eigenvalue problem.
It is called `generalized' because we have the matrix $B$ in place of the identity matrix for the standard linear eigenvalue problem.
The important point here is that now the matrices $A$ and $B$ are both independent of the energy.
We can therefore use Eq.~\eqref{eq190} to expand the Green's function.

We can confirm Eq.~\eqref{eq190} as follows.
The first row gives the trivial identity
\begin{align}
-\lambda P|\psi\ket+\left(\lambda P|\psi\ket\right) =0,
\end{align}
whereas the second row gives
\begin{align}
P|\psi\ket+\left[\Ham_\mathrm{d}+\lambda(I_N-\Theta)\right]\left(\lambda P|\psi\ket\right)=0,
\end{align}
which is equivalent to Eq.~\eqref{eq150}.
This is analogous to the technique of splitting a second-order differential equation into a set of two first-order differential equations.\cite{Tisseur01}

We can see the equivalence between Eqs.~\eqref{eq150} and~\eqref{eq190} more clearly using the two matrices\cite{Tisseur01}
\begin{align}
X(\lambda)&\define
\begin{pmatrix}
-\Ham_\mathrm{d}-\lambda(I_N-\Theta) & I_N \\
I_N & 0
\end{pmatrix},
\\
Y_1(\lambda)&\define
\begin{pmatrix}
I_N & 0 \\
\lambda I_N & I_N
\end{pmatrix},
\\
Y_2(\lambda)&\define
\begin{pmatrix}
I_N & \lambda I_N \\
0 & I_N
\end{pmatrix}.
\end{align}
 Straightforward algebra shows
\begin{align}
X(\lambda)(A-\lambda B)Y_1(\lambda)
=Y_2(\lambda)(A-\lambda B)X(\lambda)
=\begin{pmatrix}
Z(\lambda) & 0 \\
0 & I_N
\end{pmatrix}.
\end{align}
Since the determinants of the matrices $X(\lambda)$, $Y_1(\lambda)$ and $Y_2(\lambda)$ are nonzero constant, we can invert them to obtain
\begin{align}
X(\lambda)^{-1}&=\begin{pmatrix}
0 & I_N \\
I_N & \Ham_\mathrm{d}+\lambda(I_N-\Theta)
\end{pmatrix},
\\
Y_1(\lambda)^{-1}&=\begin{pmatrix}
I_N & 0 \\
-\lambda I_N & I_N
\end{pmatrix},
\\
Y_2(\lambda)^{-1}&=\begin{pmatrix}
I_N & -\lambda I_N \\
0 & I_N
\end{pmatrix}.
\end{align}
We then have
\begin{align}
Y_1(\lambda)^{-1}(A-\lambda B)^{-1}X(\lambda)^{-1}
=X(\lambda)^{-1}(A-\lambda B)^{-1}Y_2(\lambda)^{-1}
=\begin{pmatrix}
Z(\lambda)^{-1} & 0 \\
0 & I_N
\end{pmatrix},
\end{align}
which confirms that the singularities of $(A-\lambda B)^{-1}$ are the same as those of $Z(\lambda)^{-1}$.
Note that because of Eq.~\eqref{eq180}, the Green's function of the effective Hamiltonian is now given by\cite{Tisseur01}
\begin{align}\label{eq340}
PG(E(\lambda))P=\Geff(E(\lambda))
&=\frac{1}{E(\lambda)I_N-\Heff(E(\lambda))}
\nonumber\\
&=-\lambda Z(\lambda)^{-1}
\nonumber\\
&=-\lambda\begin{pmatrix}
I_N & 0 
\end{pmatrix}
Y_1(\lambda)^{-1}(A-\lambda B)^{-1}X(\lambda)^{-1}
\begin{pmatrix}
I_N\\
0
\end{pmatrix}
\nonumber\\
&=-\lambda\begin{pmatrix}
I_N & 0 
\end{pmatrix}
(A-\lambda B)^{-1}
\begin{pmatrix}
0\\
I_N
\end{pmatrix}.
\end{align}

To summarize the present section, we can solve the $N$-dimensional eigenvalue equation~\eqref{eq100} of the energy-dependent effective Hamiltonian $\Heff(E)$ by solving the $2N$-dimensional generalized eigenvalue equation~\eqref{eq190} of the energy-independent matrices $A$ and $B$.
Since the $2N$-dimensional generalized eigenvalue problem generally yields $2N$ eigenstates, we have $2N$ eigenstates for the $N$-dimensional energy-dependent eigenvalue problem~\eqref{eq100}.
(Some eigenvalues can become infinite for special values of the system parameters; see Appendix~H of Ref.~\cite{Sasada11})
We will use these $2N$ eigenstates to expand the Green's function.

Note that although the matrices $A$ and $B$ are both Hermitian (more precisely, real symmetric), the present generalized eigenvalue problem is still non-Hermitian (more precisely, real asymmetric) because $A$ and $B$ do not commute with each other in general: $[A,B]\neq 0$.
If $B$ is invertible in particular, the eigenvalue equation~\eqref{eq190} reduces to the standard eigenvalue equation $(B^{-1}A-\lambda I)|\Psi\ket=0$ for the non-Hermitian (real asymmetric) matrix $B^{-1}A$.

\section{Generalized eigenvalue problem}
\label{sec7}

In order to see how we can use the $2N$ eigenstates for the Green's function expansion, let us give a tutorial review of the generalized eigenvalue problem in the context of Eq.~\eqref{eq190}.
In the present section, we will drop the projection operator $P$ for brevity;
we always work in the $P$ subspace here.
Suppose that the eigenvalue equation~\eqref{eq150} has an eigenvalue $\lambda_n$ with the right-eigenvector $|\psi_n\ket$ and the left-eigenvector $\bra\tilde{\psi}_n|$:
\begin{align}
Z(\lambda_n)|\psi_n\ket= \bra\tilde{\psi}_n|Z(\lambda_n)=0
\end{align}
for $n=1,2,\ldots,2N$.
Then the $2N$-dimensional column vector
\begin{align}\label{eq570}
|\Psi_n\ket\define
\begin{pmatrix}
|\psi_n\ket \\
\lambda_n|\psi_n\ket
\end{pmatrix}
\end{align}
and the $2N$-dimensional row vector
\begin{align}\label{eq580}
\bra\tilde{\Psi}_n|\define
\begin{pmatrix}
\bra\tilde{\psi}_n| & \lambda_n\bra\tilde{\psi}_n|
\end{pmatrix}
\end{align}
are the right- and left-eigenvectors of the generalized eigenvalue equation~\eqref{eq190} with the eigenvalue $\lambda_n$:
\begin{align}\label{eq380}
(A-\lambda_n B)|\Psi_n\ket=\bra\tilde{\Psi}_n|(A-\lambda_nB)=0
\end{align}
for $n=1,2,\ldots,2N$.
We can indeed confirm Eq.~\eqref{eq380} by using
\begin{align}
A-\lambda B
=X(\lambda)^{-1}
\begin{pmatrix}
Z(\lambda) & 0 \\
0 & I
\end{pmatrix}
Y_1(\lambda)^{-1}
=Y_2(\lambda)^{-1}
\begin{pmatrix}
Z(\lambda) & 0 \\
0 & I
\end{pmatrix}
X(\lambda)^{-1}
\end{align}
because
\begin{align}
Y_1(\lambda_n)^{-1}|\Psi_n\ket&=\begin{pmatrix}
|\psi_n\ket \\
0
\end{pmatrix},
\\
\bra\tilde{\Psi}_n|Y_2(\lambda_n)^{-1}&=\begin{pmatrix}
\bra\tilde{\psi}_n| &
0
\end{pmatrix}.
\end{align}

We now show that the vectors $\{|\Psi_n\ket|n=1,2,\ldots,2N\}$ and $\{\bra\tilde{\Psi}_n||n=1,2,\ldots,2N\}$ constitute a bi-orthonormal set under the metric given by $B$.
Let us normalize the vector $|\psi_n\ket$ and $\bra\tilde{\psi}_n|$ so that $|\Psi_n\ket$ and $\bra\tilde{\Psi}_n|$ may satisfy
\begin{align}\label{eq630}
\bra\tilde{\Psi}_n|B|\Psi_n\ket=1.
\end{align}
Equation~\eqref{eq380} then is followed by
\begin{align}
0=\bra\tilde{\Psi}_n|(A-\lambda_nB)|\Psi_n\ket=\bra\tilde{\Psi}_n|A|\Psi_n\ket-\lambda_n,
\end{align}
or
\begin{align}
\bra\tilde{\Psi}_n|A|\Psi_n\ket=\lambda_n.
\end{align}
Assume that the eigenvalues have no degeneracy $\lambda_n\neq\lambda_m$ for $n\neq m$.
Then we have
\begin{align}
\bra\tilde{\Psi}_m|(A-\lambda_nB)|\Psi_n\ket=
\bra\tilde{\Psi}_m|(A-\lambda_mB)|\Psi_n\ket=0,
\end{align}
or
\begin{align}
(\lambda_m-\lambda_n)\bra\tilde{\Psi}_m|B|\Psi_n\ket=0,
\end{align}
and therefore
\begin{align}\label{eq680}
\bra\tilde{\Psi}_m|B|\Psi_n\ket=0
\end{align}
for $m\neq n$. 

Equations~\eqref{eq630} with~\eqref{eq680} indicate that the vectors $\{|\Psi_n\ket|n=1,2,\ldots,2N\}$ and $\{\bra\tilde{\Psi}_n||n=1,2,\ldots,2N\}$ constitute a bi-orthonormal pair under the metric $B$.
By constructing the $2N\times 2N$ matrices
\begin{align}\label{eq690}
U&\define
\begin{pmatrix}
|\Psi_1\ket &
|\Psi_2\ket &
\cdots &
|\Psi_{2N}\ket
\end{pmatrix}
\nonumber\\
&=\begin{pmatrix}
|\psi_1\ket &
|\psi_2\ket &
\cdots &
|\psi_{2N}\ket
\\
\lambda_1|\psi_1\ket &
\lambda_2|\psi_2\ket &
\cdots &
\lambda_{2N}|\psi_{2N}\ket
\end{pmatrix},
\\\label{eq700}
\tilde{U}&\define
\begin{pmatrix}
\bra\tilde{\Psi}_1| \\
\bra\tilde{\Psi}_2| \\
\vdots \\
\bra\tilde{\Psi}_{2N}|
\end{pmatrix}
=\begin{pmatrix}
\bra \tilde{\psi}_1| & \lambda_1\bra \tilde{\psi}_1| \\
\bra \tilde{\psi}_2| & \lambda_2\bra \tilde{\psi}_2| \\
\vdots & \vdots \\
\bra \tilde{\psi}_{2N}| & \lambda_{2N}\bra \tilde{\psi}_{2N}| \\
\end{pmatrix},
\end{align}
we have
\begin{align}
\tilde{U}AU&=\Lambda,
\\\label{eq530}
\tilde{U}BU&=I_{2N},
\\\label{eq540}
\tilde{U}(A-\lambda B)U&=\Lambda-\lambda I_{2N},
\end{align}
where $\Lambda$ is the $2N\times 2N$ diagonal matrix with the diagonal element $\Lambda_{nn}=\lambda_n$ and $I_{2N}$ here is the $2N\times 2N$ identity matrix.

In the present case, in particular, the matrices $A$ and $B$ are real symmetric as given in Eq.~\eqref{eq410} and~\eqref{eq420}; 
note that both $H_\mathrm{d}$ in Eq.~\eqref{eq40} and $\Theta$ in Eq.~\eqref{eq350} can be taken to be real symmetric because there are no magnetic fields and hence we do not break the time-reversal symmetry explicitly.
The first equation of Eq.~\eqref{eq380} is therefore followed by its transpose
\begin{align}
(|\Psi_n\ket)^T(A-\lambda_nB)=0,
\end{align}
which, compared with the second equation of Eq.~\eqref{eq380}, yields the relation
\begin{align}
\bra\tilde{\Psi}_n|=|\Psi_n\ket^T
\end{align}
and hence
\begin{align}\label{eq755}
\bra\tilde{\psi}_n|=|\psi_n\ket^T.
\end{align}
The last relation is the standard one found in the literature;\cite{Zeldovich60,Hokkyo65,Romo68,Moiseyev11} see Eq.~\eqref{eq201}.

\section{Eigenstate expansion of the Green's function}
\label{sec8}

We now have $2N$ eigenstates in the $2N$-dimensional space which are bi-orthogonal to each other with a metric $B$.
Using the eigenstates as bases, we will first expand the Green's function in the $2N$-dimensional space.
We will then map the expansion back into the $N$-dimensional space.

Equation~\eqref{eq540} is followed by
\begin{align}
U^{-1}(A-\lambda B)^{-1}\tilde{U}^{-1}=(\Lambda-\lambda I_{2N})^{-1},
\end{align}
or
\begin{align}\label{eq760}
(A-\lambda B)^{-1}&=U(\Lambda-\lambda I_{2N})^{-1}\tilde{U}.
\end{align}
This is the expansion of the Green's function in the $2N$-dimensional space.

Substituting Eq.~\eqref{eq760} into Eq.~\eqref{eq340}, we have\cite{Klaiman11}
\begin{align}\label{eq800}
PG(E(\lambda))P=
\Geff(E(\lambda))&=
-\lambda
\begin{pmatrix}
I_N & 0 
\end{pmatrix}
U(\Lambda-\lambda I_{2N})^{-1}\tilde{U}
\begin{pmatrix}
0\\
I_N
\end{pmatrix}
\nonumber\\
&=
\sum_{n=1}^{2N}
P|\psi_n\ket
\frac{\lambda\lambda_n}{\lambda-\lambda_n}
\bra \tilde{\psi}_n|P,
\end{align}
where we used the expressions~\eqref{eq690} and~\eqref{eq700} in the last line.
This is the result that we presented in Eq.~\eqref{eq140} above  and was first given in Ref.~\cite{Klaiman11} for a one-dimensional open quantum system.
We stress again that the expansion is given only by the eigenstates with point spectra.
It is remarkable that we do not have any background integrals.
We also emphasize that this expansion is time-reversal symmetric;
the resonant and anti-resonant states contribute in a time-reversal symmetric way.

Appendix~\ref{appC} shows that the new expansion~\eqref{eq800} leads to the expansion in our previous work (Eqs.~(4) and~(56) in Ref.~\cite{Sasada11}):
\begin{align}\label{eq810}
P(G^\mathrm{R}(E)+G^\mathrm{A}(E))P=\sum_{n=1}^{2N}
P|\phi_n\ket\frac{1}{E-E_n}\bra\tilde{\phi}_n|P,
\end{align}
where $G^R(E)$ and $G^A(E)$ are the retarded and advanced Green's functions of the whole Hamiltonian,
\begin{align}
G^\mathrm{R/A}(E)=\frac{1}{EI_\infty-H\pm i\delta}
\end{align}
with an infinitesimal $\delta$.
This is another piece of evidence that the expansion~\eqref{eq800} is time-reversal symmetric;
it contains both the retarded and advanced components of the Green's function.
We will see in Sec.~\ref{sec11} that the retarded component is chosen when we consider the initial-condition problem, while the advanced component is chosen when we consider the terminal-condition problem.
Incidentally, we denoted in Eq.~\eqref{eq810} the eigenstates by $|\phi_n\ket$ intentionally because its normalization is different from the one of $|\psi_n\ket$ in Eq.~\eqref{eq800} as follows:
\begin{align}\label{eq841}
|\phi_n\ket&=\sqrt{1-{\lambda_n}^2}|\psi_n\ket,
\\\label{eq842}
\bra\tilde{\phi}_n|&=\sqrt{1-{\lambda_n}^2}\bra\tilde{\psi}_n|;
\end{align}
see Appendix~\ref{appD} for the derivation.

Although the expansion~\eqref{eq800} is done in the $P$ subspace, we can always relate the eigenstates in the $P$ subspace with those in the $Q$ subspace as shown in Eqs.~\eqref{eq300} and~\eqref{eq320}.
More specifically, we can utilize Eqs.~\eqref{eqa110}--\eqref{eqa130} in order to obtain expansions outside the $P$ subspace.
For example, we have
\begin{align}\label{eq820}
QG(E(\lambda))P&=Q\frac{1}{E-QHQ}QHP\Geff(E(\lambda))
\nonumber\\
&=-\sum_{n=1}^{2N}\sum_{i\alpha}{t_{i\alpha}}\int_{-\pi}^\pi \frac{dk_{i\alpha}}{2\pi} Q|k_{i\alpha}\ket
\frac{\sqrt{2}\sin k_{i\alpha}}{E(\lambda)+2\cos k_{i\alpha}}
\bra d_i|\psi_n\ket
\frac{\lambda\lambda_n}{\lambda-\lambda_n}\bra\tilde{\psi}_n|P,
\end{align}
where we used the expression~\eqref{eqb200}.
We can thereby express the Green's function in the whole space in the form
\begin{align}\label{eq825}
G(E(\lambda)) &\define\frac{1}{E(\lambda)I_\infty-H}
\nonumber\\
&= Q\frac{1}{E(\lambda)-QHQ}Q \nonumber\\
&+\left(I_N+Q\frac{1}{E(\lambda)-QHQ}QHP\right)\Geff(E(\lambda))\left(I_N+PHQ\frac{1}{E(\lambda)-QHQ}Q\right) 
\nonumber\\
&=\sum_{i\alpha} \int_{-\pi}^\pi \frac{dk_{i\alpha}}{2\pi}  Q|k_{i\alpha}\ket\frac{1}{E(\lambda)+2\cos k_{i\alpha}}\bra k_{i\alpha}|Q \nonumber\\
&+\sum_{n=1}^{2N}
\left(I_N-\sum_{i\alpha}{t_{i\alpha}}\int_{-\pi}^\pi \frac{dk_{i\alpha}}{2\pi} Q|k_{i\alpha}\ket
\frac{\sqrt{2}\sin k_{i\alpha}}{E(\lambda)+2\cos k_{i\alpha}}\bra d_i|P
\right)
\nonumber\\
&\times\left(P|\psi_n\ket\frac{\lambda\lambda_n}{\lambda-\lambda_n}\bra\tilde{\psi}_n|P\right)
\nonumber\\
&\times
\left(I_N
-\sum_{i\alpha}t_{i\alpha}\int_{-\pi}^\pi \frac{dk_{i\alpha}}{2\pi} P|d_i\ket
\frac{\sqrt{2}\sin k_{i\alpha}}{E(\lambda)+2\cos k_{i\alpha}}\bra k_{i\alpha}|Q
\right) 
\end{align}
The first term on the right-hand side is the expansion~\eqref{eqb190} of the Green's function of the \textit{un}perturbed lead Hamiltonian in the $Q$ subspace. 
The second term is an expansion of the contributions that involve the $P$ subspace.

The expansion~\eqref{eq825} is different from the conventional expansion~\eqref{eq230} in the following two notable points: 
(i) the second term is still written as a sum over all discrete eigenstates;
(ii) the integrals are taken over the \textit{un}perturbed states and hence can be carried out rigorously for specific matrix elements;
see Sec.~\ref{sec13}, for example. 
The remarkable difference is due to the fact that we have essentially factorized the $P$ subspace from the $Q$ subspace.
Thanks to this factorization, the expansion~\eqref{eq825} is given in terms of the perturbed states $|\psi_n\ket $ in the $P$ subspace and the \textit{un}perturbed states $|k_{i\alpha}\ket$ in the $Q$ subspace.

\section{New resolution of unity}
\label{sec9}

\begin{figure}
\includegraphics[width=\textwidth]{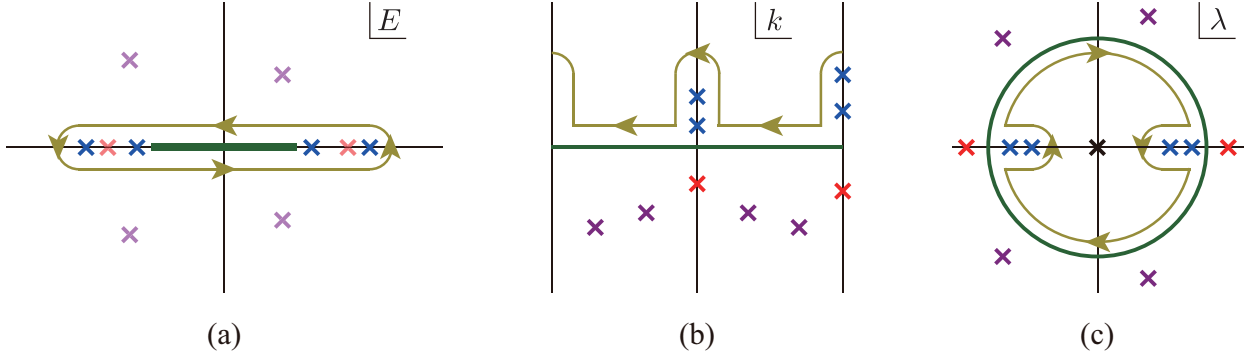}
\caption{(a) The integration contour $C_1$ in the complex energy plane. (b) The contour mapped onto the complex wave-number plane. (c) The contour mapped onto the complex $\lambda$ plane, which is referred to as $C_2$.
The contours are all marked by yellow curves. The blue crosses indicate the bound states, the red crosses the anti-bound states and the purple crosses the resonant and anti-resonant states. The green curves indicate the scattering states. 
In (a), the contour is on the first Riemann sheet, whereas the anti-bound states, the resonant states and the anti-resonant states are marked as thin crosses to indicate that they are all on the second Riemann sheet.
The scattering states  double back in (a).
The black cross in (c) is the pole at the origin, which corresponds to a point infinitely far away in (a) and (b).}
\label{fig7}
\end{figure}

We now prove the new resolution of unity presented in Eq.~\eqref{eq179}.
We begin the proof with the conventional resolution of unity for the whole system, namely Eq.~\eqref{eq220}.
We can cast the right-hand side of Eq.~\eqref{eq220} into the form
\begin{align}\label{eq845}
I_\infty=\frac{1}{2\pi i}\int_{C_1}\frac{1}{EI_\infty-H}dE,
\end{align}
where the integration contour $C_1$ is specified in Fig.~\ref{fig7}(a).

We project Eq.~\eqref{eq845} onto the $P$ subspace, having
\begin{align}\label{eq830}
I_N&=\frac{1}{2\pi i}\int_{C_1}P\frac{1}{EI_\infty-H}PdE
\nonumber\\
&=\frac{1}{2\pi i}\int_{C_1}\Geff(E)dE,
\end{align}
where we used Eq.~\eqref{eq260}.
We further transform this integral over $E$ to an integral over $\lambda$, which yields
\begin{align}\label{eq840}
I_N&=\frac{1}{2\pi i}\int_{C_2}\Geff(E(\lambda))\left(-1+\frac{1}{\lambda^2}\right)d\lambda
\nonumber\\
&=\frac{1}{2\pi i}\sum_{n=1}^{2N}\int_{C_2}P|\psi_n\ket\frac{\lambda_n}{\lambda-\lambda_n}\bra\tilde{\psi}_n|P
\left(-\lambda+\frac{1}{\lambda}\right)d\lambda,
\end{align}
where the integration contour $C_2$ is specified in Fig.~\ref{fig7}(c) and we used the expansion~\eqref{eq800} of the Green's function in the second line.
We can indeed obtain the same expression by putting $t=0$ in Eq.~\eqref{eq880} below in Sec.~\ref{sec10}.

The only pole that contributes to the contour $C_2$ is the one at $\lambda=0$.
Since we circle around the pole in the clockwise direction, we have
\begin{align} \label{eqIN}
I_N=\sum_{n=1}^{2N}P|\psi_n\ket\bra\tilde{\psi}_n|P,
\end{align}
which proves Eq.~\eqref{eq220}.
It is again remarkable that this resolution of unity is free of any background integrals.
We also stress again that the summation contains the resonant and anti-resonant states in a time-reversal symmetric way.

We can extend the argument to the resolution of unity in the whole space, $I_\infty$.
Using the same procedure for the Green's function in the whole space, Eq.~\eqref{eq825}, or putting $t=0$ in Eq.~\eqref{eq895} below in Sec.~\ref{sec10}, we have
\begin{align}\label{eq881}
I_\infty
&=\frac{1}{2\pi i}\int_{C_1}dE\,Q\frac{1}{E-QHQ}Q
\nonumber\\
&+\sum_{n=1}^{2N} \frac{1}{2\pi i}
\int_{C_2}d\lambda\left(-\lambda+\frac{1}{\lambda}\right) \nonumber\\
&\times\left(I_N+Q\frac{1}{E(\lambda)-QHQ}QHP\right)
\nonumber\\
&\times\left(P|\psi_n\ket\frac{\lambda_n}{\lambda-\lambda_n}\bra\tilde{\psi}_n|P\right)
\nonumber\\
&\times \left(I_N+PHQ\frac{1}{E(\lambda)-QHQ}Q\right)
\end{align}
In the second term on the right-hand side, the poles that appear in addition to the ones in Eq.~\eqref{eq840} are at $\lambda=e^{\pm ik_0}$ as can be read off from Eq.~\eqref{eq820}.
These poles are on the unit circle in the complex $\lambda$ plane, and hence do not contribute to the integral along the contour $C_2$.
Only the contribution due to the pole at $\lambda=0$ remains again, which gives the same expansion as Eq.~\eqref{eqIN}.
The first term, on the other hand, gives the resolution of unity in the $Q$ subspace, which can be written in terms of the scattering eigenstates $|k_{i\alpha}\ket$ of the \textit{un}perturbed Hamiltonian $QHQ$, and hence can be generally carried out for specific matrix elements.
We therefore arrive at
\begin{align}\label{eqIinfty}
I_\infty
= \sum_{n=1}^{2N}P|\psi_n\ket\bra\tilde{\psi}_n|P 
+ \sum_{i\alpha}\int_{-\pi}^\pi \frac{dk_{i\alpha}}{2\pi} \, Q|k_{i\alpha}\ket\bra k_{i\alpha} |Q,
\end{align}
which is Eq.~\eqref{eq245}.

To summarize this section, we have expressed the unit operator in terms of the perturbed discrete eigenstates on the dot and the unperturbed eigenstates on the lead;
this is the critical difference of Eq.~\eqref{eqIinfty} from the conventional resolution of unity, Eq.~\eqref{eq220}, as we emphasized above.

\section{Eigenstate expansion of the time-evolution operator}
\label{sec10}

In the present section, we show the expansion of the time-evolution operator $\exp(-iHt)$ only with respect to the eigenstates with point spectra, without any background integrals.
We first cast the time-evolution operator into the integral form
\begin{align}
e^{-iHt}=\frac{1}{2\pi i}\int_{C_1}e^{-iEt}\frac{1}{EI_\infty-H}dE,
\end{align}
where the integration contour $C_1$ is again specified in Fig.~\ref{fig7}(a).

We then consider the operator in the $P$ subspace, having
\begin{align}
Pe^{-iHt}P&=\frac{1}{2\pi i}\int_{C_1}e^{-iEt}P\frac{1}{EI_\infty-H}PdE
\nonumber\\
&=\frac{1}{2\pi i}\int_{C_1}e^{-iEt}\Geff(E)dE.
\end{align}
Following the same transformation from Eq.~\eqref{eq830} to Eq.~\eqref{eq840}, we have
\begin{align}\label{eq880}
Pe^{-iHt}P&=
\frac{1}{2\pi i}\sum_{n=1}^{2N}\int_{C_2}
\exp\left[i\left(\lambda+\frac{1}{\lambda}\right)t\right]
P|\psi_n\ket\frac{\lambda_n}{\lambda-\lambda_n}\bra\tilde{\psi}_n|P
\left(-\lambda+\frac{1}{\lambda}\right)d\lambda,
\end{align}
where the integration contour $C_2$ is again specified in Fig.~\ref{fig7}(c).
This is the result that we presented in Eq.~\eqref{eq141} above.
We will calculate the matrix element explicitly in Sec.~\ref{sec11}.
We will then see that the time-reversal symmetry is broken upon choosing $t>0$ or $t<0$.

The time-evolution operator in the other subspaces can be obtained from Eqs.~\eqref{eqa110}--\eqref{eqa130}.
For example, we have
\begin{align}\label{eq890}
Qe^{-iHt}P
=&\frac{1}{2\pi i}\int_{C_1}e^{-iEt}Q\frac{1}{EI_\infty-H}PdE
\nonumber\\
=&\frac{1}{2\pi i}\int_{C_1}e^{-iEt}Q\frac{1}{E-QHQ}QHP\Geff(E)dE
\nonumber\\
=&\frac{1}{2\pi i}\sum_{n=1}^{2N}\int_{C_2}d\lambda\left(-\lambda+\frac{1}{\lambda}\right)
\exp\left[i\left(\lambda+\frac{1}{\lambda}\right)t\right]
\nonumber\\
&\qquad
\sum_{i\alpha}\int_{-\pi}^\pi \frac{dk_{i\alpha}}{2\pi}Q|k_{i\alpha}\ket\frac{-\sqrt{2}t_{i\alpha}\sin k_{i\alpha}}{E(\lambda)+2\cos k_{i\alpha}}
\bra d_i|\psi_n\ket
\frac{\lambda_n}{\lambda-\lambda_n}\bra \psi_n|P,
\end{align}
where we used Eq.~\eqref{eqb200} in the transformation from the second line to the third.
The time-evolution operator in the whole space is therefore expressed in the form 
\begin{align}\label{eq895}
e^{-iHt}=&\frac{1}{2\pi i}
\int_{C_2}d\lambda\left(-1+\frac{1}{\lambda^2}\right) \exp\left[i\left(\lambda+\frac{1}{\lambda}\right)t\right] 
\nonumber\\
&\qquad\sum_{i\alpha}\int_{-\pi}^\pi \frac{dk_{i\alpha}}{2\pi}Q|k_{i\alpha}\ket\frac{1}{E(\lambda)+2\cos k_{i\alpha}}\bra k_{i\alpha}|Q
\nonumber\\ 
&+ \frac{1}{2\pi i}
\sum_{n=1}^{2N}
\int_{C_2}d\lambda\left(-\lambda+\frac{1}{\lambda}\right)
\exp\left[i\left(\lambda+\frac{1}{\lambda}\right)t\right]
\nonumber\\
&\times\left(I_N
-\sum_{i\alpha}t_{i\alpha}\int_{-\pi}^\pi \frac{dk_{i\alpha}}{2\pi}Q|k_{i\alpha}\ket\frac{\sqrt{2}\sin k_{i\alpha}}{E(\lambda)+2\cos k_{i\alpha}}
\bra d_i|P
\right)
\nonumber\\
&\times\left(P|\psi_n\ket\frac{\lambda_n}{\lambda-\lambda_n}\bra\tilde{\psi}_n|P\right)
\nonumber\\
&\times \left(I_N
-\sum_{i\alpha}t_{i\alpha}\int_{-\pi}^\pi \frac{dk_{i\alpha}}{2\pi}P|d_i\ket
\frac{\sqrt{2}\sin k_{i\alpha}}{E(\lambda)+2\cos k_{i\alpha}}\bra k_{i\alpha}|Q
\right).
\end{align}
Once again, the integrals on the right-hand side are taken over the \textit{un}perturbed scattering eigenstates in the $Q$ subspace and hence can be carried out rigorously, whereas the contributions involving the $P$ subspace are given by a sum over all discrete eigenstates in a time-reversal symmetric way.

\section{Time evolution of a dot state: Survival amplitude}
\label{sec11}

We here compute the survival amplitude and more generally the matrix element
\begin{align}\label{eq901}
\bra d_j|e^{-iHt}|d_i\ket,
\end{align}
using Eq.~\eqref{eq880}.
We will show that for $t>0$, this has exponentially decaying terms due to the resonant states, oscillatory terms due to the bound and anti-bound states and power-law decaying terms due to integrals.
We stress that the integrals are not background integrals but do appear in the coefficients of the resonant-state expansion.
For $t<0$, the exponentially decaying terms are replaced by the exponentially growing terms due to the anti-resonant states.
This is exactly where the time-reversal symmetry is broken for the first time in the sense that we have to take contributions of resonant-state poles for $t>0$ and those of anti-resonant-state poles for $t<0$;
we will discuss this point further at the end of the present section.

\begin{figure}
\centering
\includegraphics[width=0.8\textwidth]{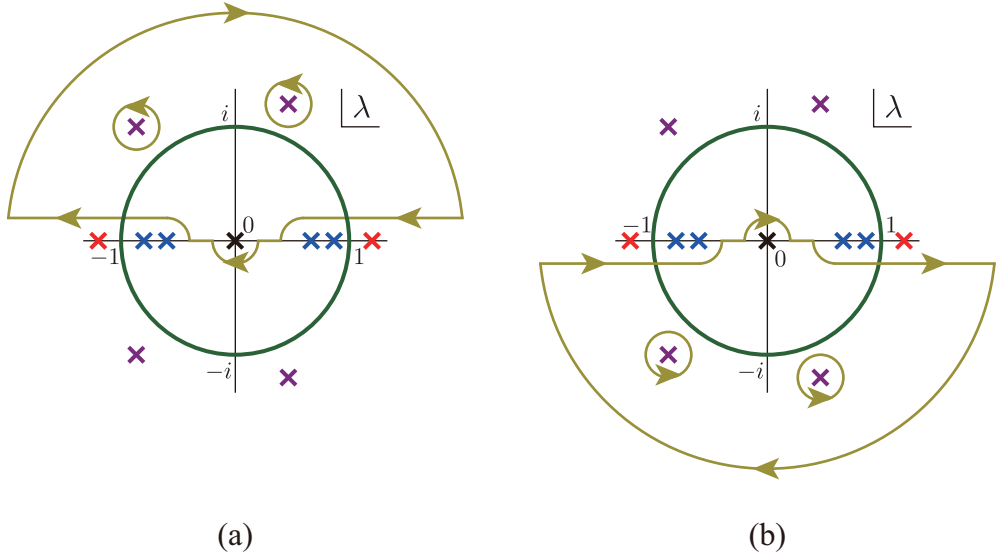}
\caption{The contours for the integration~\eqref{eq910}: (a) the contour $C_{2+}$ for $t>0$; (b) the contour $C_{2-}$ for $t<0$.}
\label{fig8}
\end{figure}

We compute Eq.~\eqref{eq901} from Eq.~\eqref{eq880} in the form
\begin{align}\label{eq910}
\bra d_j|e^{-iHt}|d_i\ket=&\frac{1}{2\pi i}\sum_{n=1}^{2N}\int_{C_2}\exp\left[i\left(\lambda+\frac{1}{\lambda}\right)t\right]
\bra d_j|\psi_n\ket\frac{\lambda_n}{\lambda-\lambda_n}\bra\tilde{\psi}_n|d_i\ket \left(-\lambda+\frac{1}{\lambda}\right)d\lambda.
\end{align}
The integration has two essential singularities because of the exponential factor in the integrand, one at $\lambda=0$ and the other at $\lambda=\infty$.
In order to avoid the contributions from the essential singularities, we have to modify the contour $C_2$ in the ways specified in Fig.~\ref{fig8}(a) and~(b) for $t>0$ and $t<0$, respectively,  and thereby taking different poles in the complex $\lambda$ plane.
This is where the time-reversal symmetry of the expansion is broken for the first time.
We can thus pinpoint the occurrence of the time-reversal symmetry breaking. 
 By the time-reversal symmetry breaking, we mean here that for $t>0$ the integration contour includes the poles of Green's function only in the upper half $\lambda$ plane (the lower half energy plane) and for $t<0$ only those in the lower half $\lambda$ plane (the upper half energy plane).

For $t>0$, the half-circle part of the contour $C_{2+}$ far away from the origin vanishes because the imaginary part of $\lambda$ diverges positively in the exponent while the imaginary part of $1/\lambda$  vanishes negatively.
On the other hand, the small half-circle part of the contour around the essential singularity at the origin $\lambda=0$ also vanishes because the imaginary part of $1/\lambda$ diverges positively in the exponent while the imaginary part of $\lambda$ vanishes negatively.
We thereby eliminate the contributions from the two essential singularities at $\lambda=0$ and $\lambda=\infty$.
Therefore, we have the contributions from all the resonant-state poles $\lambda_n$ in the upper half plane, the half-circle contributions from all bound and anti-bound states on the real axis, and the principal part of the integration over the real axis:
\begin{align}\label{eq920}
\lefteqn{
\bra d_j|e^{-iHt}|d_i\ket
}\nonumber\\
=&\sum_{n\in\mbox{\scriptsize res.}}e^{-iE_nt}\bra d_j|\psi_n\ket \bra\tilde{\psi}_n|d_i\ket\lambda_n(-2i\sin k_n)
\\\label{eq921}
+\frac{1}{2}&\sum_{\substack{n\in\mbox{\scriptsize bound}\\ n\in\mbox{\scriptsize anti-bound}}}
e^{-iE_nt}\bra d_j|\psi_n\ket \bra\tilde{\psi}_n|d_i\ket\lambda_n(-2i\sin k_n)
\\\label{eq940}
-\frac{1}{2\pi i}&\sum_{n=1}^{2N}\mathop{\mathrm{P}}\int_{-\infty}^\infty d\lambda
\exp\left[i\left(\lambda+\frac{1}{\lambda}\right)t\right]
\bra d_j|\psi_n\ket\frac{\lambda_n}{\lambda-\lambda_n}\bra\tilde{\psi}_n|d_i\ket \left(-\lambda+\frac{1}{\lambda}\right).
\end{align}
Note that each term in the the first line~\eqref{eq920} exponentially decays in time because the summation is restricted to the resonant states, whereas each term in the second line~\eqref{eq921} oscillates in time because the summation is restricted to the states with real eigenvalues.

Let us evaluate each integration in the third line~\eqref{eq940} in the saddle-point approximation, which will produce the power law $t^{-3/2}$.
The saddle points of the exponent $it(\lambda+1/\lambda)$ are at $\lambda=\pm 1$, which correspond to the band edges  $E=\mp 2$ in Fig.~\ref{fig5}(a).
Indeed, the band edges are branch points in the complex $E$ plane, which are known to produce non-Markovian dynamics without a characteristic time scale and hence cause a power-law decay in the long-time limit.\cite{Khalfin57}
The saddle points of the integral in Eq.~\eqref{eq940} always correspond to  the band edges because these are the minimum and maximum values of the energy $E(\lambda) =-( \lambda+\lambda^{-1})$, for which $dE/d\lambda=0$.

\begin{figure}
\centering
\includegraphics[width=0.3\textwidth]{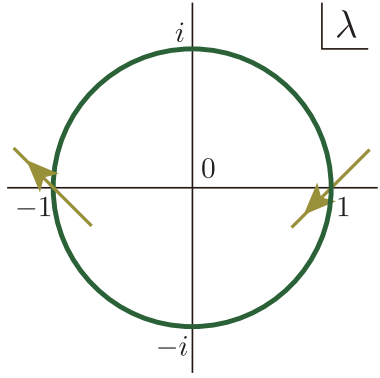}
\caption{The rotation of the integration contour around the saddle points $\lambda=\pm 1$.}
\label{fig9}
\end{figure}

We can expand the exponent around each saddle point in the form
\begin{align}\label{eq950}
it\left(\lambda+\frac{1}{\lambda}\right)=\pm2it\pm it(\lambda\mp1)^2+\mathop{\mathrm{O}}\left((\lambda\mp1)^3\right),
\end{align}
 where the upper signs correspond to the lower band edge $E=-2$ and the lower ones to the upper band edge $E=+2$.
We can convert the exponential function in the line~\eqref{eq940} into the Gaussian form $e^{-s^2}$ by choosing a new integration variable $s$ around each saddle point in the form
\begin{align}\label{eq960}
s=&\sqrt{\mp it}(\lambda\mp1) =e^{\mp i\pi/4}\sqrt{t}(\lambda \mp 1),
\end{align}
or by rotating the integration contour around the saddle points as shown in Fig.~\ref{fig9}.
This is legitimate only when the range of the Gaussian function, which is $\sim 1/\sqrt{t}$, is narrow enough not to include any bound or anti-bound states.
In other words, the evaluation here is correct in a long time scale or if any bound or anti-bound states are not close to the band edges.
It may not be correct in a shorter time scale or when a bound or anti-bound state approaches a band edge, which was indeed observed in Ref.~\citep{Garmon13}.

Coming back to the saddle-point approximation, we can approximate the integrand in the term~\eqref{eq940} around the two saddle points as
\begin{align}
\lefteqn{
\exp\left[i\left(\lambda+\frac{1}{\lambda}\right)t\right]
\frac{\lambda_n}{\lambda-\lambda_n}\left(-\lambda+\frac{1}{\lambda}\right)
}
\nonumber\\
= & -\frac{\lambda_n  e^{-s^2}}{(\lambda_n\mp1)-e^{\pm i\pi/4} s/\sqrt{t}}
\left(-2e^{\pm i\pi/4} \frac{s}{\sqrt{t}}+\mathop{\mathrm{O}}(s^2)\right)
\nonumber\\
= & \frac{2e^{\pm i\pi/4}\lambda_n}{\lambda_n\mp 1}\left(1+\frac{e^{\pm i\pi/4}}{\lambda_n\mp1}\frac{s}{\sqrt{t}}\right) \frac{s}{\sqrt{t}}e^{-s^2}
+\mathop{\mathrm{O}}(s^3e^{-s^2}).
\end{align}
in the long-time limit $t\to\infty$.
Because the  integral of $s e^{-s^2}$ vanishes, the greatest contribution in the long-time limit comes from
\begin{align}\label{eq270}
&e^{\pm 2it}\frac{2e^{\pm i\pi/4}\lambda_n}{\lambda_n \mp 1}\frac{e^{\pm i\pi/4}}{\lambda_n \mp 1}
\int_{-\infty}^\infty e^{-s^2}\left(\frac{s}{\sqrt{t}}\right)^2
\frac{ds}{e^{\mp i\pi/4}\sqrt{t}}
\nonumber\\
=&e^{\pm 2it}\frac{\sqrt{\pi}e^{\pm 3 i\pi/4}}{t^{3/2}}\frac{\lambda_n}{(\lambda_n \mp 1)^2}
\nonumber\\
=&-e^{\pm 2it}\frac{\sqrt{\pi}e^{\pm 3 i\pi/4}}{t^{3/2}}\frac{1}{E_n\pm 2}.
\end{align}
We thereby summarize the terms~\eqref{eq920}--\eqref{eq940} as
\begin{align}\label{eq275}
\lefteqn{
\bra d_j|e^{-iHt}|d_i\ket
}\nonumber\\
=&-\sum_{n\in\mbox{\scriptsize res.}}e^{-iE_nt}\bra d_j|\psi_n\ket \bra\tilde{\psi}_n|d_i\ket 2ie^{ik_n}\sin k_n
\\
&-\frac{1}{2}\sum_{\substack{n\in\mbox{\scriptsize bound}\\ n\in\mbox{\scriptsize anti-bound}}}e^{-iE_nt}\bra d_j|\psi_n\ket \bra\tilde{\psi}_n|d_i\ket 2ie^{ik_n}\sin k_n
\\
&+\frac{1}{t^{3/2}}\sum_{n=1}^{2N}\sum_{\sigma=\pm 1}e^{2\sigma it}\frac{\sqrt{\pi}e^{3\sigma i\pi/4}}{2\pi i}\bra d_j|\psi_n\ket \bra\tilde{\psi}_n|d_i\ket\frac{1}{2\sigma+E_n}
\end{align}
in the long-time limit $t\to\infty$,  where $\sigma=1$ indicates the contribution from the lower band edge $E=-2$ and $\sigma=-1$ from the upper band edge $E=+2$.
We thus have exponentially decaying terms, oscillatory terms and power-law decaying terms.
The exponent $3/2$ of the power-law decay coincides with the one given in Ref.~\cite{Garmon13} for the long-time limit.

Finally for $t<0$, we modify the contour $C_2$ into the one in Fig.~\ref{fig8}(b).
The contributions from the half-circle contour far away from the origin as well as the one close to the origin vanish.
We thus eliminate the contributions from the two essential singularities again.
We therefore have the contributions from all the anti-resonant states this time instead of the resonant states.
Employing the same argument as above, we end up with 
\begin{align}\label{eq276}
\lefteqn{
\bra d_j|e^{-iHt}|d_i\ket
}\nonumber\\
=&-\sum_{n\in\mbox{\scriptsize anti-res.}}e^{-iE_nt}\bra d_j|\psi_n\ket \bra\tilde{\psi}_n|d_i\ket 2ie^{ik_n}\sin k_n
\\
&-\frac{1}{2}\sum_{\substack{n\in\mbox{\scriptsize bound}\\ n\in\mbox{\scriptsize anti-bound}}}e^{-iE_nt}\bra d_j|\psi_n\ket \bra\tilde{\psi}_n|d_i\ket 2ie^{ik_n}\sin k_n
\\\label{eq278}
&-\frac{1}{t^{3/2}}\sum_{n=1}^{2N}\sum_{\sigma=\pm 1}e^{2\sigma it}\frac{\sqrt{\pi}e^{3\sigma i\pi/4}}{2\pi i}\bra d_j|\psi_n\ket \bra\tilde{\psi}_n|d_i\ket\frac{1}{2\sigma+E_n}
\end{align}
in the long-time limit $t\to-\infty$.

An advantage in the present framework of computing the survival probability is that we can clearly see which contribution produces which time dependence;
the resonant and anti-resonant states cause the exponentially decaying and growing terms, respectively, the bound and anti-bound state cause the oscillatory terms, and the branch points cause the power-law terms.

We again stress that choosing $t>0$ or $t<0$ breaks the time-reversal symmetry in the sense that we have to take different poles for $t>0$ and $t<0$;
to be able to pinpoint the instance of time-reversal symmetry breaking is another advantage of the present framework. 

The two cases of $t>0$ and $t<0$ above correspond to the initial condition problem and the terminal condition problem for the Schr\"{o}dinger equation, respectively.
The computation of the time-evolution operator is equivalent to integrating the Schr\"{o}dinger equation. 
Since the Schr\"{o}dinger equation is a first-order differential equation with respect to time, we need to specify one boundary condition in order to obtain a physical solution.
The boundary condition in time can be either the initial condition or the terminal condition. 
In the initial-condition problem,  we seek a solution for positive times after the initial condition. 
This is equivalent to computing the time evolution operator for $t>0$ and applying it to an initial ket vector. 
In the terminal-condition problem, on the other hand, we seek a solution for negative times before the terminal condition. 
This is equivalent to computing the time evolution operator for $t<0$ and applying it to a final bra vector. 
In short, choosing $t>0$ or $t<0$ respectively corresponds to setting the boundary condition either as the initial condition or the terminal condition, and thereby breaks the time-reversal symmetry between the decaying resonant states and the growing anti-resonant states.
This view is to some extent shared by Peierls\cite{Peierls79} and Price\cite{Price06}.

We will discuss the time-reversal symmetry breaking more quantitatively in Sec.~\ref{sec13}. We will show that the broken symmetry between the resonant states and the anti-resonant states becomes exact only in the case in which the boundary condition (the initial and terminal condition) itself is time-reversal symmetric.
We will demonstrate that without the symmetry of the boundary condition, we can even see pole contributions of growing ant-resonant states in the time-evolution from an initial condition.

\section{Time evolution of a dot state: Escaping amplitude}
\label{sec12}

We next compute the escaping amplitude from a dot state,
$\bra k_{j\beta}|e^{-iHt}|d_i\ket$
and $\bra x_{j\beta}|e^{-iHt}|d_i\ket$.
We will show that there is an additional oscillatory term in the former matrix element due to the plane wave $|k_{j\beta}\ket$.

 For the calculation of the matrix element $\bra k_{j\beta}|e^{-iHt}|d_i\ket$, we use Eq.~\eqref{eq890} to have 
\begin{align} \label{eqEscapingA}
\bra k_{j\beta}|e^{-iHt}|d_i\ket
=&\frac{1}{2\pi i}\sum_{n=1}^{2N}
\int_{C_2}\exp\left[i\left(\lambda+\frac{1}{\lambda}\right)t\right]
\nonumber\\
&\times
\frac{-\sqrt{2}t_{j\beta}\sin k_{j\beta}}{E(\lambda)+2\cos k_{j\beta}} 
\bra d_j|\psi_n\ket
\frac{\lambda_n}{\lambda-\lambda_n}\bra\tilde{\psi}_n|d_i\ket 
\left(-\lambda+\frac{1}{\lambda}\right)d\lambda
\nonumber\\
=&
\frac{1}{2\pi i}\frac{t_{j\beta}}{\sqrt{2}i}\sum_{n=1}^{2N}
\int_{C_2}\exp\left[i\left(\lambda+\frac{1}{\lambda}\right)t\right]
\nonumber\\
&\times
\left(\frac{\lambda}{\lambda-e^{ik_{j\beta}}}-\frac{\lambda}{\lambda-e^{-ik_{j\beta}}}\right)
\bra d_j|\psi_n\ket
\frac{\lambda_n}{\lambda-\lambda_n}\bra\tilde{\psi}_n|d_i\ket 
\left(-\lambda+\frac{1}{\lambda}\right)d\lambda.
\end{align}
We therefore have an additional pole contribution from either $\lambda=e^{ik_{j\beta}}$ or $e^{-ik_{j\beta}}$ in the integration over $\lambda$.
Let us assume $k_{j\beta}>0$, which means an out-going wave.
The pole $\lambda=e^{ik_{j\beta}}$ is on the upper half of the unit circle $|\lambda|=1$ and the other pole $\lambda=e^{-ik_{j\beta}}$ on the lower half.
The former contributes for $t>0$ and the latter contributes for $t<0$.

For $t>0$, we therefore have
\begin{align}
\lefteqn{
\bra k_{j\beta}|e^{-iHt}|d_i\ket
}
\\
=
&\sqrt{2} t_{j\beta}\sin k_{j\beta}\sum_{n:\mathrm{res}}e^{-iE_nt}\frac{2i\lambda_n\sin k_n}{E_n+2\cos k_{j\beta}}\bra d_j|\psi_n\ket\bra\tilde{\psi}_n|d_i\ket 
\\
+
&\frac{1}{\sqrt{2}} t_{j\beta}\sin k_{j\beta}\sum_{n:\mathrm{b,ab}}e^{-iE_nt}\frac{2i\lambda_n\sin k_n}{E_n+2\cos k_{j\beta}}\bra d_j|\psi_n\ket\bra\tilde{\psi}_n|d_i\ket 
\\
-&\sqrt{2} t_{j\beta}\sin k_{j\beta}\sum_{n=1}^{2N}
e^{2it\cos k_{j\beta}}\frac{e^{ik_{j\beta}} \lambda_n}{e^{ik_{j\beta}}-\lambda_n}\bra d_j|\psi_n\ket\bra\tilde{\psi}_n|d_i\ket
\\\label{eq1160}
-\frac{1}{2\pi i}&\sqrt{2} t_{j\beta}\sin k_{j\beta}\sum_{n=1}^{2N}\mathrm{\mathop{P}}\int_{-\infty}^\infty\exp\left[i\left(\lambda+\frac{1}{\lambda}\right)t\right]
\nonumber\\
&\qquad\times\frac{1}{\lambda+1/\lambda-2\cos k_{j\beta}}
\bra d_j|\psi_n\ket
\frac{\lambda_n}{\lambda-\lambda_n}\bra\tilde{\psi}_n|d_i\ket 
\left(-\lambda+\frac{1}{\lambda}\right)d\lambda.
\end{align}
We evaluate the integral in the last line~\eqref{eq1160} again in the saddle-point approximation.
After the transformation to the new integration variable in Eq.~\eqref{eq960}, the integrand takes the form
\begin{align}
&
\exp\left[i\left(\lambda+\frac{1}{\lambda}\right)t\right]
\frac{1}{\lambda+1/\lambda-2\cos k_{j\beta}}
\frac{\lambda_n}{\lambda-\lambda_n}\left(-\lambda+\frac{1}{\lambda}\right)
\nonumber\\
&=\frac{e^{-s^2}}{\pm 2-2\cos k_{j\beta}+\mathop{\mathrm{O}}(s^2)}
\nonumber\\
&\times\left[\frac{2e^{\pm i\pi/4}\lambda_n}{\lambda_n\mp 1}\left(1+\frac{e^{\pm i\pi/4}}{\lambda_n\mp1}\frac{s}{\sqrt{t}}\right) \frac{s}{\sqrt{t}}+\mathop{\mathrm{O}}(s^3)\right]
\end{align}
in the long-time limit $t\to\infty$.
The greatest contribution is similar to Eq.~\eqref{eq270}:
\begin{align}
-\frac{1}{\pm 2-2\cos k_{j\beta}}\frac{\sqrt{\pi}e^{\pm 3 i\pi/4}}{t^{3/2}}\frac{1}{E_n\pm 2}.
\end{align}
We thereby summarize all lines as
\begin{align}
\lefteqn{
\bra k_{j\beta}|e^{-iHt}|d_i\ket=\sqrt{2} t_{j\beta}\sin k_{j\beta}
}
\\\label{eq1218}
&\times\left[\sum_{n:\mathrm{res}}e^{-iE_nt}\frac{2ie^{ik_n}\sin k_n}{E_n+2\cos k_{j\beta}}\bra d_j|\psi_n\ket\bra\tilde{\psi}_n|d_i\ket \right.
\\\label{eq1219}
&+
\frac{1}{2}\sum_{n:\mathrm{b,ab}}e^{-iE_nt}\frac{2ie^{ik_n}\sin k_n}{E_n+2\cos k_{j\beta}}\bra d_j|\psi_n\ket\bra\tilde{\psi}_n|d_i\ket 
\\\label{eq1220}
&-\sum_{n=1}^{2N}
e^{2it\cos k_{j\beta}}\frac{e^{ik_{j\beta}} e^{ik_n}}{e^{ik_{j\beta}}-e^{ik_n}}\bra d_j|\psi_n\ket\bra\tilde{\psi}_n|d_i\ket
\\\label{eq1221}
&\left.+\frac{1}{t^{3/2}}\sum_{n=1}^{2N}\sum_{\sigma=\pm 1}e^{2\sigma it}
\frac{\sqrt{\pi}e^{3\sigma i\pi/4}}{2\pi i}\bra d_j|\psi_n\ket \bra\tilde{\psi}_n|d_i\ket\frac{1}{2\sigma-2\cos k_{j\beta}}\frac{1}{2\sigma+E_n} \right]
\end{align}
in the long-time limit $t\to\infty$.
For $t<0$, we similarly have
\begin{align}
\lefteqn{
\bra k_{j\beta}|e^{-iHt}|d_i\ket=\sqrt{2} t_{j\beta}\sin k_{j\beta}
}
\\\label{eq1268}
&\times\left[\sum_{n:\mathrm{ar}}e^{-iE_nt}\frac{2ie^{ik_n}\sin k_n}{E_n+2\cos k_{j\beta}}\bra d_j|\psi_n\ket\bra\tilde{\psi}_n|d_i\ket \right.
\\\label{eq1269}
&+
\frac{1}{2}\sum_{n:\mathrm{b,ab}}e^{-iE_nt}\frac{2ie^{ik_n}\sin k_n}{E_n+2\cos k_{j\beta}}\bra d_j|\psi_n\ket\bra\tilde{\psi}_n|d_i\ket 
\\\label{eq1270}
&-\sum_{n=1}^{2N}
e^{2it\cos k_{j\beta}}\frac{e^{-ik_{j\beta}}e^{ik_n}}{e^{-ik_{j\beta}}-e^{ik_n}}\bra d_j|\psi_n\ket\bra\tilde{\psi}_n|d_i\ket
\\\label{eq1271}
&\left.-\frac{1}{t^{3/2}}\sum_{n=1}^{2N}\sum_{\sigma=\pm 1}e^{2\sigma it}
\frac{\sqrt{\pi}e^{3\sigma i\pi/4}}{2\pi i}\bra d_j|\psi_n\ket \bra\tilde{\psi}_n|d_i\ket\frac{1}{2\sigma-2\cos k_{j\beta}}\frac{1}{E_n+2\sigma}\right]
\end{align}
in the long-time limit $t\to-\infty$.
The terms~\eqref{eq1218} and~\eqref{eq1268} decay and grow exponentially, respectively, the terms~\eqref{eq1219} and~\eqref{eq1269} oscillate in time, and the terms~\eqref{eq1221} and~\eqref{eq1271} decay as $t^{-3/2}$, which were also present in the survival amplitude~\eqref{eq275}--\eqref{eq278}.
The terms~\eqref{eq1220} and~\eqref{eq1270} are the additional oscillatory terms due to the plane wave $|k_{j\beta}\ket$.

For the calculation of the matrix element $\bra x_{j\beta}|e^{-iHt}|d_i\ket$, we use Eq.~\eqref{eqEscapingA} to have
\begin{align}\label{eq14000}
\bra x_{j\beta}|e^{-iHt}|d_i \ket 
&= \int_{-\pi}^{\pi}\frac{dk_{j\beta}}{2\pi}\bra x_{j\beta} |k_{j\beta}\ket\bra k_{j\beta}|e^{-iHt}|d_i\ket
\nonumber\\
&= \frac{1}{2\pi i} \frac{t_{j\beta}}{i}\sum_{n=1}^{2N} \int_{C_2}d\lambda\left(-\lambda+\frac{1}{\lambda}\right)\exp\left[i\left(\lambda+\frac{1}{\lambda}\right)t\right]
\nonumber\\
& \times\bra d_j|\psi_n\ket\frac{\lambda_n}{\lambda-\lambda_n}\bra\tilde{\psi}_n|d_i\ket
\nonumber\\
& \times \int_{-\pi}^\pi \frac{dk_{j\beta}}{2\pi} \sin(k_{j\beta} x_{j\beta})
\left(\frac{\lambda}{\lambda-e^{ik_{j\beta}}}-\frac{\lambda}{\lambda-e^{-ik_{j\beta}}}\right),
\end{align}
where we used $\bra x_{j\beta} |k_{j\beta}\ket = \sqrt{2}\sin(k_{j\beta} x_{j\beta})$ in Eq.~\eqref{plainwave}.
We can easily carry out the integral with respect to $k_{j\beta}$ by changing the integration variable to $\lambda_0=e^{ik_{j\beta}}$:
\begin{align}\label{eq14100}
&\int_{-\pi}^\pi \frac{dk_{j\beta}}{2\pi} \sin(k_{j\beta} x_{j\beta})
\left(\frac{\lambda}{\lambda-e^{ik_{j\beta}}}-\frac{\lambda}{\lambda-e^{-ik_{j\beta}}}\right)
\nonumber\\
&=\int_{C_0}  \frac{d\lambda_0}{2\pi i\lambda_0} \frac{{\lambda_0}^{x} - {\lambda_0}^{-x}}{2 i}
\left(\frac{\lambda}{\lambda-\lambda_0}-\frac{\lambda}{\lambda-\lambda_0^{-1}}\right),
\end{align}
where $C_0$ is the contour of the counterclockwise unit circle and we left out the subscript of $x_{j\beta}$ for brevity.
Since $\lambda$ runs on the contour $C_2$ specified in Fig.~\ref{fig7}(c), which is inside $C_0$ in the complex $\lambda_0$ plane, the pole at $\lambda_0=\lambda$ is inside the unit circle but the one at $\lambda_0=\lambda^{-1}$ is outside it. 
For evaluating the term ${\lambda_0}^x$ in Eq.~\eqref{eq14100}, we take the pole $\lambda_0=\lambda$, while for the term ${\lambda_0}^{-x}$, we take the pole $\lambda_0=\lambda^{-1}$ because $x\geq 1$, each of which yields the residue $i\lambda^x/2$.
Equation~\eqref{eq14100} therefore reduces to $i\lambda^x$, which gives the following simplified form of the matrix element~\eqref{eq14000}:
\begin{align} \label{eq14200}
\bra x_{j\beta}|e^{-iHt}|d_i \ket 
=\frac{t_{j\beta}}{2\pi i}
\sum_{n=1}^{2N} \int_{C_2}\exp\left[i\left(\lambda+\frac{1}{\lambda}\right)t\right]
\bra d_j|\psi_n\ket\frac{\lambda_n}{\lambda-\lambda_n}\bra\tilde{\psi}_n|d_i\ket\left(-\lambda+\frac{1}{\lambda}\right)\lambda^{x_{j\beta}}d\lambda.
\end{align}
This is indeed the same as Eq.~\eqref{eq910} except for the factor $t_{j\beta}\lambda^{x_{j\beta}}$, which is consistent with Eq.~\eqref{eqb240}.
Since the factor $\lambda^x$ reduces to $\sigma^x$ in the saddle-point approximation~\eqref{eq950}--\eqref{eq270}, we accordingly modify Eqs.~\eqref{eq275}--\eqref{eq278} to have
\begin{align}
\lefteqn{
\bra x_{j\beta}|e^{-iHt}|d_i\ket
}\nonumber\\
=&-t_{j\beta}\sum_{n\in\mbox{\scriptsize res.}}e^{-iE_nt}\bra d_j|\psi_n\ket \bra\tilde{\psi}_n|d_i\ket 2ie^{ik_n(x_{j\beta}+1)}\sin k_n
\\
&-\frac{t_{j\beta}}{2}\sum_{\substack{n\in\mbox{\scriptsize bound}\\ n\in\mbox{\scriptsize anti-bound}}}e^{-iE_nt}\bra d_j|\psi_n\ket \bra\tilde{\psi}_n|d_i\ket 2ie^{ik_n(x_{j\beta}+1)}\sin k_n
\\
&+\frac{t_{j\beta}}{t^{3/2}}\sum_{n=1}^{2N}\sum_{\sigma=\pm 1}e^{2\sigma it}\frac{\sqrt{\pi}e^{3\sigma i\pi/4}}{2\pi i}\bra d_j|\psi_n\ket \bra\tilde{\psi}_n|d_i\ket\frac{\sigma^{x_{j\beta}}}{2\sigma+E_n}
\end{align}
in the limit $t\to\infty$ and
\begin{align}
\lefteqn{
\bra x_{j\beta}|e^{-iHt}|d_i\ket
}\nonumber\\
=&-t_{j\beta}\sum_{n\in\mbox{\scriptsize anti-res.}}e^{-iE_nt}\bra d_j|\psi_n\ket \bra\tilde{\psi}_n|d_i\ket 2ie^{ik_n(x_{j\beta}+1)}\sin k_n
\\
&-\frac{t_{j\beta}}{2}\sum_{\substack{n\in\mbox{\scriptsize bound}\\ n\in\mbox{\scriptsize anti-bound}}}e^{-iE_nt}\bra d_j|\psi_n\ket \bra\tilde{\psi}_n|d_i\ket 2ie^{ik_n(x_{j\beta}+1)}\sin k_n
\\
&-\frac{t_{j\beta}}{t^{3/2}}\sum_{n=1}^{2N}\sum_{\sigma=\pm 1}e^{2\sigma it}\frac{\sqrt{\pi}e^{3\sigma i\pi/4}}{2\pi i}\bra d_j|\psi_n\ket \bra\tilde{\psi}_n|d_i\ket\frac{\sigma^{x_{j\beta}}}{2\sigma+E_n}
\end{align}
in the limit $t\to-\infty$.

\section{Time-reversal symmetry breaking and growth from the initial condition}
\label{sec13}


We have shown in the two preceding sections that:
(i) for the time evolution from the initial condition, the resonant states contribute, giving decays;
(ii) for the time evolution to the terminal condition, the anti-resonant states contribute, giving growths.
One may then pose the following question.
Suppose that we trace back the time evolution from a terminal condition $|\psi(0)\ket$ to a moment in the past, $t=-t_0<0$, and find a state $|\psi(-t_0)\ket$.
The time evolution from this new initial condition $|\psi(-t_0)\ket$ then must be a growth into the state $|\psi(0)\ket$.
This would seem to contradict the above statement.

The answer to the question is as follows.
The above statement exactly applies only to the case in which  the boundary condition chosen as the initial and terminal condition is time-reversal symmetric.
Because we go back in time from the state $|\psi(0)\ket$, the state $|\psi(-t_0)\ket$ consists of anti-resonant states of exponentially large amplitudes and resonant states of exponentially small amplitudes.
In other words, it is not a time-reversal symmetric state but  an asymmetric state which was engineered so that it may be dominated by anti-resonant states.
This is why the anti-resonant states give exponentially growing contributions to the time evolution from the initial condition $|\psi(-t_0)\ket$.
%
Mathematically, these contributions are originated when we perform an integration over $k_{j\beta}$ in terms such as Eqs.~\eqref{eq1220} and~\eqref{eq1270}  with a suitable function of $k_{j\beta}$. 
We will show this explicitly in the present section.

Consider the survival amplitude~\eqref{eq901},  $\bra d_j|e^{-iHt}|d_i\ket$, again with $t>0$.
According to the expansions~\eqref{eq910}, we have terms of the form
\begin{align}\label{eq1410}
\bra d_j|e^{-iHt}|d_i\ket&=
\sum_{n=1}^{2N}
p_n \bra d_j|\psi_n\ket  \bra \tilde{\psi}_n|d_i\ket
\end{align}
with appropriate numbers $\{p_n\}$.
We showed in Sec.~\ref{sec11} that for $t>0$, the contributions of the resonant-state poles take over and the contributions  of the anti-resonant-state poles vanish in the summation.

Let us then take the time reversal (the complex conjugate) of Eq.~\eqref{eq1410}.
Noting that the Hamiltonian is a real matrix because of the time-reversal symmetry, we have
\begin{align}\label{eq1420}
\bra d_j| e^{-iH(-t)}|d_i\ket&=
\bra d_j|e^{-iHt}|d_i\ket^\ast
\nonumber\\
&=
\sum_{n=1}^{2N}
{p_n}^\ast  \bra d_j|\psi_n\ket^\ast  \bra \tilde{\psi}_n|d_i\ket^\ast
\nonumber\\
&=
\sum_{n=1}^{2N}
p_{\bar{n}}  \bra d_j|\psi_{\bar{n}}\ket  \bra \tilde{\psi}_{\bar{n}}|d_i\ket,
\end{align}
where $\psi_{\bar{n}}$ is the corresponding anti-resonant state if $\psi_n$ is a resonant state;
we used Eq.~\eqref{eq:n-nbar} here.
We showed in Sec.~\ref{sec11} too that for $(-t)<0$, the contributions of the anti-resonant-state poles take over instead of the resonant-state poles.
In particular, Eq.~\eqref{eq1420} indicates that if one resonant state has a specific contribution in Eq.~\eqref{eq1410}, the corresponding anti-resonant state has a contribution of the same magnitude in Eq.~\eqref{eq1420}.
Note that this is because the initial and terminal conditions are time-reversal symmetric states: $|d_i\ket^\ast=|d_i\ket$ and $\bra d_j|^\ast=\bra d_j|$.

In other words, if we choose a time-reversal symmetric state as initial and terminal conditions, the following statement becomes exact:
the initial-condition problem features only the decaying states while the terminal-condition problem features the growing states, and the solutions are time reversal to each other.

An amendment to this statement is in order if we choose a time-reversal \textit{asymmetric} state as the initial condition.
We will show in the following that if we choose a specifically engineered state for the initial condition, we can even observe anti-resonant contributions in the time evolution from it.

The basic argument is as follows. 
Consider an initial state where the particle is located at a site $d_i$. 
This state evolves as  $|\psi(t) \ket = e^{-i Ht} |d_i\ket$. 
Say that at $t=t_0>0$ we perform a time inversion to obtain the state $|\psi(-t_0) \ket = e^{i Ht_0} |d_i\ket$. 
The time inversion can be obtained by reversing the velocity of the particle.
We can alternatively regard this state as the solution at $t=-t_0$ of the terminal-condition problem.
We will hereafter take the state $|\psi(-t_0) \ket$ as a new initial state and consider its time evolution from it. 
In other words, we use a time-reversal asymmetric initial condition: $|\psi(-t_0)\ket\neq|\psi(-t_0)\ket^\ast$.

\begin{figure}
\centering
\includegraphics[width=0.6\textwidth]{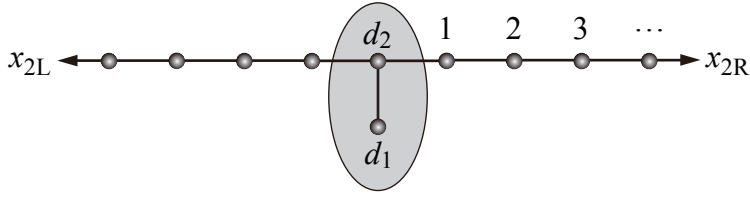}
\caption{The T-shaped quantum dot model.
The $P$ subspace is a two-dimensional space formed by the two sites $d_1$ and $d_2$ encircled by the circle, and therefore the model has four discrete eigenvalues.
The site $d_1$ is not connected to any leads directly, while the site $d_2$ is connected to two leads, which we refer to as $x_{2\mathrm{R}}$ and $x_{2\mathrm{L}}$.
In some parameter regions, there are two bound states, a resonant state and the corresponding anti-resonant state.
For the numerical demonstration here, we chose the parameter values at $t_{12}=t_{21}=t_{2\mathrm{R}}=t_{2\mathrm{L}}=1$, $\varepsilon_1=-0.85$ and $\varepsilon_2=0$.}
\label{fig:Tdot}
\end{figure}

Specifically, we will consider the `T-shaped' quantum dot model shown in Fig.~\ref{fig:Tdot} and compute the amplitude $\bra x_{2{\rm R}}|e^{-i H t}|\psi(-t_0)\ket$ that the particle is found on the right lead $x_{2\mathrm{R}}$ for $t>0$ as it is being absorbed by the dot  site $d_1$. 
As we argue now, this amplitude in fact contains contributions growing exponentially for $t>0$ due to the anti-resonant states. 
Indeed,  we have
\begin{align} \label{Tinv}
\bra x_{2{\rm R}}|e^{-i H t}|\psi(-t_0)\ket &=\bra x_{2{\rm R}}|e^{-i H t} e^{i H t_0}|d_1\ket 
\nonumber\\
&= \bra x_{2{\rm R}}|e^{-i H (t-t_0)} |d_1\ket. 
\end{align}
As discussed in section \ref{sec12}, when time $t$ in the amplitude $\bra x_{2{\rm R}}|e^{-i H t}|d_1\ket$ is negative,  the anti-resonant states give contributions that grow exponentially in the form $\exp(-i E_\mathrm{ar} t)$ with $\mathop{\mathrm{Im}} E_\mathrm{ar}>0$ as $t$ increases. In the amplitude of Eq.~\eqref{Tinv}, this occurs for $t<t_0$. Therefore, even though we have $t>0$, the anti-resonant states  give exponentially growing contributions to this amplitude until $t = t_0$.

\begin{figure}
\centering
\includegraphics[width=.5\textwidth]{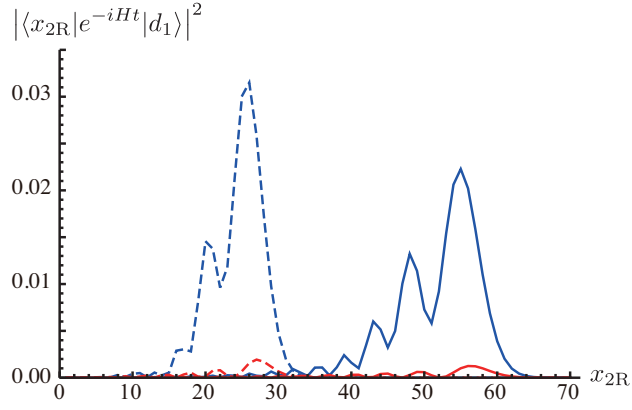}\\
(a)\\
\includegraphics[width=.5\textwidth]{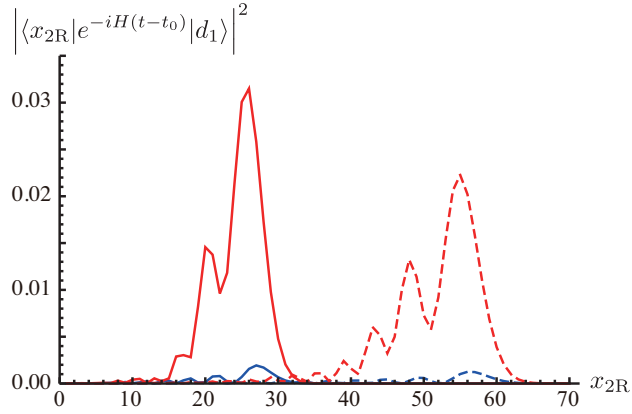}\\
(b)
\caption{
(a) Resonant and anti-resonant pole contributions to the escaping probability~\eqref{eq1450} for $t=15$ (dashed lines) and $t=30$ (solid lines). 
The wave packet was emitted from the dot site $d_1$ at time $t=0$ and moves to the right as time $t$ increases.
The resonant-state contributions are the blue lines (larger peaks) and the anti-resonant-state contributions are the red lines (smaller peaks).
(b) Resonant and anti-resonant pole contributions to the probability~\eqref{eq1460} with $t_0=30$ for $t=0$  (dashed lines) and $t=15$  (solid lines).
As time $t$  increases, both wave packets move to the left towards the contact site $d_2$ at $x_{2\mathrm{R}}=0$.
The resonant-state contributions are the blue lines (smaller peaks) and the anti-resonant-state contributions are the red lines (larger peaks).
For the parameter values specified in Fig.~\ref{fig:Tdot}, we have the resonant and anti-resonant poles at $\lambda_\mathrm{res} = 0.502834 - 1.21680 i$ and $\lambda_\mathrm{ar} = 0.502834 + 1.21680 i $, respectively.}
\label{fig:emitted} 
\end{figure}

For comparison, let us first compute the escaping probability
\begin{align}\label{eq1450}
\left|\bra x_{2\mathrm{R}} | e^{-iHt} |d_1\ket\right|^2
\end{align}
for the T-shaped model, using Eq.~\eqref{eq14200}. 
The model, in an appropriate parameter region, has only one resonant state and one anti-resonant state forming a complex-conjugate pair, along with two bound states.
We show in Fig.~\ref{fig:emitted}(a) a numerical evaluation of both the resonant-state contribution $\lambda_n=\lambda_\mathrm{res}$ and the anti-resonant-state contribution $\lambda_n = \lambda_\mathrm{ar}$ to the amplitude in Eq.~\eqref{eq14200}.
We see that for $t>0$ the resonant-state contribution is far greater than the anti-resonant-state contribution.
The resonant-state contribution forms most of the wave packet emitted from the dot site $d_1$. 
As time increases, the wave packet shifts in the positive $x_{2\mathrm{R}}$ direction away from the dot site $d_1$ and the contact site $d_2$ at $x_{2\mathrm{R}}=0$. 
For any given specific location through which the wave packet passes, there is a time period during which the amplitude decreases exponentially; 
this is directly related to the exponential shape of the wave packet shown in Fig.~\ref{fig:emitted} and corresponds to the exponential decay due to the resonant-state pole. 

In contrast, in the probability
\begin{align}\label{eq1460}
\left|\bra x_{2\mathrm{R}} | e^{-iHt} |\psi(-t_0)\ket\right|^2
\end{align}
for the time-inverted state $|\psi(-t_0)\ket=e^{+iHt_0}|d_1\ket$, the roles of the resonant-state and anti-resonant-state contributions are exchanged as shown in Fig.~\ref{fig:emitted}(b). 
The wave packet now moves in the negative $x_{2\mathrm{R}}$ direction towards the contact site $d_2$ at $x_{2\mathrm{R}}=0$.  
As the wave packet passes through a given location on the lead, the amplitude grows exponentially. 
All exponential growth stops at  $t=t_0$ when the wave packet has been absorbed by the dot site $d_1$. 
Subsequently the wave packet is re-emitted as shown in Fig.~\ref{fig:emitted}(a).


\section{Resonant scattering of a wave packet} 
\label{sec14}
In the present section we describe time-reversal symmetry breaking in resonant scattering of a wave packet. 
At $t=0$ we specify a time-reversal symmetric wave packet located on a lead.  
We will show that again time-reversal symmetry is broken depending on whether we regard the wave packet at $t=0$ as either a terminal or an initial condition. 
We will again consider the T-shaped quantum dot model described in the previous section. 

At $t=0$ we have a Gaussian wave packet (Fig.~\ref{fig:Gaussian1}) on the left lead $x_{2\mathrm{L}}$, given by
\begin{align}\label{init-term-cond}
\bra x_{2\beta}  |\varphi(0)\ket\ = \delta_{\beta, \mathrm{L}} A e^{-(x_{2\mathrm{L}}-x_0)^2/\sigma^2} e^{i x_{2\mathrm{L}} k_0}\theta(x_{2\mathrm{L}}) 
\end{align}
where $A$ is the normalization constant, $x_0$ is the location of the peak, $\sigma$ is the width, $k_0$ is the initial momentum and $\theta(x)$ is the step function, equal to $1$ for $x\ge 1$ and $0$ otherwise. We will consider the case $k_0=0$, for which the wave  packet is  time-reversal symmetric.
\begin{figure}
\centerline{\includegraphics[width=.5\textwidth]{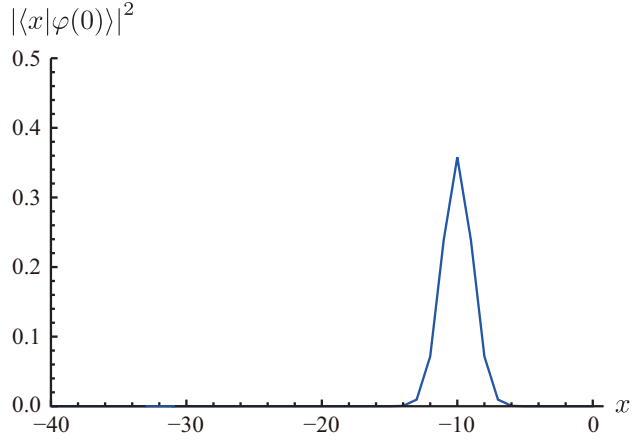}}
\caption{Gaussian wave packet 
$|\bra x|\varphi(t)\ket|^2$ 
at $t=0$ for the T-shaped model of Fig.~\ref{fig:Tdot}. 
In this figure as well as in Fig.~\ref{fig:Gaussian2} the positions $x_{2\mathrm{L}}$ on the left lead are represented by negative values of $x=-x_{2\mathrm{L}}$, while the positions $x_{2\mathrm{R}}$ on the right lead by positive $x=x_{2\mathrm{R}}$. 
The contact site $d_2$ is at $x=0$.}
\label{fig:Gaussian1} 
\end{figure}

The wave packet evolves as
\begin{align}
\bra x_{2\beta} |\varphi(t)\ket\ = \sum_{x_{2\mathrm{L}}'=1}^\infty \bra x_{2\beta}| e^{-i H t}|x_{2\mathrm{L}}'\ket\bra x_{2\mathrm{L}}'|\varphi(0)\ket.
\end{align}
In order to isolate the contribution from each discrete eigenvalue (the point spectra) of the Hamiltonian,
we will  decompose the  transition amplitude $\bra x_{2\beta}| e^{-i H t}|x_{2\mathrm{L}}'\ket$ into the following terms: 
\begin{align}
\bra x_{2\beta}| e^{-i H t}|x_{2\mathrm{L}}'\ket = \bra x_{2\beta}| e^{-i H t}|x_{2\mathrm{L}}'\ket_0 + \sum_{n=1}^{2 N} \bra x_{2\beta}| e^{-i H t}|x_{2\mathrm{L}}'\ket_n,
\end{align}
where the first term is the free time-evolution (involving only the left lead, not the dot), corresponding to the first term in the right-hand side of Eq.~\eqref{eq895}, whereas the other terms correspond to the states with $n=1,2,\cdots, 2N$ in Eq.~\eqref{eq895}. 
The free-evolving term is only non-zero when $x_{2\beta}$ is on the left lead $(\beta=\mathrm{L})$ and is given by 
\begin{align}\label{eq:free-evol}
\bra x_{2\beta} | e^{-i H t}|x_{2\mathrm{L}}'\ket_0 =\delta_{\beta,\mathrm{L}} \int_{-\pi}^{\pi} \frac{dk}{2\pi} 2 \sin(k x_{2\beta} ) \sin(k x_{2\mathrm{L}}') e^{2 i t \cos k}.
\end{align}
The other terms are given by 
\begin{align} 
\bra x_{2\beta}| e^{-i H t}|x_{2\mathrm{L}}'\ket_n 
=&\frac{1}{2\pi i}
\int_{C_2}\exp\left[i\left(\lambda+\frac{1}{\lambda}\right)t\right] \left(-\lambda+\frac{1}{\lambda}\right)
\nonumber\\
&\times \int_{-\pi}^{\pi} \frac{dk_{2\beta}}{2\pi} \int_{-\pi}^{\pi} \frac{dk_{2\mathrm{L}}'}{2\pi} 2 \sin(k_{2\beta} x_{2\beta} ) \sin(k_{2\mathrm{L}}' x_{2\mathrm{L}}')  \nonumber\\
&\times
\frac{-\sqrt{2}t_{2\beta}\sin k_{2\beta}}{E(\lambda)+2\cos k_{2\beta}} 
\bra d_2|\psi_n\ket
\frac{\lambda_n}{\lambda-\lambda_n}\bra\tilde{\psi}_n|d_2\ket \frac{-\sqrt{2}t_{2L}\sin k_{2\mathrm{L}}'}{E(\lambda)+2\cos k_{2\mathrm{L}}'} 
d\lambda.
\nonumber\\
\end{align}
We can evaluate the integrals over $k_{2\beta}$ and $k_{2\mathrm{L}}'$ similarly to Eq.~\eqref{eq14200} as follows: 
\begin{align} \label{eq:wpevol}
\bra x_{2\beta}| e^{-i H t}|x_{2\mathrm{L}}'\ket_n 
=&\frac{1}{2\pi i}
\int_{C_2}\exp\left[i\left(\lambda+\frac{1}{\lambda}\right)t\right] \left(-\lambda+\frac{1}{\lambda}\right) \lambda^{(x_{2\beta}+x_{2\mathrm{L}}')}
\nonumber\\
&\times 
\bra d_2|\psi_n\ket
\frac{\lambda_n}{\lambda-\lambda_n}\bra\tilde{\psi}_n|d_2\ket\ d\lambda.
\end{align}
Note that this expression is independent of the lead ($\beta =\mathrm{L}$ or $\beta =\mathrm{R}$). 
Therefore the $n$ components of the transition amplitude (with $n\ne 0$) are symmetric around the dot site $d_1$; 
they are either incoming or outgoing scattered wave packets. 

We show in Fig.~\ref{fig:Gaussian2} the  components of the wave packets
\begin{align} \label{eq:wp_n}
\bra x_{2\beta}|\varphi(t)\ket_n \equiv \sum_{x_{2\mathrm{L}}'} \bra x_{2\beta}|e^{-iHt}|x_{2\mathrm{L}}' \ket_n \bra x_{2\mathrm{L}}'| \varphi(0)\ket
\end{align}
at different times.
The negative times represent the time evolution towards  the terminal condition \eqref{init-term-cond} at $t=0$. 
The positive times represent the time evolution starting at $t=0$, which is now regarded as an initial condition. 
\begin{figure}[p]
\begin{center}
\includegraphics[width=.49\textwidth]{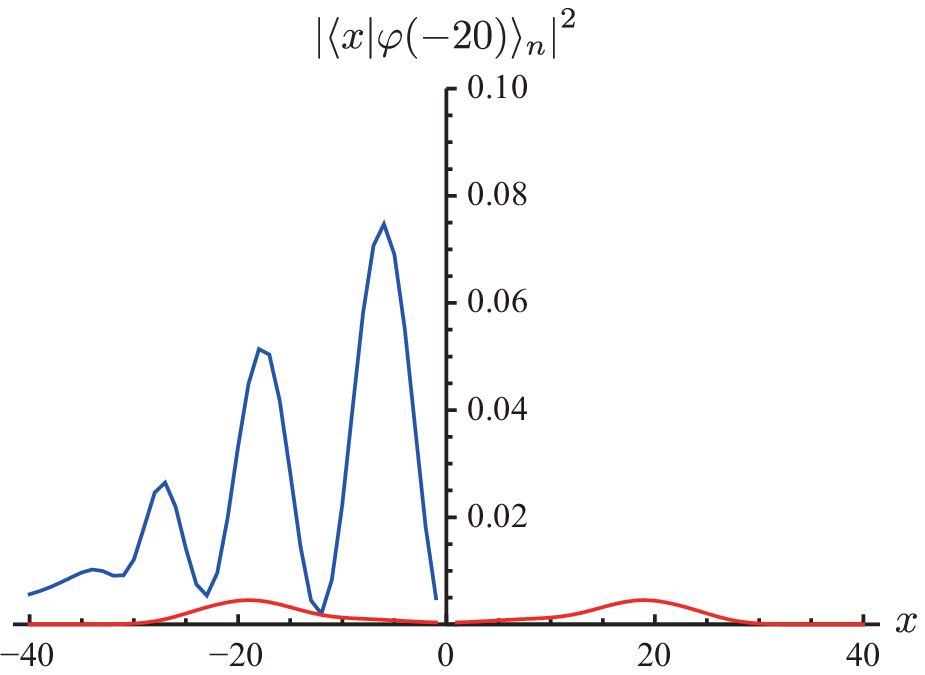}
\hfill
\includegraphics[width=.49\textwidth]{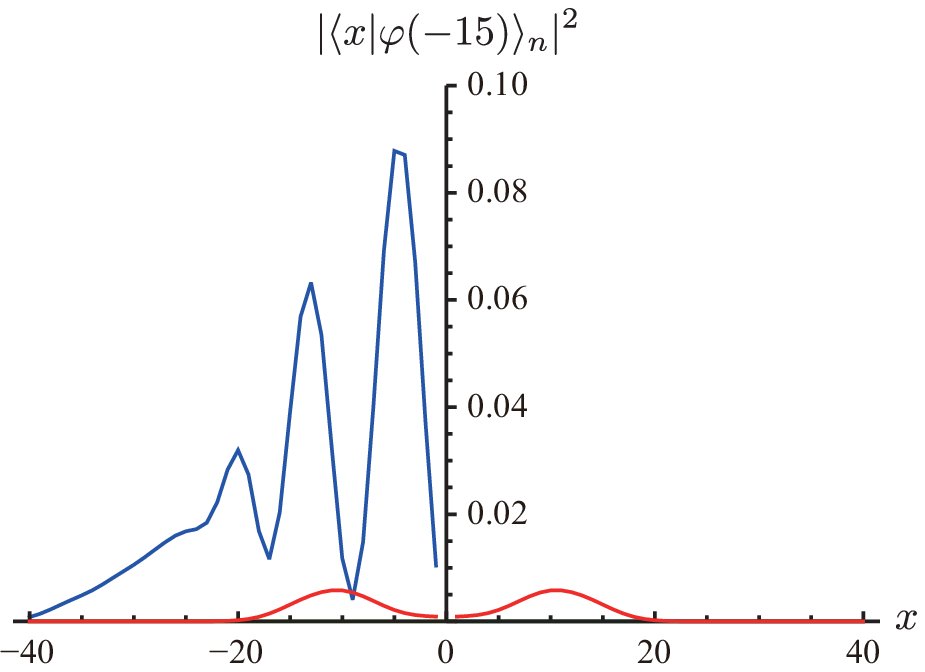}

(a)\hspace{0.48\textwidth}(b)

\vspace{\baselineskip}

\includegraphics[width=.49\textwidth]{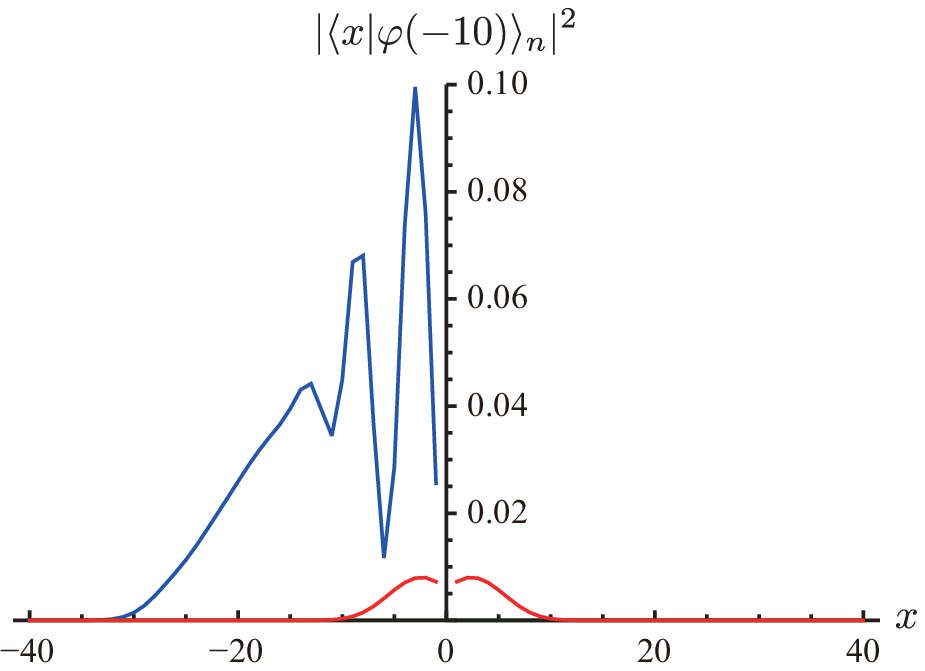}
\hfill
\includegraphics[width=.49\textwidth]{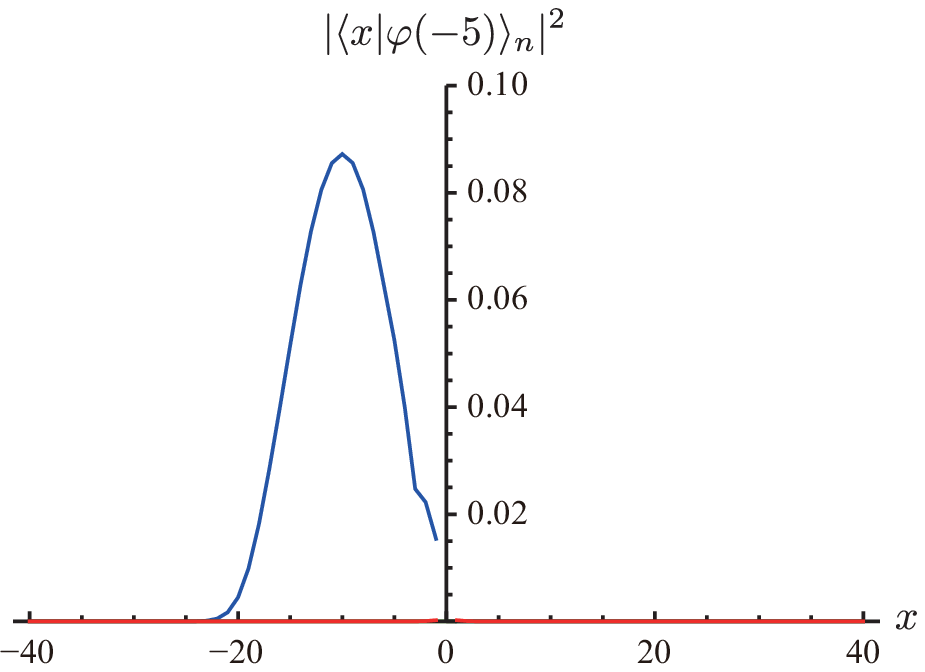}

(c)\hspace{0.48\textwidth}(d)
\end{center}
\caption{\textit{Continued to the next page.}}
\end{figure}
\addtocounter{figure}{-1}
\begin{figure}[p]
\begin{center}
\includegraphics[width=.49\textwidth]{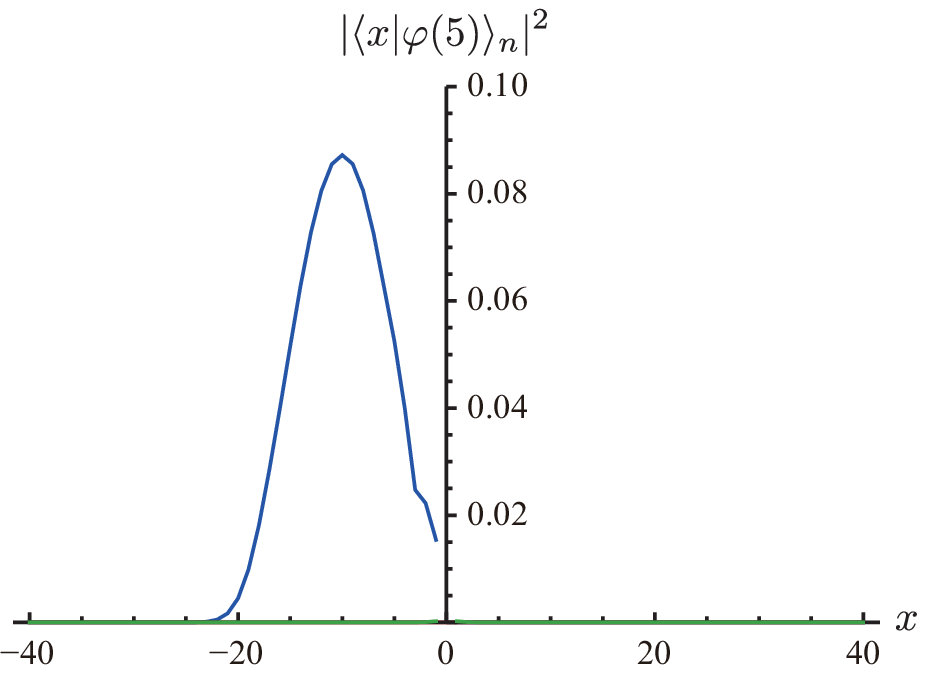}
\hfill
\includegraphics[width=.49\textwidth]{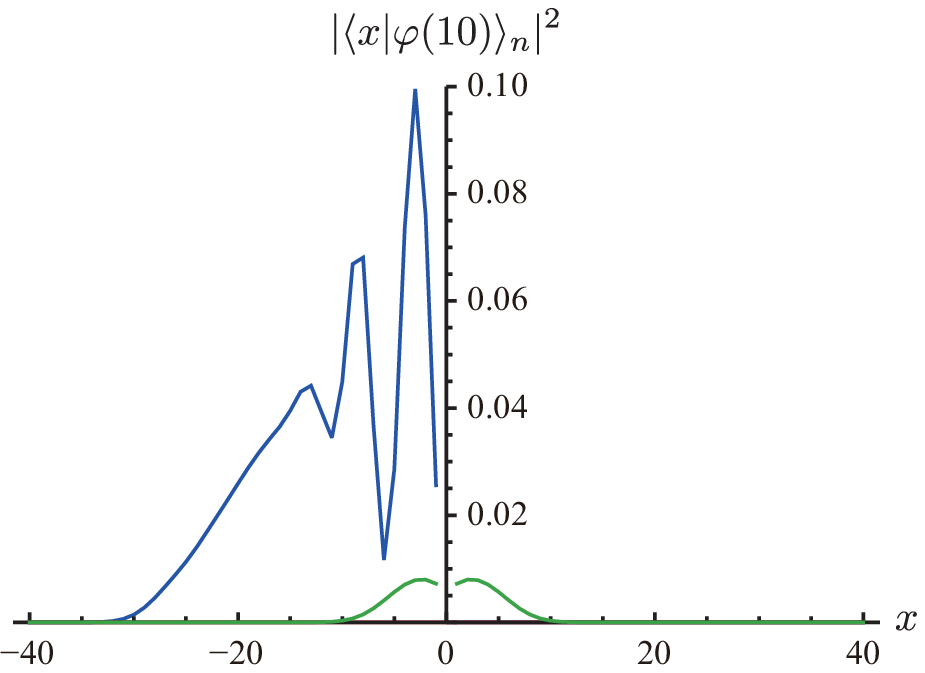}

(e)\hspace{0.48\textwidth}(f)

\vspace{\baselineskip}

\includegraphics[width=.49\textwidth]{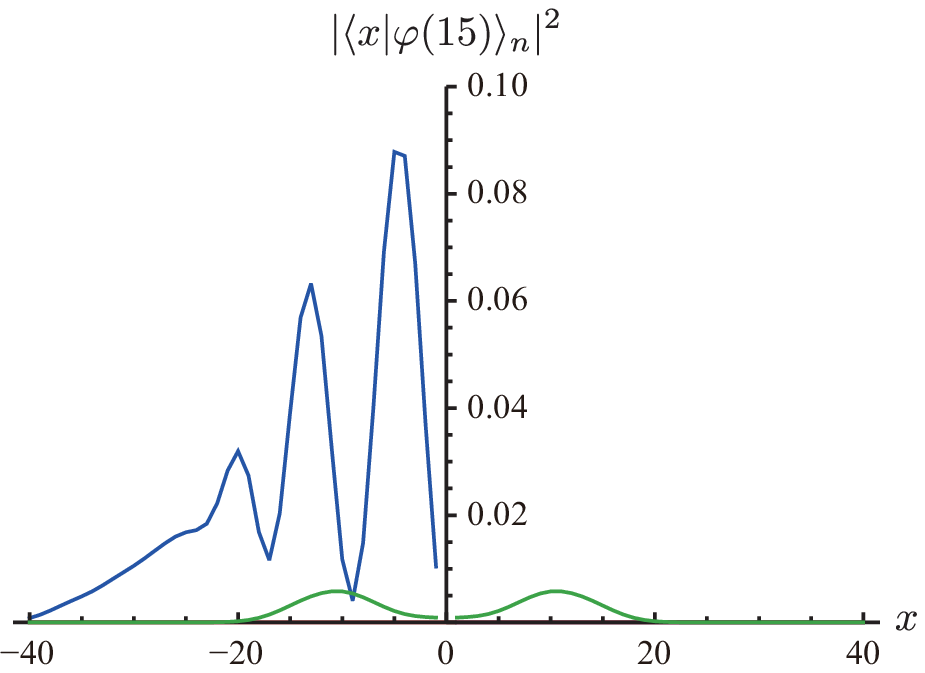}
\hfill
\includegraphics[width=.49\textwidth]{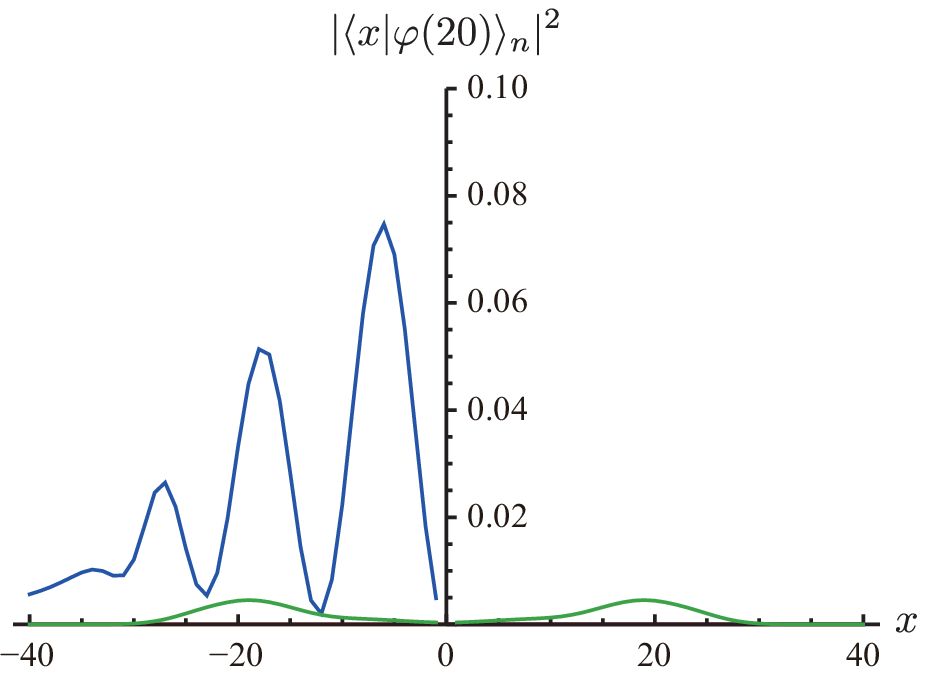}

(g)\hspace{0.48\textwidth}(h)
\end{center}
\caption{We here show dominant components of the time-evolved wave packet of Fig.~\ref{fig:Gaussian1} at eight different times: (a) $t=-20$, (b) $t=-15$, (c) $t=-10$, (d) $t=-5$, (e) $t=5$, (f) $t=10$, (g) $t=15$ and (h) $t=20$.  The large (blue) wave packets are due to the free time-evolution ($\bra x_{2\beta}|\varphi(t)\ket_0$ in Eq.~\eqref{eq:wp_n}). 
The small red wave packets are due to the anti-resonant state for negative times and the green ones  are due to the resonant state for positive times. 
The anti-resonant wave packets move inward toward $x=0$ as $t$  increases, while the resonant ones move outward. To obtain the resonant and anti-resonant wave packets we numerically evaluated $\bra x_{2\beta}|\varphi(t)\ket_n$ in Eq.~\eqref{eq:wp_n} for $n=\mbox{ res.}$ and $n=\mbox{anti-res.}$, respectively.
The anti-resonant wave packets at any time $-t$ are identical to the resonant wave packets at time $t$;
for example, compare (b) and (g).}
\label{fig:Gaussian2} 
\end{figure}

The free time-evolution  for $t>0$ is as follows  (see Fig.~\ref{fig:Gaussian2}(e)--(h)).
At $t=0$ the free wave packet  of the initial condition in Fig.~\ref{fig:Gaussian1} starts spreading out in both left and right directions  as $t$ increases  (Fig.~\ref{fig:Gaussian2}(e)). The right-hand side of the wave packet is then reflected by the dot site at $x=0$, producing interference with the portion of the wave packet that is not reflected yet (Fig.~\ref{fig:Gaussian2}(f)). 
 The interference pattern further continues to spread towards negative $x$ as shown in Fig.~\ref{fig:Gaussian2}(g)--(h).

The  free time-evolution for $t<0$ towards $t=0$ is the exact inverse process; see Fig.~\ref{fig:Gaussian2}(a)--(d).
Starting with the interference pattern, the  portion of reflected wave packet moves towards the dot   site at $x=0$ as in Fig.~\ref{fig:Gaussian2}(a)--(c), bounces off the dot,  forms a spread Gaussian wave packet in Fig.~\ref{fig:Gaussian2}(d), and ends up with the wave packet of the terminal condition in Fig.~\ref{fig:Gaussian1} at $t=0$.

Out of the other contributions to the time evolution, the anti-resonant-state contribution dominates for $t<0$  (small (red) wave packets) and the resonant-state contribution dominates for $t>0$  (small (green) wave packets).  Note that both the resonant and anti-resonant wave packets obey causality.  For example, for $t>0$ the resonant wave packets only appear after the incident wave packet reaches the dot site $d_1$ (within the quantum uncertainty), because the wave packet needs to be absorbed by the quantum dot before emission can occur.  Similarly, the anti-resonant wave packets  exist only {\textit{before}} the dot site $d_1$ ejects the reflected portion of the free wave packet to form the Gaussian wave packet at $t=0$. 

\section{Summary and discussion}
\label{sec15}

We presented  a new resolution of unity for a class of tight-binding open quantum systems and used it for expansions of the Green's function and the time evolution operator.
All of our expansions in the $P$ subspace are expressed in terms of the discrete states (the states with point spectra), not containing integrals over the continuum states (the states with a continuum spectrum).
Although they contain integrals over the continuum states outside the $P$ subspace, the integrations are taken over \textit{un}perturbed states.
This makes possible to factor out the expansion in the $P$ subspace and to keep the form of the summation over all discrete states.
Because of this feature of the expansion, we can clearly see which contribution produces which time dependence in time evolution;
the resonant states cause exponential decay, the anti-resonant states cause exponentially growth, the bound and anti-bound states cause oscillations, and the branch points cause power-law decay.

The most remarkable feature of the present expansions is that they observe the time-reversal symmetry because the resonant and anti-resonant states always come into the expansions as a pair.
The time-reversal symmetry is broken only as we try to compute matrix elements specifying the sign of $t$.
We can conceptually express this in the following way:
when we track the time evolution from an initial condition, we are forced to choose the resonant states and hence we observe decays;
when we track the time evolution towards a terminal condition, we are forced to choose the anti-resonant states and hence we observe growths.
The present expansion therefore symbolically shows that  the time evolution itself does not break the time-reversal symmetry, but the choice of whether we solve the initial-condition problem or the terminal-condition problem does,
together with the condition's symmetry or anti-symmetry under time inversion. 
We will detail this point in a separate publication.\cite{Ordonez14}.

The present argument is general in the sense that it does not depend on the scattering potential, but specific in the sense that the quadratic eigenvalue problem~\eqref{eq150} holds only for the tight-binding models.
A possibly related expansion in the one-dimensional continuum space has been formulated by Garc\'{i}a-Calder\'{o}n.\cite{GarciaCalderon10}
The coordinate representation of the Green's function in that study may be upgraded to an operator representation by extending the present formulation.

Directions of further possible generalizations include consideration of systems with massless linear dispersions as well as interacting systems.
The dispersion of light, $E\propto |k|$, has a singularity at $k=0$, which can yield an anomaly in the expansion.
Another interesting linear dispersion is the Dirac dispersion $E\propto k$, which does not have a singularity at $k=0$.
Particle-particle interactions will be essential in equilibration of the system and hence will be particularly important in discussing the entropy production and its connection to  spontaneous time-reversal symmetry breaking.

Another ambitious generalization is the possible extension to the dynamics of the density matrix, which is governed by the Liouville-von Neumann equation
\begin{align}
i\frac{d}{dt}\rho(t)=[H,\rho(t)].
\end{align}
We may then be able to argue  the monotonic time dependence of the entropy
\begin{align}
S(t)=-\mathop{\mathrm{Tr}}\rho(t)\ln\rho(t).
\end{align}
This approach can be quite different from a widely spread view of the time-reversal symmetry breaking.
When we reduce the microscopic degrees of freedom and specify the state of a system only in terms of macroscopic variables, we cannot trace back the time evolution of the system from a terminal condition.
This (possibly spuriously) suggests that coarse-graining is the reason of the time-reversal symmetry breaking.
Our approach may indicate that the time-reversal symmetry can be broken even in the level of microscopic description of the time-evolution.

\acknowledgments

One of the authors (N.H.) is deeply indebted to Dr.~S.~Klaiman for introducing to him the concept of the quadratic eigenvalue problem and to Prof.~A.~Leggett for introducing to him the articles by H.~Price.
He also greatly appreciates the Clark Way Harrison Visiting Professorship of Washington University in St.~Louis.
Both authors express sincere gratitude to Dr.~T.~Petrosky for helpful comments.
The paper is partially supported by JSPS Grant-in-Aid for Scientific Research (B) No.~22340110, a Research Grant from the Yamada Science Foundation, a Research Grant in the Natural Sciences from the Mitsubishi foundation, as well as the Holcomb Awards Committee and Woods Lecture Series at Butler University.

\appendix

\section{Effective Hamiltonian}
\label{appA}

We here review a derivation of the effective Hamiltonian~\eqref{eq250}.\cite{Livshits56,Feshbach58,Feshbach62,Rotter91,Albeverio96,Petrosky96,Petrosky97,Fyodorov97,Dittes00,Pichugin01,Sadreev03,Okolowicz03,Kunz06,Kunz08,Rotter09,Sasada11}
We can derive it for a general set of projection operators $P$ and $Q$ with $P+Q=I$.
We operate them on the stationary Schr\"{o}dinger equation
\begin{align}
H|\psi\ket=E|\psi\ket,
\end{align}
obtaining
\begin{align}\label{eqa20}
PHP|\psi\ket+PHQ|\psi\ket=EP|\psi\ket,
\\\label{eqa30}
QHP|\psi\ket+QHQ|\psi\ket=EQ|\psi\ket.
\end{align}
We formally solve Eq.~\eqref{eqa30} with respect to $Q|\psi\ket$ to have
\begin{align}\label{eqa40}
Q|\psi\ket=\frac{1}{E-QHQ}QHP|\psi\ket.
\end{align}
By substituting Eq.~\eqref{eqa40} into Eq.~\eqref{eqa20}, we have
\begin{align}
PHP|\psi\ket+PHQ\frac{1}{E-QHQ}QHP|\psi\ket=EP|\psi\ket,
\end{align}
which we can cast into the form of the Schr\"{o}dinger equation for the states in the $P$ subspace:
\begin{align}
\Heff(E)(P|\psi\ket)=E(P|\psi\ket)
\end{align}
with
\begin{align}\label{eqa70}
\Heff(E)=PHP+PHQ\frac{1}{E-QHQ}QHP.
\end{align}
This is the effective Hamiltonian $\Heff(E)$ given by Eq.~\eqref{eq250}.

\section{Green's function in the $P$ and $Q$ subspaces}
\label{appB}

We here prove the following equalities:
\begin{align}\label{eqa100}
P\frac{1}{E-H}P&=P\frac{1}{E-\Heff(E)}P,
\\\label{eqa110}
Q\frac{1}{E-H}P&=Q\frac{1}{E-QHQ}QHP\frac{1}{E-\Heff(E)}P,
\\\label{eqa120}
P\frac{1}{E-H}Q&=P\frac{1}{E-\Heff(E)}PHQ\frac{1}{E-QHQ}Q,
\\\label{eqa130}
Q\frac{1}{E-H}Q&=Q\frac{1}{E-QHQ}Q+ Q\frac{1}{E-QHQ}QHP\frac{1}{E-\Heff(E)}PHQ\frac{1}{E-QHQ}Q.
\end{align}
We use these equalities in Eqs.~\eqref{eq260}, \eqref{eq820}, \eqref{eq825}, \eqref{eq881}, \eqref{eq890} and~\eqref{eq895}.

In order to prove Eqs.~\eqref{eqa100}--\eqref{eqa130}, we split the full Hamiltonian $H$ into the two parts
\begin{align}
H_0&=PHP+QHQ=H_\mathrm{d}+H_\mathrm{leads},
\\
H_1&=PHQ+QHP=H_\mathrm{contacts}
\end{align}
and consider the resolvent expansion
\begin{align}
\frac{1}{E-H}=\frac{1}{E-H_0}+\frac{1}{E-H_0}H_1\frac{1}{E-H_0}+\frac{1}{E-H_0}H_1\frac{1}{E-H_0}H_1\frac{1}{E-H_0}+\cdots.
\end{align}
We here note
\begin{align}\label{eqa430}
\frac{1}{E-H_0}&=\sum_{n=0}^\infty \frac{(PHP+QHQ)^n}{E^{n+1}}
\nonumber\\
&=\sum_{n=0}^\infty \frac{(PHP)^n+(QHQ)^n}{E^{n+1}}
\nonumber\\
&=P\frac{1}{E-PHP}P+Q\frac{1}{E-QHQ}Q,
\end{align}
where we used $PQ=QP=0$.

First, we compute
\begin{align}
P\frac{1}{E-H}P=&P\frac{1}{E-H_0}P+P\frac{1}{E-H_0}H_1\frac{1}{E-H_0}P
\nonumber\\
&+P\frac{1}{E-H_0}H_1\frac{1}{E-H_0}H_1\frac{1}{E-H_0}P+\cdots.
\end{align}
Because of Eq.~\eqref{eqa430}, we have
\begin{align}
P\frac{1}{E-H}P=&P\frac{1}{E-PHP}P+P\frac{1}{E-PHP}PH_1\frac{1}{E-H_0}P
\nonumber\\
&+P\frac{1}{E-PHP}PH_1\frac{1}{E-H_0}H_1\frac{1}{E-H_0}P+\cdots
\nonumber\\
=&P\frac{1}{E-PHP}P+P\frac{1}{E-PHP}PHQ\frac{1}{E-H_0}P
\nonumber\\
&+P\frac{1}{E-PHP}PHQ\frac{1}{E-H_0}H_1\frac{1}{E-H_0}P+\cdots.
\nonumber\\
=&P\frac{1}{E-PHP}P+P\frac{1}{E-PHP}PHQ\frac{1}{E-QHQ}P
\nonumber\\
&+P\frac{1}{E-PHP}PHQ\frac{1}{E-QHQ}H_1\frac{1}{E-H_0}P+\cdots.
\end{align}
We realize that the second term on the right-hand side vanishes.
By the same token, all odd terms with respect to $H_1$ vanish.
We therefore arrive at
\begin{align}
P\frac{1}{E-H}P=&P\frac{1}{E-PHP}P+P\frac{1}{E-PHP}PHQ\frac{1}{E-QHQ}QHP\frac{1}{E-PHP}P+\cdots
\nonumber\\
=&P\frac{1}{E-PHP}P+P\frac{1}{E-PHP}\Sigma(E)\frac{1}{E-PHP}P
\nonumber\\
&+P\frac{1}{E-PHP}\Sigma(E)\frac{1}{E-PHP}\Sigma(E)\frac{1}{E-PHP}P+\cdots
\nonumber\\
=&P\frac{1}{E-PHP-\Sigma(E)}P,
\end{align}
where
\begin{align}
\Sigma(E)=PHQ\frac{1}{E-QHQ}QHP
\end{align}
is often called the self-energy of the leads.
By setting
\begin{align}
\Heff(E)=PHP+\Sigma(E)
\end{align}
according to Eq.~\eqref{eqa70}, we have Eq.~\eqref{eqa100}.

Next, we compute 
\begin{align}
Q\frac{1}{E-H}P=&Q\frac{1}{E-H_0}P+Q\frac{1}{E-H_0}H_1\frac{1}{E-H_0}P
\nonumber\\
&+Q\frac{1}{E-H_0}H_1\frac{1}{E-H_0}H_1\frac{1}{E-H_0}P+\cdots.
\end{align}
This time, all even terms with respect to $H_1$ vanish.
We therefore have
\begin{align}
Q\frac{1}{E-H}P=&Q\frac{1}{E-QHQ}QHP\frac{1}{E-PHP}P
\nonumber\\
&+Q\frac{1}{E-QHQ}QHP\frac{1}{E-PHP}PHQ\frac{1}{E-QHQ}QHP\frac{1}{E-PHP}P
+\cdots
\nonumber\\
=&Q\frac{1}{E-QHQ}QHP\frac{1}{E-PHP}P
\nonumber\\
&+Q\frac{1}{E-QHQ}QHP\frac{1}{E-PHP}\Sigma(E)\frac{1}{E-PHP}P
\nonumber\\
&+Q\frac{1}{E-QHQ}QHP\frac{1}{E-PHP}\Sigma(E)\frac{1}{E-PHP}\Sigma(E)\frac{1}{E-PHP}P+\cdots
\nonumber\\
=&Q\frac{1}{E-QHQ}QHP\frac{1}{E-PHP-\Sigma(E)}P,
\end{align}
which is Eq.~\eqref{eqa110}.
We can prove the other two equalities~\eqref{eqa120} and~\eqref{eqa130} similarly.

In the specific case of the present system in Fig.~\ref{fig4}(a), we can explicitly obtain the Green's function for $QHQ$,
\begin{align}\label{eqC160}
Q\frac{1}{E-QHQ}Q.
\end{align}
Note that the partial Hamiltonian $QHQ$ is composed of semi-infinite chains:
\begin{align}
QHQ=H_\mathrm{leads}=\sum_{i=1}^N\sum_{\alpha=1}^{n_i}H_{i\alpha}^\mathrm{lead}.
\end{align}
Its eigenstate $|k\ket$ is the direct product of the eigenstate of each chain, which is given by the plane wave
\begin{align}\label{plainwave}
\bra x_{i\alpha}|k_{i\alpha}\ket=\sqrt{2}\sin(k_{i\alpha}x_{i\alpha}).
\end{align}
We therefore have
\begin{align}\label{eqb190}
Q\frac{1}{E-QHQ}Q=
\sum_{i\alpha} \int_{-\pi}^\pi \frac{dk_{i\alpha}}{2\pi}  Q|k_{i\alpha}\ket\frac{1}{E+2\cos k_{i\alpha}}\bra k_{i\alpha}|Q,
\end{align}
and hence
\begin{align}\label{eqb195}
\bra x|Q\frac{1}{E-QHQ}Q|y\ket
&=\int_{-\pi}^\pi \frac{2\sin(k_0x)\sin(k_0y)}{E+2\cos k_0}\frac{dk_0}{2\pi}
\nonumber\\
&=\int_{C_0} \frac{d\lambda_0}{4\pi i\lambda_0}
\frac{({\lambda_0}^x-{\lambda_0}^{-x})({\lambda_0}^y-{\lambda_0}^{-y})}{\lambda_0-{\lambda_0}^{-1}}
\left(\frac{\lambda}{\lambda-\lambda_0}-\frac{\lambda}{\lambda-{\lambda_0}^{-1}}\right)
\end{align}
for the sites $x$ and $y$ on one lead, where the contour $C_0$ is the contour of the counterclockwise unit circle.
We converted the energy variable from $E$ to $\lambda$ as well as from $e^{ik_0}$ to $\lambda_0$ in the second line.
The integrand of Eq.~\eqref{eqb195} has poles at $\lambda_0=0, \lambda, \lambda^{-1}, \infty$.

Particularly when we use the contour $C_2$ in Fig.~\ref{fig7}(c) for the integration over $\lambda$, we can specify that the pole $\lambda_0=\lambda$ is inside the unit circle while the pole $\lambda_0=\lambda^{-1}$ is outside it.
Equation~\eqref{eqb195} is then generally integrable.
Let us consider the case $x>y\geq 1$, for example. 
The terms with ${\lambda_0}^{x+y}$ and ${\lambda_0}^{x-y}$ in the numerator do not have poles at $\lambda_0=0$.
For them, we encircle the pole $\lambda_0=\lambda$ inside the unit circle counterclockwise, having
\begin{align}\label{eqb196}
2\pi i \frac{1}{4\pi i\lambda}\frac{\lambda^x(\lambda^y-\lambda^{-y})}{\lambda-\lambda^{-1}}(-\lambda)
=-\frac{1}{2}\lambda^x\frac{\lambda^y-\lambda^{-y}}{\lambda-\lambda^{-1}}.
\end{align}
Note that the zeros of the denominator $\lambda_0-{\lambda_0}^{-1}$ in Eq.~\eqref{eqb195} are not poles because the numerator has canceling zeros in.
On the other hand, the terms with ${\lambda_0}^{-x+y}$ and ${\lambda_0}^{-x-y}$ in the numerator have higher poles at $\lambda_0=0$.
In order to avoid computing the residue at $\lambda_0$, we rather encircle the pole $\lambda_0=\lambda^{-1}$ outside the unit circle clockwise, having
\begin{align}
-2\pi i\frac{1}{4\pi i\lambda^{-1}}\frac{-\lambda^x(\lambda^{-y}-\lambda^y)}{\lambda^{-1}-\lambda}(-\lambda^{-1})
=-\frac{1}{2}\lambda^x\frac{\lambda^y-\lambda^{-y}}{\lambda-\lambda^{-1}},
\end{align}
which is equal to Eq.~\eqref{eqb196}.
We thereby arrive at
\begin{align}\label{eqb197}
\bra x|Q\frac{1}{E(\lambda)-QHQ}Q|y\ket
&=-\lambda^x\frac{\lambda^y-\lambda^{-y}}{\lambda-\lambda^{-1}}
\end{align}
for $x>y\geq1$.
A similar calculation reveals that it holds for $x=y\geq 1$ too.
We can use the expression~\eqref{eqb197} for the first term on the right-hand side of Eq.~\eqref{eqa130}.

We can also  write down the following matrix element which can appear in Eq.~\eqref{eqa110}:
\begin{align}\label{eqb200}
Q\frac{1}{E-QHQ}QHP&=
\sum_{i\alpha}\int_{-\pi}^\pi \frac{dk_{i\alpha}}{2\pi}
Q|k_{i\alpha}\ket\frac{1}{E+2\cos k_{i\alpha}}\bra k_{i\alpha}|1_{i\alpha}\ket(-t_{i\alpha})\bra d_i|P
\nonumber\\
&=-\sum_{i\alpha}t_{i\alpha}\int_{-\pi}^\pi \frac{dk_{i\alpha}}{2\pi}Q|k_{i\alpha}\ket\frac{\sqrt{2}\sin k_{i\alpha}}{E+2\cos k_{i\alpha}}\bra d_i|P.
\end{align}
We then have
\begin{align}\label{eqb210}
\bra x_{i\alpha}|Q\frac{1}{E(\lambda)-QHQ}QHP|d_i\ket
&=-t_{i\alpha}\int_{-\pi}^\pi \frac{dk_{i\alpha}}{2\pi}\frac{2\sin(k_{i\alpha}x_{i\alpha})\sin k_{i\alpha}}{E+2\cos k_{i\alpha}}.
\nonumber\\
&={-t_{i\alpha}}\int_{C_0} \frac{d\lambda_0}{4\pi i\lambda_0}\left({\lambda_0}^{x}-{\lambda_0}^{-x}\right)
\left(\frac{\lambda}{\lambda-\lambda_0}-\frac{\lambda}{\lambda-{\lambda_0}^{-1}}\right),
\end{align}
where we converted the energy variable from $E$ to $\lambda$ and $e^{ik_{i\alpha}}$ to $\lambda_0$.
We also dropped the subscript of $x_{i\alpha}$ in the second line for brevity.
Particularly when we use the contour $C_2$ in Fig.~\ref{fig7}(c) for the integration over $\lambda$, we use the same algebra as in Eqs.~\eqref{eqb196} and~\eqref{eqb197}, arriving at
\begin{align}\label{eqb240}
\bra x|Q\frac{1}{E(\lambda)-QHQ}QHP|d_i\ket
&= t_{i\alpha}\lambda^{x}.
\end{align}
We use this in Eq.~\eqref{eq320}.

Finally, let us calculate the matrix element of
\begin{align}
PHQ\frac{1}{E-QHQ}QHP
&=\sum_{i\alpha}\int_{-\pi}^\pi \frac{dk_{i\alpha}}{2\pi}
P|d_i\ket(-t_{i\alpha})\bra 1_{i\alpha}|k_{i\alpha}
\frac{1}{E+2\cos k_{i\alpha}}
\bra k_{i\alpha}|1_{i\alpha}\ket(-t_{i\alpha})\bra d_i|P,
\end{align}
which is the second term of the effective Hamiltonian~\eqref{eqa70}, and equivalently the self-energy term of Eq.~\eqref{eq70}.
In a similar way to the above, we have
\begin{align}\label{eqb280}
&\bra d_j| PHQ\frac{1}{E(\lambda)-QHQ}QHP|d_i\ket
=\delta_{ij}\sum_{\alpha}
(t_{i\alpha})^2\int_{-\pi}^\pi \frac{dk_{i\alpha}}{2\pi}\frac{2\sin^2k_{i\alpha}}{E+2\cos k_{i\alpha}}
\nonumber\\
&\qquad
=\delta_{ij}\sum_{\alpha}
(t_{i\alpha})^2\int_{C_0} \frac{d\lambda_{0}}{4\pi i\lambda_0}
\left({\lambda_0}-{\lambda_0}^{-1}\right)
\left(\frac{\lambda}{\lambda-\lambda_0}-\frac{\lambda}{\lambda-{\lambda_0}^{-1}}\right)
\nonumber\\
&\qquad
=-\delta_{ij}\sum_{\alpha}(t_{i\alpha})^2\lambda
\nonumber\\
&\qquad
=-\lambda \bra d_j | PHQHP | d_i\ket.
\end{align}
We use this in Eq.~\eqref{eq71}.

\section{Derivation of our previous expansion}
\label{appC}

We here derive our previous expansion~\eqref{eq810} (Eqs.~(4) and~(56) in Ref.~\cite{Sasada11}) from our new expansion~\eqref{eq800}.
Let us note here that if the value of $\lambda=e^{ik}$ gives $E+i\delta$ with a real value of $E$, the value of $1/\lambda=e^{-ik}$ gives $E-i\delta$.
This means that if $\Geff(E(\lambda))$ gives the retarded Green's function $G^\mathrm{R}(E)$ in the $P$ subspace, $\Geff(E(1/\lambda))$ should give the advanced Green's function $G^\mathrm{A}$ in the $P$ subspace.
We will indeed derive
\begin{align}\label{eqc10}
\Geff(E(\lambda))+\Geff(E(1/\lambda))=P|\phi_n\ket\frac{1}{E-E_n}\bra\tilde{\phi}_n|P,
\end{align}
which corresponds to the expansion~\eqref{eq810}.

We first transform the new expansion~\eqref{eq800} to
\begin{align}\label{eqc890}
\Geff(E(\lambda))&=
\sum_{n=1}^{2N}
P|\psi_n\ket
\frac{1-\lambda\lambda_n}{E-E_n}
\bra \tilde{\psi}_n|P
\nonumber\\
&=
\sum_{n=1}^{2N}
P|\phi_n\ket
\frac{{\lambda_n}^{-1}-\lambda}{{\lambda_n}^{-1}-\lambda_n}\frac{1}{E-E_n}
\bra \tilde{\phi}_n|P,
\end{align}
where we used the identity
\begin{align}\label{eqc900}
E-E_n&=-\left(\lambda+\frac{1}{\lambda}-\lambda_n-\frac{1}{\lambda_n}\right)
\nonumber\\
&=-(\lambda-\lambda_n)\left(1-\frac{1}{\lambda\lambda_n}\right)
\nonumber\\
&=\frac{\lambda-\lambda_n}{\lambda\lambda_n}\left(1-\lambda\lambda_n\right)
\end{align}
in the first line and used the relations~\eqref{eq841} and~\eqref{eq842} in the second line.
We further transform it by noting that Eq.~\eqref{eqc890} should be invariant under the substitution $\lambda\to 1/\lambda$ and $\lambda_n\to 1/\lambda_n$.
This is because the Green's function is originally a function of the energy only and because the energy and the eigenenergy are invariant under the substitution above.
This substitution in Eq.~\eqref{eqc890} leads to
\begin{align}\label{eqc891}
\Geff(E(\lambda))&=
\sum_{n=1}^{2N}
P|\phi_n\ket
\frac{\lambda_n-\lambda^{-1}}{\lambda_n-{\lambda_n}^{-1}}\frac{1}{E-E_n}
\bra \tilde{\phi}_n|P.
\end{align}
We therefore have
\begin{align}\label{eqc892}
\Geff(E(1/\lambda))&=
\sum_{n=1}^{2N}
P|\phi_n\ket
\frac{\lambda_n-\lambda}{\lambda_n-{\lambda_n}^{-1}}\frac{1}{E-E_n}
\bra \tilde{\phi}_n|P.
\end{align}
Summing Eqs.~\eqref{eqc890} and~\eqref{eqc892}, we have Eq.~\eqref{eqc10}, and hence Eq.~\eqref{eq810}.

\section{Normalization of the states with point spectra}
\label{appD}

We here compare the normalization set by Eq.~\eqref{eq630} with the standard normalization of the bound and resonant states and derive Eqs.~\eqref{eq841} and~\eqref{eq842}, where $|\psi_n\ket$ and $\bra\tilde{\psi}_n|$ are the eigenstates normalized under Eq.~\eqref{eq630} while $|\phi_n\ket$ and $\bra\tilde{\phi}_n|$ are the eigenstates normalized in the standard way.
By using the expressions~\eqref{eq420}, \eqref{eq570} and~\eqref{eq580} in Eq.~\eqref{eq630}, we can explicitly write down the normalization in the present formulation as follows:
\begin{align}\label{eqD10}
\left(1-{\lambda_n}^2\right)\bra\tilde{\psi}_n|P|\psi_n\ket
+{\lambda_n}^2\bra\tilde{\psi}_n|P\Theta P|\psi_n\ket=1.
\end{align}

On the other hand, the standard normalization is of course given by
\begin{align}\label{eqD30}
\sum_{i=1}^N\bra d_i|\tilde{\phi}_n\ket\bra\phi_n|d_i\ket
+\sum_{i=1}^N\sum_{\alpha=1}^{n_i}\sum_{x_{i\alpha}=1}^\infty
\bra x_{i\alpha}|\tilde{\phi}_n\ket\bra\phi_n|x_{i\alpha}\ket=1,
\end{align}
where we denoted the eigenstates with $|\phi_n\ket$ because of the difference in the normalization.
The normalization~\eqref{eqD30} can be cast into the form
\begin{align}
\bra\tilde{\phi}_n|P|\phi_n\ket+\sum_{x=1}^\infty{\lambda_n}^{2x}\bra\tilde{\phi}_n|P\Theta P|\phi_n\ket=1.
\end{align}
The transformation of the second term on the left-hand side is confirmed by using Eqs.~\eqref{eq320} and~\eqref{eq350}.
Although $|\lambda_n|$ is greater than unity for the resonant and anti-resonant states, it is customary to sum up the geometric series nonetheless\cite{Zeldovich60,Hokkyo65,Romo68,GarciaCalderon10}, often by introducing a convergence factor.
This procedure gives the same result for all eigenstates with point spectra:
\begin{align}
\bra\tilde{\phi}_n|P|\phi_n\ket+\frac{{\lambda_n}^2}{1-{\lambda_n}^2}\bra\tilde{\phi}_n|P\Theta P|\phi_n\ket=1.
\end{align}
Comparing Eqs.~\eqref{eqD10} and~\eqref{eqD30}, we obtain the relations
\begin{align}
|\phi_n\ket&=\sqrt{1-{\lambda_n}^2}|\psi_n\ket,
\\
\bra\tilde{\phi}_n|&=\sqrt{1-{\lambda_n}^2}\bra\tilde{\psi}_n|
\end{align}
which are  Eqs.~\eqref{eq841} and~\eqref{eq842}.
Note that the normalization in the present framework does not require the summation of the divergent geometric series, which was once a subject of debate.\cite{Zeldovich60,Hokkyo65,Romo68}

\bibliography{hatano}

\begin{thebibliography}{60}%
\makeatletter
\providecommand \@ifxundefined [1]{%
 \@ifx{#1\undefined}
}%
\providecommand \@ifnum [1]{%
 \ifnum #1\expandafter \@firstoftwo
 \else \expandafter \@secondoftwo
 \fi
}%
\providecommand \@ifx [1]{%
 \ifx #1\expandafter \@firstoftwo
 \else \expandafter \@secondoftwo
 \fi
}%
\providecommand \natexlab [1]{#1}%
\providecommand \enquote  [1]{``#1''}%
\providecommand \bibnamefont  [1]{#1}%
\providecommand \bibfnamefont [1]{#1}%
\providecommand \citenamefont [1]{#1}%
\providecommand \href@noop [0]{\@secondoftwo}%
\providecommand \href [0]{\begingroup \@sanitize@url \@href}%
\providecommand \@href[1]{\@@startlink{#1}\@@href}%
\providecommand \@@href[1]{\endgroup#1\@@endlink}%
\providecommand \@sanitize@url [0]{\catcode `\\12\catcode `\$12\catcode
  `\&12\catcode `\#12\catcode `\^12\catcode `\_12\catcode `\%12\relax}%
\providecommand \@@startlink[1]{}%
\providecommand \@@endlink[0]{}%
\providecommand \url  [0]{\begingroup\@sanitize@url \@url }%
\providecommand \@url [1]{\endgroup\@href {#1}{\urlprefix }}%
\providecommand \urlprefix  [0]{URL }%
\providecommand \Eprint [0]{\href }%
\providecommand \doibase [0]{http://dx.doi.org/}%
\providecommand \selectlanguage [0]{\@gobble}%
\providecommand \bibinfo  [0]{\@secondoftwo}%
\providecommand \bibfield  [0]{\@secondoftwo}%
\providecommand \translation [1]{[#1]}%
\providecommand \BibitemOpen [0]{}%
\providecommand \bibitemStop [0]{}%
\providecommand \bibitemNoStop [0]{.\EOS\space}%
\providecommand \EOS [0]{\spacefactor3000\relax}%
\providecommand \BibitemShut  [1]{\csname bibitem#1\endcsname}%
\let\auto@bib@innerbib\@empty
\bibitem [{\citenamefont {Nakanishi}(1958)}]{Nakanishi58}%
  \BibitemOpen
  \bibfield  {author} {\bibinfo {author} {\bibfnamefont {N.}~\bibnamefont
  {Nakanishi}},\ }\bibfield  {title} {\enquote {\bibinfo {title} {A theory of
  clothed unstable particles},}\ }\href@noop {} {\bibfield  {journal} {\bibinfo
   {journal} {Prog. Theor. Phys.}\ }\textbf {\bibinfo {volume} {19}},\ \bibinfo
  {pages} {607--621} (\bibinfo {year} {1958})}\BibitemShut {NoStop}%
\bibitem [{\citenamefont {Prigogine}\ \emph {et~al.}(1973)\citenamefont
  {Prigogine}, \citenamefont {George}, \citenamefont {Henin},\ and\
  \citenamefont {Rosenfeld}}]{Prigogine73}%
  \BibitemOpen
  \bibfield  {author} {\bibinfo {author} {\bibfnamefont {I.}~\bibnamefont
  {Prigogine}}, \bibinfo {author} {\bibfnamefont {C.}~\bibnamefont {George}},
  \bibinfo {author} {\bibfnamefont {F.}~\bibnamefont {Henin}}, \ and\ \bibinfo
  {author} {\bibfnamefont {L.}~\bibnamefont {Rosenfeld}},\ }\bibfield  {title}
  {\enquote {\bibinfo {title} {A unified formulation of dynamics and
  thermodynamics},}\ }\href@noop {} {\bibfield  {journal} {\bibinfo  {journal}
  {Chem. Scr.}\ }\textbf {\bibinfo {volume} {4}},\ \bibinfo {pages} {5}
  (\bibinfo {year} {1973})}\BibitemShut {NoStop}%
\bibitem [{\citenamefont {Sudarshan}, \citenamefont {Chiu},\ and\ \citenamefont
  {Gorini}(1978)}]{Sudarshan78}%
  \BibitemOpen
  \bibfield  {author} {\bibinfo {author} {\bibfnamefont {E.~C.~G.}\
  \bibnamefont {Sudarshan}}, \bibinfo {author} {\bibfnamefont {C.~B.}\
  \bibnamefont {Chiu}}, \ and\ \bibinfo {author} {\bibfnamefont
  {V.}~\bibnamefont {Gorini}},\ }\bibfield  {title} {\enquote {\bibinfo {title}
  {Decaying states as complex energy eigenvectors in generalized quantum
  mechanics},}\ }\href {\doibase 10.1103/PhysRevD.18.2914} {\bibfield
  {journal} {\bibinfo  {journal} {Phys. Rev. D}\ }\textbf {\bibinfo {volume}
  {18}},\ \bibinfo {pages} {2914--2929} (\bibinfo {year} {1978})}\BibitemShut
  {NoStop}%
\bibitem [{\citenamefont {Bohm}, \citenamefont {Gadella},\ and\ \citenamefont
  {Mainland}(1989)}]{Bohm89}%
  \BibitemOpen
  \bibfield  {author} {\bibinfo {author} {\bibfnamefont {A.}~\bibnamefont
  {Bohm}}, \bibinfo {author} {\bibfnamefont {M.}~\bibnamefont {Gadella}}, \
  and\ \bibinfo {author} {\bibfnamefont {G.~B.}\ \bibnamefont {Mainland}},\
  }\bibfield  {title} {\enquote {\bibinfo {title} {Gamov vectors and decaying
  states},}\ }\href@noop {} {\bibfield  {journal} {\bibinfo  {journal} {Am. J.
  Phys.}\ }\textbf {\bibinfo {volume} {57}},\ \bibinfo {pages} {1103--1108}
  (\bibinfo {year} {1989})}\BibitemShut {NoStop}%
\bibitem [{\citenamefont {{Petrosky}}, \citenamefont {{Prigogine}},\ and\
  \citenamefont {{Tasaki}}(1991)}]{Tasaki91}%
  \BibitemOpen
  \bibfield  {author} {\bibinfo {author} {\bibfnamefont {T.}~\bibnamefont
  {{Petrosky}}}, \bibinfo {author} {\bibfnamefont {I.}~\bibnamefont
  {{Prigogine}}}, \ and\ \bibinfo {author} {\bibfnamefont {S.}~\bibnamefont
  {{Tasaki}}},\ }\bibfield  {title} {\enquote {\bibinfo {title} {{Quantum
  theory of non-integrable systems}},}\ }\href {\doibase
  10.1016/0378-4371(91)90257-D} {\bibfield  {journal} {\bibinfo  {journal}
  {Physica A Statistical Mechanics and its Applications}\ }\textbf {\bibinfo
  {volume} {173}},\ \bibinfo {pages} {175--242} (\bibinfo {year}
  {1991})}\BibitemShut {NoStop}%
\bibitem [{\citenamefont {Petrosky}\ and\ \citenamefont
  {Prigogine}(1996)}]{Petrosky96}%
  \BibitemOpen
  \bibfield  {author} {\bibinfo {author} {\bibfnamefont {T.}~\bibnamefont
  {Petrosky}}\ and\ \bibinfo {author} {\bibfnamefont {I.}~\bibnamefont
  {Prigogine}},\ }\bibfield  {title} {\enquote {\bibinfo {title} {Poincare
  resonance and the extension of classical dynamics},}\ }\href@noop {}
  {\bibfield  {journal} {\bibinfo  {journal} {Chaos Solitons Fractals}\
  }\textbf {\bibinfo {volume} {7}},\ \bibinfo {pages} {441--498} (\bibinfo
  {year} {1996})}\BibitemShut {NoStop}%
\bibitem [{\citenamefont {Petrosky}\ and\ \citenamefont
  {Prigogine}(1997)}]{Petrosky97}%
  \BibitemOpen
  \bibfield  {author} {\bibinfo {author} {\bibfnamefont {T.}~\bibnamefont
  {Petrosky}}\ and\ \bibinfo {author} {\bibfnamefont {I.}~\bibnamefont
  {Prigogine}},\ }\bibfield  {title} {\enquote {\bibinfo {title} {The liouville
  space extension of quantum mechanics},}\ }in\ \href@noop {} {\emph {\bibinfo
  {booktitle} {Advances in Chemical Physics, Volume 99}}},\ \bibinfo {editor}
  {edited by\ \bibinfo {editor} {\bibfnamefont {I.}~\bibnamefont {Prigogine}}\
  and\ \bibinfo {editor} {\bibfnamefont {S.}~\bibnamefont {Rice}}}\ (\bibinfo
  {publisher} {John Wiley and Sons},\ \bibinfo {year} {1997})\ pp.\ \bibinfo
  {pages} {1--120}\BibitemShut {NoStop}%
\bibitem [{\citenamefont {Hatano}(2010)}]{Hatano10}%
  \BibitemOpen
  \bibfield  {author} {\bibinfo {author} {\bibfnamefont {N.}~\bibnamefont
  {Hatano}},\ }\bibfield  {title} {\enquote {\bibinfo {title} {Resonant states
  of open quantum systems},}\ }\href@noop {} {\bibfield  {journal} {\bibinfo
  {journal} {Prog. Thoer. Phys. Suppl.}\ }\textbf {\bibinfo {volume} {184}},\
  \bibinfo {pages} {497--515} (\bibinfo {year} {2010})}\BibitemShut {NoStop}%
\bibitem [{\citenamefont {Klaiman}\ and\ \citenamefont
  {Hatano}(2011)}]{Klaiman11}%
  \BibitemOpen
  \bibfield  {author} {\bibinfo {author} {\bibfnamefont {S.}~\bibnamefont
  {Klaiman}}\ and\ \bibinfo {author} {\bibfnamefont {N.}~\bibnamefont
  {Hatano}},\ }\bibfield  {title} {\enquote {\bibinfo {title} {Resonance theory
  for discrete models: Methodology and isolated resonances},}\ }\href@noop {}
  {\bibfield  {journal} {\bibinfo  {journal} {J. Chem. Phys.}\ }\textbf
  {\bibinfo {volume} {134}},\ \bibinfo {pages} {154111 (9pp)} (\bibinfo {year}
  {2011})}\BibitemShut {NoStop}%
\bibitem [{\citenamefont {Livshits}(1956)}]{Livshits56}%
  \BibitemOpen
  \bibfield  {author} {\bibinfo {author} {\bibfnamefont {M.~S.}\ \bibnamefont
  {Livshits}},\ }\bibfield  {title} {\enquote {\bibinfo {title} {The
  application of non-self-adjoint operators to scattering theory (in
  russian)},}\ }\href@noop {} {\bibfield  {journal} {\bibinfo  {journal} {Zh.
  {\`{E}}ksper. Teoret. Fiz.}\ }\textbf {\bibinfo {volume} {31}},\ \bibinfo
  {pages} {121--131} (\bibinfo {year} {1956})},\ \bibinfo {note} {[English
  translation in Sov. Phys. JETP, 4, 91--98 (1957)]}\BibitemShut {NoStop}%
\bibitem [{\citenamefont {Feshbach}(1958)}]{Feshbach58}%
  \BibitemOpen
  \bibfield  {author} {\bibinfo {author} {\bibfnamefont {H.}~\bibnamefont
  {Feshbach}},\ }\bibfield  {title} {\enquote {\bibinfo {title} {A unified
  theory of nuclear reactions},}\ }\href@noop {} {\bibfield  {journal}
  {\bibinfo  {journal} {Ann. Phys. (New York)}\ }\textbf {\bibinfo {volume}
  {5}},\ \bibinfo {pages} {357--390} (\bibinfo {year} {1958})}\BibitemShut
  {NoStop}%
\bibitem [{\citenamefont {Feshbach}(1962)}]{Feshbach62}%
  \BibitemOpen
  \bibfield  {author} {\bibinfo {author} {\bibfnamefont {H.}~\bibnamefont
  {Feshbach}},\ }\bibfield  {title} {\enquote {\bibinfo {title} {A unified
  theory of nuclear reactions ii},}\ }\href@noop {} {\bibfield  {journal}
  {\bibinfo  {journal} {Ann. Phys. (New York)}\ }\textbf {\bibinfo {volume}
  {19}},\ \bibinfo {pages} {287--313} (\bibinfo {year} {1962})}\BibitemShut
  {NoStop}%
\bibitem [{\citenamefont {Rotter}(1991)}]{Rotter91}%
  \BibitemOpen
  \bibfield  {author} {\bibinfo {author} {\bibfnamefont {I.}~\bibnamefont
  {Rotter}},\ }\bibfield  {title} {\enquote {\bibinfo {title} {A continuum
  shell model for the open quantum mechanical nuclear system},}\ }\href@noop {}
  {\bibfield  {journal} {\bibinfo  {journal} {Rep. Prog. Phys.}\ }\textbf
  {\bibinfo {volume} {54}},\ \bibinfo {pages} {635--682} (\bibinfo {year}
  {1991})}\BibitemShut {NoStop}%
\bibitem [{\citenamefont {Rotter}(2009)}]{Rotter09}%
  \BibitemOpen
  \bibfield  {author} {\bibinfo {author} {\bibfnamefont {I.}~\bibnamefont
  {Rotter}},\ }\bibfield  {title} {\enquote {\bibinfo {title} {A non-hermitian
  hamiltonian operator and the physics of open quantum systems},}\ }\href@noop
  {} {\bibfield  {journal} {\bibinfo  {journal} {J. Phys. A: Math. Theor.}\
  }\textbf {\bibinfo {volume} {42}},\ \bibinfo {pages} {153001 (51pp)}
  (\bibinfo {year} {2009})}\BibitemShut {NoStop}%
\bibitem [{\citenamefont {Nakano}, \citenamefont {Hatano},\ and\ \citenamefont
  {Petrosky}(2011)}]{Nakano11}%
  \BibitemOpen
  \bibfield  {author} {\bibinfo {author} {\bibfnamefont {R.}~\bibnamefont
  {Nakano}}, \bibinfo {author} {\bibfnamefont {N.}~\bibnamefont {Hatano}}, \
  and\ \bibinfo {author} {\bibfnamefont {T.}~\bibnamefont {Petrosky}},\
  }\bibfield  {title} {\enquote {\bibinfo {title} {Nontrivial eigenvalues of
  the liouvillian of an open quantum system},}\ }\href@noop {} {\bibfield
  {journal} {\bibinfo  {journal} {Int. J. Theor. Phys.}\ }\textbf {\bibinfo
  {volume} {50}},\ \bibinfo {pages} {1134--1142} (\bibinfo {year}
  {2011})}\BibitemShut {NoStop}%
\bibitem [{\citenamefont {Hatano}(2013)}]{Hatano13}%
  \BibitemOpen
  \bibfield  {author} {\bibinfo {author} {\bibfnamefont {N.}~\bibnamefont
  {Hatano}},\ }\bibfield  {title} {\enquote {\bibinfo {title} {Equivalence of
  the effective hamiltonian approach and the siegert boundary condition for
  resonant states},}\ }\href@noop {} {\bibfield  {journal} {\bibinfo  {journal}
  {Fortschr. Phys.}\ }\textbf {\bibinfo {volume} {61}},\ \bibinfo {pages}
  {238--249} (\bibinfo {year} {2013})}\BibitemShut {NoStop}%
\bibitem [{\citenamefont {Ordonez}, \citenamefont {Petrosky},\ and\
  \citenamefont {Prigogine}(2001)}]{Ordonez01}%
  \BibitemOpen
  \bibfield  {author} {\bibinfo {author} {\bibfnamefont {G.}~\bibnamefont
  {Ordonez}}, \bibinfo {author} {\bibfnamefont {T.}~\bibnamefont {Petrosky}}, \
  and\ \bibinfo {author} {\bibfnamefont {I.}~\bibnamefont {Prigogine}},\
  }\bibfield  {title} {\enquote {\bibinfo {title} {Quantum transitions and
  dressed unstable states},}\ }\href@noop {} {\bibfield  {journal} {\bibinfo
  {journal} {Phys. Rev. A}\ }\textbf {\bibinfo {volume} {63}},\ \bibinfo
  {pages} {052106 (23pp)} (\bibinfo {year} {2001})}\BibitemShut {NoStop}%
\bibitem [{\citenamefont {Petrosky}, \citenamefont {Ordonez},\ and\
  \citenamefont {Prigogine}(2001)}]{Petrosky01}%
  \BibitemOpen
  \bibfield  {author} {\bibinfo {author} {\bibfnamefont {T.}~\bibnamefont
  {Petrosky}}, \bibinfo {author} {\bibfnamefont {G.}~\bibnamefont {Ordonez}}, \
  and\ \bibinfo {author} {\bibfnamefont {I.}~\bibnamefont {Prigogine}},\
  }\bibfield  {title} {\enquote {\bibinfo {title} {Space-time formulation of
  quantum transitions},}\ }\href {\doibase 10.1103/PhysRevA.64.062101}
  {\bibfield  {journal} {\bibinfo  {journal} {Phys. Rev. A}\ }\textbf {\bibinfo
  {volume} {64}},\ \bibinfo {pages} {062101} (\bibinfo {year}
  {2001})}\BibitemShut {NoStop}%
\bibitem [{\citenamefont
  {Garc{\'{i}}a-Calder{\'{o}}n}(2010)}]{GarciaCalderon10}%
  \BibitemOpen
  \bibfield  {author} {\bibinfo {author} {\bibfnamefont {G.}~\bibnamefont
  {Garc{\'{i}}a-Calder{\'{o}}n}},\ }\bibfield  {title} {\enquote {\bibinfo
  {title} {Theory of resonant states: an exact analytical approach for open
  quantum systems},}\ }\href@noop {} {\bibfield  {journal} {\bibinfo  {journal}
  {Adv. Quant. Chem.}\ }\textbf {\bibinfo {volume} {60}},\ \bibinfo {pages}
  {407--455} (\bibinfo {year} {2010})}\BibitemShut {NoStop}%
\bibitem [{\citenamefont {{Kim}}(2014)}]{Kim14}%
  \BibitemOpen
  \bibfield  {author} {\bibinfo {author} {\bibfnamefont {S.}~\bibnamefont
  {{Kim}}},\ }\bibfield  {title} {\enquote {\bibinfo {title} {{Complex spectral
  analysis and test function spaces}},}\ }\href@noop {} {\bibfield  {journal}
  {\bibinfo  {journal} {ArXiv e-prints}\ } (\bibinfo {year} {2014})},\ \Eprint
  {http://arxiv.org/abs/1210.6714} {arXiv:1210.6714 [quant-ph]} \BibitemShut
  {NoStop}%
\bibitem [{\citenamefont {Gamow}(1928)}]{Gamow28}%
  \BibitemOpen
  \bibfield  {author} {\bibinfo {author} {\bibfnamefont {G.}~\bibnamefont
  {Gamow}},\ }\bibfield  {title} {\enquote {\bibinfo {title} {Zur
  quantentheorie des atomkernes (on quantum theory of atomic nuclei)},}\
  }\href@noop {} {\bibfield  {journal} {\bibinfo  {journal} {Z. Phys. A}\
  }\textbf {\bibinfo {volume} {51}},\ \bibinfo {pages} {204--212} (\bibinfo
  {year} {1928})}\BibitemShut {NoStop}%
\bibitem [{\citenamefont {Siegert}(1939)}]{Siegert39}%
  \BibitemOpen
  \bibfield  {author} {\bibinfo {author} {\bibfnamefont {A.~J.~F.}\
  \bibnamefont {Siegert}},\ }\bibfield  {title} {\enquote {\bibinfo {title} {On
  the derivation of the dispersion formula for nuclear reactions},}\
  }\href@noop {} {\bibfield  {journal} {\bibinfo  {journal} {Phys. Rev.}\
  }\textbf {\bibinfo {volume} {56}},\ \bibinfo {pages} {750--752} (\bibinfo
  {year} {1939})}\BibitemShut {NoStop}%
\bibitem [{\citenamefont {Peierls}(1959)}]{Peierls59}%
  \BibitemOpen
  \bibfield  {author} {\bibinfo {author} {\bibfnamefont {R.~E.}\ \bibnamefont
  {Peierls}},\ }\bibfield  {title} {\enquote {\bibinfo {title} {Complex
  eigenvalues in scattering theory},}\ }\href@noop {} {\bibfield  {journal}
  {\bibinfo  {journal} {Proc. Roy. Soc. London A}\ }\textbf {\bibinfo {volume}
  {253}},\ \bibinfo {pages} {16--36} (\bibinfo {year} {1959})}\BibitemShut
  {NoStop}%
\bibitem [{\citenamefont {le~Couteur}(1960)}]{leCouteur60}%
  \BibitemOpen
  \bibfield  {author} {\bibinfo {author} {\bibfnamefont {K.~J.}\ \bibnamefont
  {le~Couteur}},\ }\bibfield  {title} {\enquote {\bibinfo {title} {The
  structure of a non-relativistic s-matrix},}\ }\href@noop {} {\bibfield
  {journal} {\bibinfo  {journal} {Proc. Roy. Soc. London A}\ }\textbf {\bibinfo
  {volume} {256}},\ \bibinfo {pages} {115--127} (\bibinfo {year}
  {1960})}\BibitemShut {NoStop}%
\bibitem [{\citenamefont {Zel'dovich}(1960)}]{Zeldovich60}%
  \BibitemOpen
  \bibfield  {author} {\bibinfo {author} {\bibfnamefont {Y.~B.}\ \bibnamefont
  {Zel'dovich}},\ }\bibfield  {title} {\enquote {\bibinfo {title} {On the
  theory of unstable states},}\ }\href@noop {} {\bibfield  {journal} {\bibinfo
  {journal} {Zh. {\`{E}}ksper. Teoret. Fiz.}\ }\textbf {\bibinfo {volume}
  {39}},\ \bibinfo {pages} {776--780} (\bibinfo {year} {1960})},\ \bibinfo
  {note} {[English translation in Sov. Phys. JETP {\textbf{12}}, 542--545
  (1961)]}\BibitemShut {NoStop}%
\bibitem [{\citenamefont {Hokkyo}(1965)}]{Hokkyo65}%
  \BibitemOpen
  \bibfield  {author} {\bibinfo {author} {\bibfnamefont {N.}~\bibnamefont
  {Hokkyo}},\ }\bibfield  {title} {\enquote {\bibinfo {title} {A remark on the
  norm of the unstable state -- a role of adjoint wave functions in
  non-self-adjoint quantum systems --},}\ }\href@noop {} {\bibfield  {journal}
  {\bibinfo  {journal} {Prog. Theor. Phys.}\ }\textbf {\bibinfo {volume}
  {33}},\ \bibinfo {pages} {1116--1128} (\bibinfo {year} {1965})}\BibitemShut
  {NoStop}%
\bibitem [{\citenamefont {Romo}(1968)}]{Romo68}%
  \BibitemOpen
  \bibfield  {author} {\bibinfo {author} {\bibfnamefont {W.~J.}\ \bibnamefont
  {Romo}},\ }\bibfield  {title} {\enquote {\bibinfo {title} {Inner product for
  resonant states and shell-model applications},}\ }\href@noop {} {\bibfield
  {journal} {\bibinfo  {journal} {Nucl. Phys. A}\ }\textbf {\bibinfo {volume}
  {116}},\ \bibinfo {pages} {618--636} (\bibinfo {year} {1968})}\BibitemShut
  {NoStop}%
\bibitem [{\citenamefont {Berggren}(1970)}]{Berggren70}%
  \BibitemOpen
  \bibfield  {author} {\bibinfo {author} {\bibfnamefont {T.}~\bibnamefont
  {Berggren}},\ }\bibfield  {title} {\enquote {\bibinfo {title} {On a
  probabilistic interpretation of expansion coefficients in the
  non-relativistic quantum theory of resonant states},}\ }\href@noop {}
  {\bibfield  {journal} {\bibinfo  {journal} {Phys. Lett.}\ }\textbf {\bibinfo
  {volume} {33B}},\ \bibinfo {pages} {547--549} (\bibinfo {year}
  {1970})}\BibitemShut {NoStop}%
\bibitem [{\citenamefont {Gyarmati}\ and\ \citenamefont
  {Vertse}(1971)}]{Gyarmati71}%
  \BibitemOpen
  \bibfield  {author} {\bibinfo {author} {\bibfnamefont {B.}~\bibnamefont
  {Gyarmati}}\ and\ \bibinfo {author} {\bibfnamefont {T.}~\bibnamefont
  {Vertse}},\ }\bibfield  {title} {\enquote {\bibinfo {title} {On the
  normalization of gamov functions},}\ }\href@noop {} {\bibfield  {journal}
  {\bibinfo  {journal} {Nucl. Phys. A}\ }\textbf {\bibinfo {volume} {160}},\
  \bibinfo {pages} {523--528} (\bibinfo {year} {1971})}\BibitemShut {NoStop}%
\bibitem [{\citenamefont {Landau}\ and\ \citenamefont
  {Lifshitz}(1977)}]{Landau77}%
  \BibitemOpen
  \bibfield  {author} {\bibinfo {author} {\bibfnamefont {L.~D.}\ \bibnamefont
  {Landau}}\ and\ \bibinfo {author} {\bibfnamefont {E.~M.}\ \bibnamefont
  {Lifshitz}},\ }\href@noop {} {\emph {\bibinfo {title} {Quantum Mechanics
  (Non-relativistic Theory), 3rd edition}}}\ (\bibinfo  {publisher} {Pergamon
  Press},\ \bibinfo {address} {Oxford},\ \bibinfo {year} {1977})\ p.\ \bibinfo
  {pages} {{\S}134}\BibitemShut {NoStop}%
\bibitem [{\citenamefont {Romo}(1980)}]{Romo80}%
  \BibitemOpen
  \bibfield  {author} {\bibinfo {author} {\bibfnamefont {W.~J.}\ \bibnamefont
  {Romo}},\ }\bibfield  {title} {\enquote {\bibinfo {title} {A study of the
  completeness properties of resonant states},}\ }\href@noop {} {\bibfield
  {journal} {\bibinfo  {journal} {J. Math. Phys.}\ }\textbf {\bibinfo {volume}
  {21}},\ \bibinfo {pages} {311--326} (\bibinfo {year} {1980})}\BibitemShut
  {NoStop}%
\bibitem [{\citenamefont {Berggren}(1982)}]{Berggren82}%
  \BibitemOpen
  \bibfield  {author} {\bibinfo {author} {\bibfnamefont {T.}~\bibnamefont
  {Berggren}},\ }\bibfield  {title} {\enquote {\bibinfo {title} {Completeness
  relations, mittag-leffeler expansions and the perturbation theory of resonant
  states},}\ }\href@noop {} {\bibfield  {journal} {\bibinfo  {journal} {Nucl.
  Phys. A}\ }\textbf {\bibinfo {volume} {389}},\ \bibinfo {pages} {261--284}
  (\bibinfo {year} {1982})}\BibitemShut {NoStop}%
\bibitem [{\citenamefont {Berggren}(1996)}]{Berggren96}%
  \BibitemOpen
  \bibfield  {author} {\bibinfo {author} {\bibfnamefont {T.}~\bibnamefont
  {Berggren}},\ }\bibfield  {title} {\enquote {\bibinfo {title} {Expectation
  value of an operator in a resonant state},}\ }\href@noop {} {\bibfield
  {journal} {\bibinfo  {journal} {Phys. Lett. B}\ }\textbf {\bibinfo {volume}
  {373}},\ \bibinfo {pages} {1--4} (\bibinfo {year} {1996})}\BibitemShut
  {NoStop}%
\bibitem [{\citenamefont {de~la Madrid}, \citenamefont
  {Garc{\'{i}}a-Calder{\'{o}}n},\ and\ \citenamefont {Muga}(2005)}]{Madrid05}%
  \BibitemOpen
  \bibfield  {author} {\bibinfo {author} {\bibfnamefont {R.}~\bibnamefont
  {de~la Madrid}}, \bibinfo {author} {\bibfnamefont {G.}~\bibnamefont
  {Garc{\'{i}}a-Calder{\'{o}}n}}, \ and\ \bibinfo {author} {\bibfnamefont
  {J.}~\bibnamefont {Muga}},\ }\bibfield  {title} {\enquote {\bibinfo {title}
  {Resonance expansions in quantum mechanics},}\ }\href@noop {} {\bibfield
  {journal} {\bibinfo  {journal} {Czech. J. Phys.}\ }\textbf {\bibinfo {volume}
  {55}},\ \bibinfo {pages} {1141--1150} (\bibinfo {year} {2005})}\BibitemShut
  {NoStop}%
\bibitem [{\citenamefont {Hatano}\ \emph {et~al.}(2008)\citenamefont {Hatano},
  \citenamefont {Sasada}, \citenamefont {Nakamura},\ and\ \citenamefont
  {Petrosky}}]{Hatano08}%
  \BibitemOpen
  \bibfield  {author} {\bibinfo {author} {\bibfnamefont {N.}~\bibnamefont
  {Hatano}}, \bibinfo {author} {\bibfnamefont {K.}~\bibnamefont {Sasada}},
  \bibinfo {author} {\bibfnamefont {H.}~\bibnamefont {Nakamura}}, \ and\
  \bibinfo {author} {\bibfnamefont {T.}~\bibnamefont {Petrosky}},\ }\bibfield
  {title} {\enquote {\bibinfo {title} {Some properties of the resonant state in
  quantum mechanics and its computation},}\ }\href@noop {} {\bibfield
  {journal} {\bibinfo  {journal} {Prog. Theor. Phys.}\ }\textbf {\bibinfo
  {volume} {119}},\ \bibinfo {pages} {187--222} (\bibinfo {year}
  {2008})}\BibitemShut {NoStop}%
\bibitem [{\citenamefont {Hatano}, \citenamefont {Kawamoto},\ and\
  \citenamefont {Feinberg}(2009)}]{Hatano09}%
  \BibitemOpen
  \bibfield  {author} {\bibinfo {author} {\bibfnamefont {N.}~\bibnamefont
  {Hatano}}, \bibinfo {author} {\bibfnamefont {T.}~\bibnamefont {Kawamoto}}, \
  and\ \bibinfo {author} {\bibfnamefont {J.}~\bibnamefont {Feinberg}},\
  }\bibfield  {title} {\enquote {\bibinfo {title} {Probabilistic interpretation
  of resonant states},}\ }\href@noop {} {\bibfield  {journal} {\bibinfo
  {journal} {Pramana J. Phys.}\ }\textbf {\bibinfo {volume} {73}},\ \bibinfo
  {pages} {553--564} (\bibinfo {year} {2009})}\BibitemShut {NoStop}%
\bibitem [{\citenamefont {Goldzak}, \citenamefont {Gilary},\ and\ \citenamefont
  {Moiseyev}(2010)}]{Goldzak10}%
  \BibitemOpen
  \bibfield  {author} {\bibinfo {author} {\bibfnamefont {T.}~\bibnamefont
  {Goldzak}}, \bibinfo {author} {\bibfnamefont {I.}~\bibnamefont {Gilary}}, \
  and\ \bibinfo {author} {\bibfnamefont {N.}~\bibnamefont {Moiseyev}},\
  }\bibfield  {title} {\enquote {\bibinfo {title} {Evaluation of partial widths
  and branching ratios from resonance wave functions},}\ }\href@noop {}
  {\bibfield  {journal} {\bibinfo  {journal} {Phys. Rev. A}\ }\textbf {\bibinfo
  {volume} {82}},\ \bibinfo {pages} {052105 (10pp)} (\bibinfo {year}
  {2010})}\BibitemShut {NoStop}%
\bibitem [{\citenamefont {Sasada}, \citenamefont {Hatano},\ and\ \citenamefont
  {Ordonez}(2011)}]{Sasada11}%
  \BibitemOpen
  \bibfield  {author} {\bibinfo {author} {\bibfnamefont {K.}~\bibnamefont
  {Sasada}}, \bibinfo {author} {\bibfnamefont {N.}~\bibnamefont {Hatano}}, \
  and\ \bibinfo {author} {\bibfnamefont {G.}~\bibnamefont {Ordonez}},\
  }\bibfield  {title} {\enquote {\bibinfo {title} {Resonant spectrum analysis
  of the conductance of an open quantum system and three types of fano
  parameter},}\ }\href@noop {} {\bibfield  {journal} {\bibinfo  {journal} {J.
  Phys. Soc. Jpn.}\ }\textbf {\bibinfo {volume} {80}},\ \bibinfo {pages}
  {104707 (27pp)} (\bibinfo {year} {2011})}\BibitemShut {NoStop}%
\bibitem [{\citenamefont {Moiseyev}(2011)}]{Moiseyev11}%
  \BibitemOpen
  \bibfield  {author} {\bibinfo {author} {\bibfnamefont {N.}~\bibnamefont
  {Moiseyev}},\ }\href@noop {} {\emph {\bibinfo {title} {Non-Hermitian Quantum
  Mechanics}}}\ (\bibinfo  {publisher} {Cambridge University Press},\ \bibinfo
  {address} {Cambrdige},\ \bibinfo {year} {2011})\BibitemShut {NoStop}%
\bibitem [{\citenamefont {Fisher}\ and\ \citenamefont {Lee}(1981)}]{Fisher81}%
  \BibitemOpen
  \bibfield  {author} {\bibinfo {author} {\bibfnamefont {D.~S.}\ \bibnamefont
  {Fisher}}\ and\ \bibinfo {author} {\bibfnamefont {P.~A.}\ \bibnamefont
  {Lee}},\ }\bibfield  {title} {\enquote {\bibinfo {title} {Relation between
  conductivity and transmission matrix},}\ }\href@noop {} {\bibfield  {journal}
  {\bibinfo  {journal} {Phys. Rev. B}\ }\textbf {\bibinfo {volume} {23}},\
  \bibinfo {pages} {6851--6854} (\bibinfo {year} {1981})}\BibitemShut {NoStop}%
\bibitem [{\citenamefont {Datta}(1995)}]{Datta95}%
  \BibitemOpen
  \bibfield  {author} {\bibinfo {author} {\bibfnamefont {S.}~\bibnamefont
  {Datta}},\ }\href@noop {} {\emph {\bibinfo {title} {Electronic Transport in
  Mesoscopic Systems}}}\ (\bibinfo  {publisher} {Cambridge University Press},\
  \bibinfo {address} {Cambridge},\ \bibinfo {year} {1995})\BibitemShut
  {NoStop}%
\bibitem [{\citenamefont {Landauer}(1957)}]{Landauer57}%
  \BibitemOpen
  \bibfield  {author} {\bibinfo {author} {\bibfnamefont {R.}~\bibnamefont
  {Landauer}},\ }\bibfield  {title} {\enquote {\bibinfo {title} {Spatial
  variation of currents and fields due to localized scatterers in metallic
  conduction},}\ }\href@noop {} {\bibfield  {journal} {\bibinfo  {journal} {IBM
  J. Res. Dev.}\ }\textbf {\bibinfo {volume} {1}},\ \bibinfo {pages} {223--231}
  (\bibinfo {year} {1957})}\BibitemShut {NoStop}%
\bibitem [{\citenamefont {Newton}(1960)}]{Newton60}%
  \BibitemOpen
  \bibfield  {author} {\bibinfo {author} {\bibfnamefont {R.~G.}\ \bibnamefont
  {Newton}},\ }\bibfield  {title} {\enquote {\bibinfo {title} {Analytic
  properties of radial wave functions},}\ }\href@noop {} {\bibfield  {journal}
  {\bibinfo  {journal} {J. Math. Phys.}\ }\textbf {\bibinfo {volume} {1}},\
  \bibinfo {pages} {319--347} (\bibinfo {year} {1960})}\BibitemShut {NoStop}%
\bibitem [{\citenamefont {Newton}(1982)}]{Newton82}%
  \BibitemOpen
  \bibfield  {author} {\bibinfo {author} {\bibfnamefont {R.~G.}\ \bibnamefont
  {Newton}},\ }\href@noop {} {\emph {\bibinfo {title} {Scattering Theory of
  Waves and Particles, 2nd edition}}}\ (\bibinfo  {publisher}
  {Springer-Verlag},\ \bibinfo {address} {New York},\ \bibinfo {year} {1982})\
  Chap.\ \bibinfo {chapter} {Sec. 12.1.5}\BibitemShut {NoStop}%
\bibitem [{\citenamefont {Berggren}(1968)}]{Berggren68}%
  \BibitemOpen
  \bibfield  {author} {\bibinfo {author} {\bibfnamefont {T.}~\bibnamefont
  {Berggren}},\ }\bibfield  {title} {\enquote {\bibinfo {title} {On the use of
  resonant states in eigenfunction expansions of scattering and reaction
  amplitudes},}\ }\href@noop {} {\bibfield  {journal} {\bibinfo  {journal}
  {Nucl. Phys.}\ }\textbf {\bibinfo {volume} {A109}},\ \bibinfo {pages}
  {265--287} (\bibinfo {year} {1968})}\BibitemShut {NoStop}%
\bibitem [{\citenamefont {Albeverio}\ \emph {et~al.}(1996)\citenamefont
  {Albeverio}, \citenamefont {Haake}, \citenamefont {Kurasov}, \citenamefont
  {Ku{\'{s}}},\ and\ \citenamefont {{\v{S}}eba}}]{Albeverio96}%
  \BibitemOpen
  \bibfield  {author} {\bibinfo {author} {\bibfnamefont {S.}~\bibnamefont
  {Albeverio}}, \bibinfo {author} {\bibfnamefont {F.}~\bibnamefont {Haake}},
  \bibinfo {author} {\bibfnamefont {P.}~\bibnamefont {Kurasov}}, \bibinfo
  {author} {\bibfnamefont {M.}~\bibnamefont {Ku{\'{s}}}}, \ and\ \bibinfo
  {author} {\bibfnamefont {P.}~\bibnamefont {{\v{S}}eba}},\ }\bibfield  {title}
  {\enquote {\bibinfo {title} {S-matrix, resonances, and wave functions for
  transport through billiards with leads},}\ }\href@noop {} {\bibfield
  {journal} {\bibinfo  {journal} {J. Math. Phys.}\ }\textbf {\bibinfo {volume}
  {37}},\ \bibinfo {pages} {4888--4903} (\bibinfo {year} {1996})}\BibitemShut
  {NoStop}%
\bibitem [{\citenamefont {Fyodorov}\ and\ \citenamefont
  {Sommers}(1997)}]{Fyodorov97}%
  \BibitemOpen
  \bibfield  {author} {\bibinfo {author} {\bibfnamefont {Y.~V.}\ \bibnamefont
  {Fyodorov}}\ and\ \bibinfo {author} {\bibfnamefont {H.-J.}\ \bibnamefont
  {Sommers}},\ }\bibfield  {title} {\enquote {\bibinfo {title} {Statistics of
  resonance poles, phase shifts and time delays in quantum chaotic scattering:
  Random matrix approach for systems with broken time-reversal invariance},}\
  }\href@noop {} {\bibfield  {journal} {\bibinfo  {journal} {J. Math. Phys.}\
  }\textbf {\bibinfo {volume} {38}},\ \bibinfo {pages} {1918--1981} (\bibinfo
  {year} {1997})}\BibitemShut {NoStop}%
\bibitem [{\citenamefont {Dittes}(2000)}]{Dittes00}%
  \BibitemOpen
  \bibfield  {author} {\bibinfo {author} {\bibfnamefont {F.-M.}\ \bibnamefont
  {Dittes}},\ }\bibfield  {title} {\enquote {\bibinfo {title} {The decay of
  quantum systems with a small number of open channels},}\ }\href@noop {}
  {\bibfield  {journal} {\bibinfo  {journal} {Phys. Rep.}\ }\textbf {\bibinfo
  {volume} {339}},\ \bibinfo {pages} {215--316} (\bibinfo {year}
  {2000})}\BibitemShut {NoStop}%
\bibitem [{\citenamefont {Pichugin}, \citenamefont {Schanz},\ and\
  \citenamefont {{\v{S}}eba}(2001)}]{Pichugin01}%
  \BibitemOpen
  \bibfield  {author} {\bibinfo {author} {\bibfnamefont {K.}~\bibnamefont
  {Pichugin}}, \bibinfo {author} {\bibfnamefont {H.}~\bibnamefont {Schanz}}, \
  and\ \bibinfo {author} {\bibfnamefont {P.}~\bibnamefont {{\v{S}}eba}},\
  }\bibfield  {title} {\enquote {\bibinfo {title} {Effective coupling for open
  billiards},}\ }\href@noop {} {\bibfield  {journal} {\bibinfo  {journal}
  {Phys. Rev. E}\ }\textbf {\bibinfo {volume} {64}},\ \bibinfo {pages} {056227
  (7pp)} (\bibinfo {year} {2001})}\BibitemShut {NoStop}%
\bibitem [{\citenamefont {Sadreev}\ and\ \citenamefont
  {Rotter}(2003)}]{Sadreev03}%
  \BibitemOpen
  \bibfield  {author} {\bibinfo {author} {\bibfnamefont {A.~F.}\ \bibnamefont
  {Sadreev}}\ and\ \bibinfo {author} {\bibfnamefont {I.}~\bibnamefont
  {Rotter}},\ }\bibfield  {title} {\enquote {\bibinfo {title} {S-matrix theory
  for transmission through billiards in tight-binding approach},}\ }\href@noop
  {} {\bibfield  {journal} {\bibinfo  {journal} {J. Phys. A: Math. Gen.}\
  }\textbf {\bibinfo {volume} {36}},\ \bibinfo {pages} {11413--11433} (\bibinfo
  {year} {2003})}\BibitemShut {NoStop}%
\bibitem [{\citenamefont {Oko{\l}owicz}, \citenamefont {P{\l}oszajczak},\ and\
  \citenamefont {Rotter}(2003)}]{Okolowicz03}%
  \BibitemOpen
  \bibfield  {author} {\bibinfo {author} {\bibfnamefont {J.}~\bibnamefont
  {Oko{\l}owicz}}, \bibinfo {author} {\bibfnamefont {M.}~\bibnamefont
  {P{\l}oszajczak}}, \ and\ \bibinfo {author} {\bibfnamefont {I.}~\bibnamefont
  {Rotter}},\ }\bibfield  {title} {\enquote {\bibinfo {title} {Dynamics of
  quantum systems embedded in a continuum},}\ }\href@noop {} {\bibfield
  {journal} {\bibinfo  {journal} {Phys. Rep.}\ }\textbf {\bibinfo {volume}
  {374}},\ \bibinfo {pages} {271--383} (\bibinfo {year} {2003})}\BibitemShut
  {NoStop}%
\bibitem [{\citenamefont {Kunz}\ and\ \citenamefont {Shapiro}(2006)}]{Kunz06}%
  \BibitemOpen
  \bibfield  {author} {\bibinfo {author} {\bibfnamefont {H.}~\bibnamefont
  {Kunz}}\ and\ \bibinfo {author} {\bibfnamefont {B.}~\bibnamefont {Shapiro}},\
  }\bibfield  {title} {\enquote {\bibinfo {title} {Resonances in a
  one-dimensional disordered chain},}\ }\href@noop {} {\bibfield  {journal}
  {\bibinfo  {journal} {J. Phys. A: Math. Gen.}\ }\textbf {\bibinfo {volume}
  {39}},\ \bibinfo {pages} {10155--10160} (\bibinfo {year} {2006})}\BibitemShut
  {NoStop}%
\bibitem [{\citenamefont {Kunz}\ and\ \citenamefont {Shapiro}(2008)}]{Kunz08}%
  \BibitemOpen
  \bibfield  {author} {\bibinfo {author} {\bibfnamefont {H.}~\bibnamefont
  {Kunz}}\ and\ \bibinfo {author} {\bibfnamefont {B.}~\bibnamefont {Shapiro}},\
  }\bibfield  {title} {\enquote {\bibinfo {title} {Statistics of resonances in
  a semi-infinite disordered chain},}\ }\href@noop {} {\bibfield  {journal}
  {\bibinfo  {journal} {Phys. Rev. B}\ }\textbf {\bibinfo {volume} {77}},\
  \bibinfo {pages} {054203 (6pp)} (\bibinfo {year} {2008})}\BibitemShut
  {NoStop}%
\bibitem [{\citenamefont {Sasada}\ and\ \citenamefont
  {Hatano}(2008)}]{Sasada08}%
  \BibitemOpen
  \bibfield  {author} {\bibinfo {author} {\bibfnamefont {K.}~\bibnamefont
  {Sasada}}\ and\ \bibinfo {author} {\bibfnamefont {N.}~\bibnamefont
  {Hatano}},\ }\bibfield  {title} {\enquote {\bibinfo {title} {Calculation of
  the self-energy of open quantum systems},}\ }\href@noop {} {\bibfield
  {journal} {\bibinfo  {journal} {J. Phys. Soc. Jpn.}\ }\textbf {\bibinfo
  {volume} {77}},\ \bibinfo {pages} {025003 (2pp)} (\bibinfo {year}
  {2008})}\BibitemShut {NoStop}%
\bibitem [{\citenamefont {Tisseur}\ and\ \citenamefont
  {meerberge}(2001)}]{Tisseur01}%
  \BibitemOpen
  \bibfield  {author} {\bibinfo {author} {\bibfnamefont {F.}~\bibnamefont
  {Tisseur}}\ and\ \bibinfo {author} {\bibfnamefont {K.}~\bibnamefont
  {meerberge}},\ }\bibfield  {title} {\enquote {\bibinfo {title} {The quadratic
  eigenvalue problem},}\ }\href@noop {} {\bibfield  {journal} {\bibinfo
  {journal} {SIAM Review}\ }\textbf {\bibinfo {volume} {43}},\ \bibinfo {pages}
  {235--286} (\bibinfo {year} {2001})}\BibitemShut {NoStop}%
\bibitem [{\citenamefont {Khalfin}(1957)}]{Khalfin57}%
  \BibitemOpen
  \bibfield  {author} {\bibinfo {author} {\bibfnamefont {L.~A.}\ \bibnamefont
  {Khalfin}},\ }\bibfield  {title} {\enquote {\bibinfo {title} {Ccontribution
  to the decay theory of a quasi-stationary state (in russian)},}\ }\href@noop
  {} {\bibfield  {journal} {\bibinfo  {journal} {Zh. {\`{E}}ksper. Teoret.
  Fiz.}\ }\textbf {\bibinfo {volume} {33}},\ \bibinfo {pages} {1371--1382}
  (\bibinfo {year} {1957})},\ \bibinfo {note} {[English translation in Sov.
  Phys. JETP, 6, 1053--1063 (1958)]}\BibitemShut {NoStop}%
\bibitem [{\citenamefont {Garmon}\ \emph {et~al.}(2013)\citenamefont {Garmon},
  \citenamefont {Petrosky}, \citenamefont {Simine},\ and\ \citenamefont
  {Segal}}]{Garmon13}%
  \BibitemOpen
  \bibfield  {author} {\bibinfo {author} {\bibfnamefont {S.}~\bibnamefont
  {Garmon}}, \bibinfo {author} {\bibfnamefont {T.}~\bibnamefont {Petrosky}},
  \bibinfo {author} {\bibfnamefont {L.}~\bibnamefont {Simine}}, \ and\ \bibinfo
  {author} {\bibfnamefont {D.}~\bibnamefont {Segal}},\ }\bibfield  {title}
  {\enquote {\bibinfo {title} {Amplification of non-markovian decay due to
  bound state absorption into continuum},}\ }\href@noop {} {\bibfield
  {journal} {\bibinfo  {journal} {Fortcshr. Phys.}\ }\textbf {\bibinfo {volume}
  {61}},\ \bibinfo {pages} {261--275} (\bibinfo {year} {2013})}\BibitemShut
  {NoStop}%
\bibitem [{\citenamefont {Peierls}(1979)}]{Peierls79}%
  \BibitemOpen
  \bibfield  {author} {\bibinfo {author} {\bibfnamefont {R.}~\bibnamefont
  {Peierls}},\ }\href@noop {} {\emph {\bibinfo {title} {Surprises in
  Theoretical Physics}}}\ (\bibinfo  {publisher} {Princeton University Press},\
  \bibinfo {address} {Princeton},\ \bibinfo {year} {1979})\ p.\ \bibinfo
  {pages} {{\S}3.8}\BibitemShut {NoStop}%
\bibitem [{\citenamefont {Price}(2006)}]{Price06}%
  \BibitemOpen
  \bibfield  {author} {\bibinfo {author} {\bibfnamefont {H.}~\bibnamefont
  {Price}},\ }\bibfield  {title} {\enquote {\bibinfo {title} {Recent work on
  the arrow or radiation},}\ }\href@noop {} {\bibfield  {journal} {\bibinfo
  {journal} {Studies in History and Philosophy of Modern Physics}\ }\textbf
  {\bibinfo {volume} {37}},\ \bibinfo {pages} {498--527} (\bibinfo {year}
  {2006})}\BibitemShut {NoStop}%
\bibitem [{\citenamefont {Ordonez}\ and\ \citenamefont {Hatano}()}]{Ordonez14}%
  \BibitemOpen
  \bibfield  {author} {\bibinfo {author} {\bibfnamefont {G.}~\bibnamefont
  {Ordonez}}\ and\ \bibinfo {author} {\bibfnamefont {N.}~\bibnamefont
  {Hatano}},\ }\href@noop {} {\bibinfo  {journal} {unpublished}\ }\BibitemShut
  {NoStop}%
\end{thebibliography}%

%
%
%
%
%
%
%
%
%
%
%

\end{document}